\documentclass[aps, prb, twocolumn,amsmath,amssymb,floatfix,reprint,longbibliography]{revtex4-2}
\usepackage{graphicx}
\usepackage{physics}
\usepackage{times}
\usepackage{dcolumn}
\usepackage{hyperref} 
\hypersetup{colorlinks, allcolors=blue}
\allowdisplaybreaks

\begin{document}

\title{Many-body effects on superconductivity mediated by double-magnon processes in altermagnets} 

\author{Kristian M{\ae}land} 
\affiliation{\mbox{Center for Quantum Spintronics, Department of Physics, Norwegian University of Science and Technology, NO-7491 Trondheim, Norway}} 
\author{Bj{\o}rnulf Brekke} 
\affiliation{\mbox{Center for Quantum Spintronics, Department of Physics, Norwegian University of Science and Technology, NO-7491 Trondheim, Norway}} 
\author{Asle Sudb{\o}}
\email[Corresponding author: ]{asle.sudbo@ntnu.no}
\affiliation{\mbox{Center for Quantum Spintronics, Department of Physics, Norwegian University of Science and Technology, NO-7491 Trondheim, Norway}}

\begin{abstract}
Altermagnets exhibit a large electron spin splitting which can be understood as a result of strong coupling between itinerant electrons and localized spins. We consider superconductivity due to electron-magnon scattering, using strong-coupling Eliashberg theory to capture many-body effects that are not covered by a weak-coupling approach. The characteristic band structure of altermagnets puts significant constraints on the spin structure of electron scattering on the Fermi surface. We emphasize the role of spin-preserving, double-magnon scattering processes compared to conventional spin-flip processes involving a single magnon. Then, we derive the Eliashberg equations for a situation where double-magnon scattering mediates spin-polarized Cooper pairs, while both double-magnon and single-magnon scatterings contribute to many-body effects. These many-body effects impact superconducting properties in a way that differs significantly from systems where conventional spin-flip processes mediate superconductivity. To highlight the role of $d$-wave magnetism on superconductivity in altermagnets, we compare our results to those found in ferromagnetic half-metals and conventional antiferromagnetic metals.
\end{abstract}

\maketitle 

\section{Introduction}

The study of spin dynamics in systems with coupled itinerant electrons and localized spins represents a fertile ground for exploring the fundamental physics of magnetism and electron transport. Beyond fundamental research, the insights gained from spin dynamics hold immense promise for technological innovations in spin-based electronics and quantum information processing \cite{Brataas2020MagnonCurrent, Chumak2015MagnonCurrent}.
The spins of itinerant electrons and localized spins in magnetic materials interact through an $sd$ coupling $J_{\text{sd}}$ locally. This coupling allows for controlling the magnetic order through effects such as spin pumping \cite{TserkovnyakPRL2002} and spin transfer torques \cite{RalphJMM2008}. These effects facilitate interconversion of charge or spin currents \cite{Han2020MagnonCurrent} carried by electrons and spin currents carried by magnons, offering low-dissipation transport of information \cite{Brataas2020MagnonCurrent, Chumak2015MagnonCurrent}.

The effect of the $sd$ coupling depends strongly on the magnetic crystal hosting the interaction. 
Ferromagnets (FMs) are uncompensated magnets with overall spin-split electron bands. A simple magnetic structure with a single magnetic sublattice captures these properties. 
In contrast, conventional antiferromagnets (AFMs) typically have a two-sublattice magnetic structure with compensated localized spins. As a consequence, the magnetic properties allow for ultrafast spin dynamics in the absence of magnetic stray fields \cite{Baltz2018YaroslavAFMrev, Fiebig2008AFMexpUltrafast}. However, the $sd$ coupling $J_{\text{sd}}$ does not give rise to spin-split electron bands. The spin-split bands are prohibited by structural symmetry evident from $\mathcal{PT}$ or $\mathcal{T}t$ symmetry, where $\mathcal{P}$, $\mathcal{T}$ and $t$ are the inversion, time-reversal, and translation operators, respectively.

More recently, a new class of colinear magnets called altermagnets (AMs) has received attention \cite{Mazin2022PRX, Smejkal2022Sep, Smejkal2022Dec}. AMs are compensated magnets for which the structural symmetry does not prohibit spin-splitting. Instead, electron bands and magnon bands show a rotational symmetry of $d$-, $g$- or $i$-wave characters. Consequently, AMs exhibit properties in common with both FMs and AFMs, in addition to unique properties absent for FMs and AFMs.
The character of the spin-splitting of the electron bands renders AMs intriguing candidates for superconducting devices \cite{mazin2022notes, Smejkal2022Dec, OuassouPRL2023, BeenakkerPRB2023, Sun2023AndreevAM, PapajPRB2023, giil2023superconductor, Ghorashi2023AMTSC, Zhang2024FFLOAM, ZhuPRB2023, Chakraborty2023AMSCd, Bose2024AMSC}.

The effect of the structural asymmetry in AMs is strong. Altermagnetic materials such as RuO$_2$ \cite{fedchenko2024observation, Lin2024RuO2}, MnTe \cite{LeePRL2024, osumi2023observation, krempaskyNature2024} and CrSb \cite{reimers2023direct} exhibit a spin-splitting with the same order of magnitude as in FMs. The spin splitting is in the range $100$~meV -- $1$~eV, and is a direct reflection of the magnitude of the $sd$ coupling $J_{\text{sd}}$, instead of relativistic spin-orbit coupling (SOC) \cite{Smejkal2022Sep}.

Minimal models of AMs have been introduced in Refs.~\cite{Brekke2023Aug, Maier2023Sep, Roig2024Feb}.
Reference \cite{Brekke2023Aug} introduces a microscopic model for two-dimensional (2D) AMs defined on a Lieb lattice \cite{Lieb1989Mar}. The model separates localized spins and itinerant electrons as distinct degrees of freedom in an $sd$ model.
$sd$ models assume that the localized spins originate with occupied electron states far from the Fermi surface (FS).
The ordering of the localized spins yields itinerant electron bands with a sizable $d$-wave spin splitting. The model is applied to consider superconductivity mediated by quantum fluctuations of the localized spins within a weak-coupling Bardeen-Cooper-Schrieffer (BCS) approach \cite{BCS}.

Replacing phonons with magnons in search of unconventional superconductivity has a long history in the context of heavy-fermion superconductivity and high-$T_c$ superconductivity in the copper oxides \cite{Pines1993,SCspinHistory_Scalapino1999, SCAFMspinMoriya2003Jul, SCspinGapSym_Hirschfeld2011}. More recently, electron-magnon coupling (EMC) has been invoked in the search for unconventional superconductivity in heterostructures 
involving FMs \cite{KargarianFMTI, ArneFMNM, brekke2023interfacial, Hugdal2020FMTI, Hugdal2018May, Erlandsen2020TIAFM}, AFMs \cite{Hugdal2018May, Erlandsen2020TIAFM, EirikAFMNM, Fjaerbu2019arneAFMNM, ThingstadEliashberg, Sun2023Aug}, and noncolinear magnets \cite{Maeland2023PRL, Maeland2023Dec, Bostrom2023Dec}. The latter efforts have led to a wide range of predicted superconducting phases, including topological superconductivity. These previous studies focused on effective electron-electron interactions \textit{ mediated by the exchange of a single boson (magnon)}. This is also the standard approach for phonon-mediated superconductivity \cite{BCS, SFsuperconductivity}. However, the combination of $d$-wave spin splitting of the electron bands and the colinearity of the magnetic state, means that an exchange of a single magnon between two electrons with opposite momenta is energetically suppressed in an AM. Instead, Ref.~\cite{Brekke2023Aug} considered superconductivity arising from effective electron-electron interactions mediated by two magnons. We refer to these as double-magnon processes, and they do not flip the spin of the electrons. This led to spin-triplet $p$-wave superconductivity with spin-polarized Cooper pairs. By tuning the chemical potential, a large enhancement of the coupling may occur due to a combination of a large density of states (DOS) and suppression of interference effects between the two magnetic sublattices.

The large magnitude of $J_{\text{sd}}$ in AMs combined with the prediction of enhanced coupling strength motivates a strong-coupling approach to superconductivity. Eliashberg theory is a strong-coupling theory of superconductivity that treats the Green's function of the superconductor. This includes many-body effects on the normal state due to the same interactions that cause the superconducting pairing \cite{Eliashberg1960Sep, Eliashberg1961, Schrieffer1964book, Carbotte1990FreeEnergy, EliashbergRevMarsiglio2020, Chubukov2020FEspecialized}. We use anomalous Green's functions to capture the superconducting pairing caused by double-magnon processes, while regular Green's functions capture renormalization effects caused by both single- and double-magnon processes.

The coexistence of superconductivity and magnetic order has also been discussed in bulk systems where the same electrons are responsible for both the magnetic order and the superconducting order. In other words, there are no localized spins, as in the $sd$ model. This has been explored theoretically \cite{ginzburg1957FMSC, Fay1980FMSC, Linder2007FMSC, Jian2009FMSC} and observed experimentally \cite{Saxena2000FMSC, Huy2007FMSCexp} for ferromagnetic order. 
Spin-triplet $p$-wave superconductivity is considered the most likely candidate \cite{Fay1980FMSC, Jian2009FMSC, Linder2007FMSC, Saxena2000FMSC, Huy2007FMSCexp}.
Furthermore, several studies \cite{Schrieffer1989AFMSC, Frenkel1990AFMSC, Capone2006AFMSC, Ismer2010AFMSC, Romer2016AFMSC} have considered the coexistence of AFM order and superconductivity. 
There, spin-singlet $d$-wave pairing is the most likely candidate for the superconducting order parameter \cite{Romer2016AFMSC}. 
Antiferromagnetic spin fluctuations have also been proposed as the mechanism behind $d$-wave superconductivity, in close proximity to an antiferromagnetic phase, in heavy-fermion superconductors and other high-$T_c$ superconductors \cite{Pines1993,SCspinHistory_Scalapino1999, SCAFMspinMoriya2003Jul, SCspinGapSym_Hirschfeld2011}. Additionally, the coexistence of superconductivity and AM order has been discussed \cite{mazin2022notes, Smejkal2022Dec}. 
Alternatively, an $sd$ model for a bulk AFM would essentially be the same as the model studied for AFM insulator/normal metal (NM) interfaces \cite{Fjaerbu2019arneAFMNM, EirikAFMNM, ThingstadEliashberg, Sun2023Aug}.
In contrast to the interfacial models, the itinerant electrons should always couple equally to the sublattices in bulk AFM metal models. In that case, spin-singlet $d$-wave superconductivity is expected \cite{ThingstadEliashberg, Sun2023Aug}.

In this paper, we study superconductivity mediated by magnons in colinear magnets, treating localized spins and itinerant electrons as separate degrees of freedom. We focus on strong-coupling superconductivity in AMs and compare to FMs and AFMs. 
Section \ref{sec:EMCcolinear} introduces the concept of single- and double-magnon EMCs and compares their contributions to magnon-mediated superconductivity. 
In Sec.~\ref{sec:Eliashberg}, we derive the Eliashberg equations necessary to study superconductivity in a spin-split system where double-magnon processes induce spin-polarized Cooper pairing. 
Section \ref{sec:AM} extends the model in Ref.~\cite{Brekke2023Aug} for a 2D AM and compares BCS to Eliashberg predictions of the critical temperature for superconductivity.
In Sec.~\ref{sec:FM}, we consider the formation of spin-triplet $p$-wave superconductivity in a ferromagnetic half-metal in an $sd$ model. We relegate additional details of the derivations to Appendixes.
Units $\hbar = k_B = a = 1$ are used throughout, where $a$ is the distance between nearest neighbors on the lattice.

\section{Electron-magnon coupling in colinear magnets} \label{sec:EMCcolinear}
The interaction between itinerant electrons and localized spins is captured by the exchange coupling \cite{Berger1996Magnondrag, Takahashi2010MagnonCurrent, Zhang2012MagnonCurrent, KargarianFMTI, Hugdal2018May, Erlandsen2020TIAFM, ArneFMNM, Fjaerbu2019arneAFMNM,EirikAFMNM, ThingstadEliashberg, brekke2023interfacial, Brekke2023Aug, Maeland2023PRL, Sun2023Aug, Maeland2023Dec, Bostrom2023Dec, Maeland2021Sep, SkTopoSCNagaosa, ExpSkHeterostructure, ExpInterfaceExchange},
\begin{equation}
    H_{\text{sd}} = -2J_{\text{sd}} \sum_i \boldsymbol{S}_i \cdot \boldsymbol{s}_i. \label{eq:Jsd}
\end{equation}
Here, $\boldsymbol{S}_i$ is the localized spin on site $i$, and $\boldsymbol{s}_i = c_{i,\sigma}^{\dagger}\boldsymbol{\sigma}_{\sigma \sigma'} c_{i,\sigma'}/2$ is the spin of an itinerant electron on site $i$. $c_{i,\sigma}^{(\dagger)}$ destroys (creates) and electron at site $i$ with spin $\sigma$, and $\boldsymbol{\sigma}$ is a vector of Pauli matrices. Experiments have observed signatures of such spin exchange coupling across interfaces \cite{ExpInterfaceExchange, ExpSkHeterostructure, Li2016MagnonDragExp, Wu2016MagnonDragExp}. 
The interfacial coupling is expected to be weaker than an onsite coupling in a bulk material \cite{ArneFMNM, Fjaerbu2019arneAFMNM,brekke2023interfacial, Maeland2021Sep, Garate2010Jsdbulk, ExpStrongJbar, Liu2009strongJbarPRL, Brekke2023Aug}.

For a system with an ordered magnetic state arising from localized magnetic moments, we use the Holstein-Primakoff (HP) transformation \cite{Holstein1940Dec} to study spin fluctuations around the classical ground state. The HP transformation expresses spin operators in terms of creation and destruction operators for quantized spin waves, i.e., magnons. Truncating at second order in magnon operators, we approximate $S_{i,x} \approx \sqrt{S/2} (a_i+a_i^\dagger) $, $S_{i,y} \approx i \sqrt{S/2}(a_i^\dagger-a_i)$, and $S_{i,z} = S-a_i^\dagger a_i$. Here, $S$ is the spin quantum number, $a_i^{(\dagger)}$ destroys (creates) a magnon at lattice site $i$, and the prefactor in $S_{i,y}$ is the imaginary unit. If the magnetic state has more than one sublattice, separate magnon modes must be introduced for each sublattice. The HP transformation is technically an expansion in $1/S$, but as long as there are few magnons in the system, i.e., the temperature is low compared to the magnon gap, it should be a good approximation even when $S$ is not large \cite{Holstein1940Dec, Maeland2021Sep, Maeland2022QSk}.

The HP transformation applied to Eq.~\eqref{eq:Jsd} yields two types of terms. The first type is an EMC with a single magnon, where the finite spin carried by the magnon flips the electron's spin. These are terms of the type $H_{\text{em}}^{(1)} = -J_{\text{sd}}\sqrt{2S}\sum_i a_i^\dagger c_{i,\uparrow}^\dagger c_{i,\downarrow} = -J_{\text{sd}}\sqrt{2S/N} \sum_{\boldsymbol{k}\boldsymbol{k}'} a_{\boldsymbol{k}-\boldsymbol{k}'}^\dagger c_{\boldsymbol{k}',\uparrow}^\dagger c_{\boldsymbol{k},\downarrow}$. The last term has been Fourier transformed to momentum space, with momentum sums restricted to the first Brillouin zone (1BZ) of the lattice under consideration, and $N$ is the total number of unit cells. The Fourier transform (FT) is defined as $c_{i\sigma} =  \sum_{\boldsymbol{k}} c_{\boldsymbol{k}\sigma} e^{i\boldsymbol{k}\cdot\boldsymbol{r}_i} /\sqrt{N}$ for electrons and $a_{i} =  \sum_{\boldsymbol{q}} a_{\boldsymbol{q}} e^{i\boldsymbol{q}\cdot\boldsymbol{r}_i} /\sqrt{N}$ for magnons, where $\boldsymbol{r}_i$ is the location of lattice site $i$. Given a colinear magnetic configuration, we assume the electrons have the same quantization axis for spin as the localized spins.

\begin{figure}
    \centering
    \includegraphics[width = 0.8\linewidth]{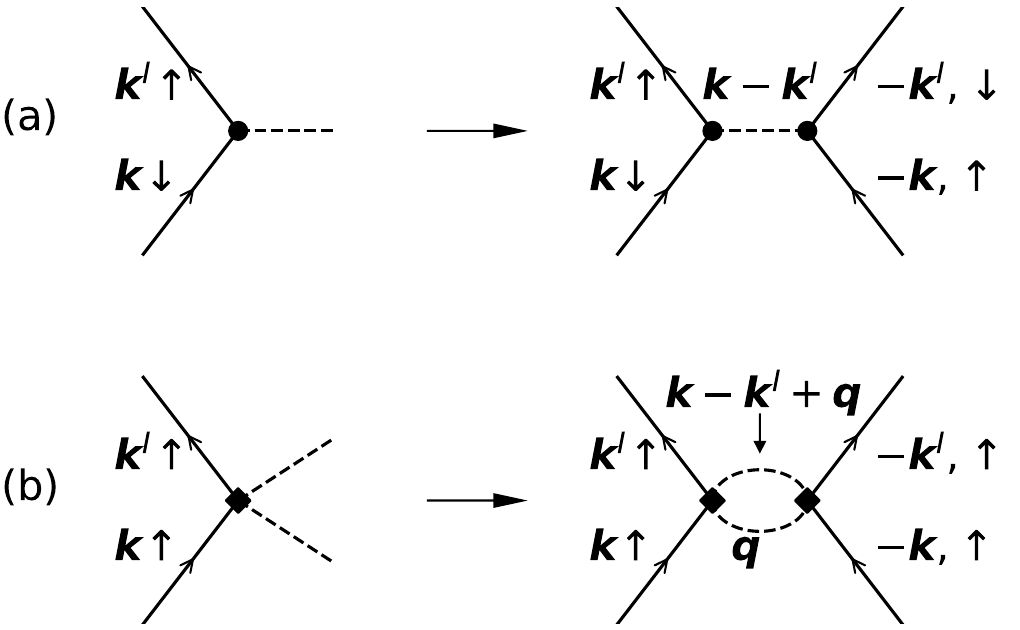}
    \caption{(a) Feynman diagram for single-magnon electron-magnon coupling (EMC). Solid lines are electrons, and dashed lines are magnons. When arising from a colinear magnetic state, the magnon flips the electron's spin. An effective electron-electron interaction is derived by connecting two such diagrams. Note that the two incoming electrons must have opposite spin. (b) Feynman diagram for double-magnon EMC, where two magnon lines connect to the same interaction vertex. In this case, the electron's spin is not flipped, permitting effective electron-electron interactions that can generate spin-polarized Cooper pairs. Here, spin up is chosen as an example. Note the free momentum $\boldsymbol{q}$ in the double-magnon-mediated electron-electron interaction.}
    \label{fig:FeynmanEMC}
\end{figure}

There are also terms involving two magnon operators. They are captured by $H_{\text{em}}^{(2)} = J_{\text{sd}}\sum_i a_i^\dagger a_i c_{i,\uparrow}^\dagger c_{i,\uparrow} = (J_{\text{sd}}/N) \sum_{\boldsymbol{k}\boldsymbol{k}'\boldsymbol{q}} a_{\boldsymbol{k}-\boldsymbol{k}'+\boldsymbol{q}}^\dagger a_{\boldsymbol{q}} c_{\boldsymbol{k}',\uparrow}^\dagger c_{\boldsymbol{k},\uparrow} $. We denote these as double-magnon processes, in contrast to single-magnon processes that contain a single magnon operator. 
Double-magnon processes originate with $S_{i,z} = S-a_i^\dagger a_i$, and do not flip the electron spin. Also, note that the interaction strength for double-magnon processes does not depend on the spin quantum number $S$. The momentum space versions of these two types of EMC are illustrated with Feynman diagrams in Fig.~\ref{fig:FeynmanEMC}, along with possible electron-electron interactions they could mediate. We will consider magnon-mediated superconductivity as an application of EMC where double-magnon processes could be important. The term double magnon has been used to describe Feynman diagrams containing four single-magnon EMC vertices such that two magnons enter the diagram \cite{Hashimoto2000DoubleMagnon}. In this paper, we consider such diagrams to be higher-order single-magnon diagrams.

\begin{figure}
    \centering
    \includegraphics[width = \linewidth]{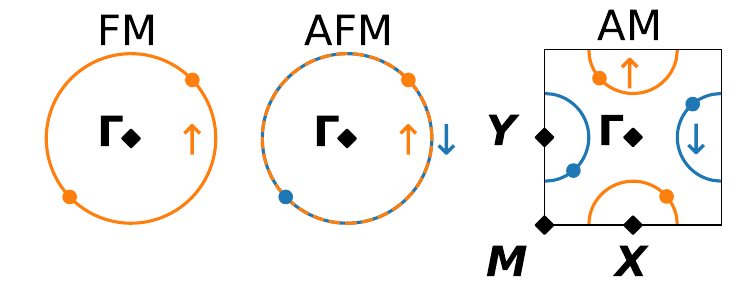}
    \caption{Sketches of Fermi surfaces (FSs) in metallic ferromagnets (FMs), antiferromagnets (AFMs) and altermagnets (AMs). Here, $\boldsymbol{\Gamma} = (0,0)$ is the center of the first Brillouin zone (1BZ). We illustrate possible zero-momentum Cooper pairs by filled circles, where orange (blue) corresponds to electrons with spin up (down). In a FM metal, the electron bands are spin split. Here, we imagine a situation where only one spin component has an FS. In that case, only spin-polarized Cooper pairs are relevant, which can occur due to double-magnon processes. The crystal symmetry of AFMs prohibits spin splitting. This allows Cooper pairs with opposite spin and opposite momenta, which can be induced by single-magnon processes. The AM has a $d$-wave spin splitting of the electron bands. Both spins have FSs, but like the FM, only spin-polarized Cooper pairs carry zero net momentum.}
    \label{fig:FS}
\end{figure}

Most studies of magnon-mediated superconductivity have ignored the double-magnon processes. The argument is that there are few magnons, such that single-magnon processes dominate. This argument relies on $\langle a_i^\dagger a_i \rangle$ being small at low temperatures due to a low concentration of magnons. However, if the magnon description requires diagonalization through a Bogoliubov transformation to obtain the long-lived magnon modes, $\langle a_i^\dagger a_i \rangle$ could contain quantum corrections originating with commutators of the diagonalized magnon operators in momentum space.
This is the case if the classical ground state is not the quantum mechanical ground state.
The $d$-wave spin splitting of the electron bands in AMs means that only spin-polarized Cooper pairs can have zero net momentum, as illustrated in Fig.~\ref{fig:FS}. Cooper pairs with zero net momentum have the greatest phase space for attractive interactions and are considered most stable \cite{SFsuperconductivity}. Hence, the single-magnon processes do not contribute to the formation of superconductivity. Reference \cite{Brekke2023Aug} explored how double-magnon scatterings can mediate an effective interaction between electrons without any spin flips. Figure \ref{fig:FS} shows a similar situation in metallic FMs, while conventional AFMs do not display any spin splitting such that single-magnon processes can mediate zero-momentum Cooper pairing without spin polarization.

Reference \cite{ArneFMNM} considers an FM/NM/FM trilayer with opposite magnetization directions in the two magnetic insulators, giving zero spin splitting in the normal metal. In that case, the FS in the NM looks like the AFM case in Fig.~\ref{fig:FS}, and the effective interaction is relatively simple. Single-magnon processes can generate superconductivity, and we can write the effective electron-electron interaction as \cite{ArneFMNM}
\begin{equation}
    H_{\text{FM},1} = \sum_{\boldsymbol{k}\boldsymbol{k}'} V_{\boldsymbol{k}\boldsymbol{k}'}^{\text{FM},1} c_{\boldsymbol{k}',\uparrow}^\dagger c_{-\boldsymbol{k}', \downarrow}^\dagger c_{-\boldsymbol{k},\uparrow}c_{\boldsymbol{k},\downarrow},
\end{equation}
with
\begin{equation}
    V_{\boldsymbol{k}\boldsymbol{k}'}^{\text{FM},1} = V^2 \frac{2\omega_{\boldsymbol{k}-\boldsymbol{k}'}}{(\epsilon_{\boldsymbol{k}}-\epsilon_{\boldsymbol{k}'})^2 - \omega_{\boldsymbol{k}-\boldsymbol{k}'}^2}.
\end{equation}
Here, $V = -J_{\text{sd}}\sqrt{2S/N}$ is a constant, $\epsilon_{\boldsymbol{k}}$ is the spin-degenerate electron spectrum, and $\omega_{\boldsymbol{q}}$ is the magnon spectrum. If we restrict the electron momenta to the FS, we get
\begin{equation}
    V_{\boldsymbol{k}\boldsymbol{k}' \in \text{FS}}^{\text{FM},1} = -\frac{2V^2}{\omega_{\boldsymbol{k}-\boldsymbol{k}'}}.
\end{equation}
The strongest coupling comes from the lowest energy magnons, which tends to happen at $\boldsymbol{k} \approx \boldsymbol{k}'$. Reference \cite{ArneFMNM} considered isotropic exchange, in which case the FM is inherently classical, and there are no quantum double-magnon processes. An electron-electron interaction mediated by double-magnon processes would require the presence of thermal magnons, i.e., Bose-Einstein distribution $n_B(\omega_{\boldsymbol{q}}) \neq 0$. Hence, it would be absent at zero temperature. 

In Appendix \ref{app:FM}, we consider a bulk 2D FM with anisotropic exchange. Such a model requires a diagonalization to find the long-lived magnon excitations. As a result, double-magnon processes are available at zero temperature due to quantum effects. We consider a half-metal, where only spin up has an FS, i.e., $\epsilon_{\boldsymbol{k},\downarrow}>0$, as illustrated in Fig.~\ref{fig:FS}. The coupling for spin-up electrons is
\begin{equation}
    H_{\text{FM},2} = \sum_{\boldsymbol{k}\boldsymbol{k}'} V_{\boldsymbol{k}\boldsymbol{k}'}^{\text{FM},2} c_{\boldsymbol{k}',\uparrow}^\dagger c_{-\boldsymbol{k}', \uparrow}^\dagger c_{-\boldsymbol{k},\uparrow}c_{\boldsymbol{k},\uparrow},
\end{equation}
with $V_{\boldsymbol{k}\boldsymbol{k}'}^{\text{FM},2} = V_{\boldsymbol{k}\boldsymbol{k}',\uparrow}$ given in Eq.~\eqref{eq:VkksFM2},
which on the FS simplifies to
\begin{equation}
    V_{\boldsymbol{k}\boldsymbol{k}' \in \text{FS}}^{\text{FM},2} = -\frac{J_{\text{sd}}^2}{2N^2} \sum_{\boldsymbol{q}} \frac{(u_{\boldsymbol{k}-\boldsymbol{k}'+\boldsymbol{q}}v_{\boldsymbol{q}} + u_{\boldsymbol{q}}v_{\boldsymbol{k}-\boldsymbol{k}'+\boldsymbol{q}})^2}{\omega_{\boldsymbol{k}-\boldsymbol{k}'+\boldsymbol{q}}+\omega_{\boldsymbol{q}}}.
\end{equation}
The factors $u_{\boldsymbol{q}}, v_{\boldsymbol{q}}$ are the coefficients in the Bogoliubov transformation. They are largest at $\boldsymbol{q}=0$, and since magnons are bosons, they can be greater than 1. Again, the strongest coupling happens for the lowest energy magnons, with $\boldsymbol{k}\approx \boldsymbol{k}'$ and $\boldsymbol{q} \approx 0$. However, unlike the single-magnon processes, the double-magnon processes involve a free sum over momentum, see also Fig.~\ref{fig:FeynmanEMC}(b). In large parts of the sum over $\boldsymbol{q}$ the coupling will be weaker than at $\boldsymbol{k}\approx \boldsymbol{k}'$ and $\boldsymbol{q} \approx 0$. As a result, the double-magnon processes should be subdominant to the single-magnon processes as long as both are operative. Schematically, we say that $1/\omega_{\boldsymbol{k}-\boldsymbol{k}'} \gg (1/N)\sum_{\boldsymbol{q}} 1/({\omega_{\boldsymbol{k}-\boldsymbol{k}'+\boldsymbol{q}}+\omega_{\boldsymbol{q}})}$ when $\boldsymbol{k}\approx \boldsymbol{k}'$, results in $V_{\boldsymbol{k}\boldsymbol{k}' \in \text{FS}}^{\text{FM},1} \gg V_{\boldsymbol{k}\boldsymbol{k}' \in \text{FS}}^{\text{FM},2}$, given the same $J_{\text{sd}}$. This means that in the studies involving FM/NM/FM trilayers \cite{ArneFMNM} or AFM systems \cite{Fjaerbu2019arneAFMNM,EirikAFMNM, ThingstadEliashberg,Sun2023Aug} any double-magnon corrections should be negligible. 
In cases where only double-magnon processes are relevant due to spin-split FSs, they could be important

Aside from generating superconductivity, double-magnon processes could also be relevant in other applications of EMC. 
This could be magnon drag \cite{Erlandsen2022MagnonDrag, Li2016MagnonDragExp, Wu2016MagnonDragExp, Lucassen2011MagnonDrag, Zhang2012MagnonCurrent, Yamaguchi2019MagnonDrag}, spin insulatronics \cite{Brataas2020MagnonCurrent}, and magnon spintronics \cite{Chumak2015MagnonCurrent, ExpInterfaceExchange, Berger1996Magnondrag, Takahashi2010MagnonCurrent}. 
Additionally, double-magnon processes can be relevant when exploring the self-energy in the normal state due to EMC \cite{Maeland2021Sep, Rost2024Jan}, as demonstrated in this paper for AMs since we solve linearized Eliashberg equations which contain the self-energy in the normal state.

We end this section by noting that long-lived magnons originating with noncolinear magnetic ground states do not have a well-defined quantization axis for spin, such that single-magnon processes can give rise to EMC that either does or does not flip the electron's spin. In that case, single-magnon processes can give rise to spin-polarized Cooper pairs and topological superconductivity \cite{Maeland2023PRL, Maeland2023Dec, Bostrom2023Dec}. From the above arguments, double-magnon processes should have little influence on the results derived in Refs.~\cite{Maeland2023PRL, Maeland2023Dec, Bostrom2023Dec}.

\section{Eliashberg equations} \label{sec:Eliashberg}
\subsection{Green's function and self-energy}
We consider a system with a spin-split electron band $\epsilon_{\boldsymbol{k},\sigma}$ crossing the FS, interacting with two magnon modes via spin-flipping single-magnon interactions and spin-preserving double-magnon interactions. 
Feynman diagrams generated by double-boson fermion-boson couplings have been explored in particle physics \cite{Aldaihan2017DoubleBoson, Drell1953DoubleBoson, Kummer1973DoubleBoson, Ferrer1999DoubleBoson}, with focus on the interaction energy between two fermions. 
They have also been considered in the case of double-bogolon-mediated superconductivity within weak- and strong-coupling approaches \cite{Villegas2019DoubleBogolon, Sun2021DoubleBogolon, Sun2021DoubleBogolonBCS, Sun2021DoubleBogolonEliashberg, Plyashechnik2023DoubleBogolon}. Bogolons are bosonic excitation on top of a Bose-Einstein condensate, and provide spin-independent interactions for both single- and double-bogolon processes. To capture the effects of spin-flipping and spin-preserving interactions in a spin split system, we present a detailed derivation of the Eliashberg equations when including single- and double-magnon EMC processes.

It is most convenient to consider Eliashberg theory in the Matsubara formalism \cite{Marsiglio1988AnalyticCont, Maeland2023CC}. We define the renormalized Green's function in imaginary time $\tau$ as
$G(\boldsymbol{k},\tau) = -\langle T_\tau \psi_{\boldsymbol{k}}(\tau) \psi_{\boldsymbol{k}}^\dagger (0) \rangle$. Here, $T_\tau$ is the time-ordering operator, the interactions are included in the expectation value \cite{BruusFlensberg, abrikosov, Schrieffer1964book}, and the Nambu spinor is $\psi^\dagger_{\boldsymbol{k}} = (d_{\boldsymbol{k}\uparrow}^\dagger, d_{\boldsymbol{k}\downarrow}^\dagger, d_{-\boldsymbol{k},\uparrow}, d_{-\boldsymbol{k},\downarrow})$. The electron operators $d_{\boldsymbol{k}\sigma}$ are potentially linear combinations of the original electron operators in a tight-binding model, resulting from a diagonalization.

\begin{figure}
    \centering
    \includegraphics[width = \linewidth]{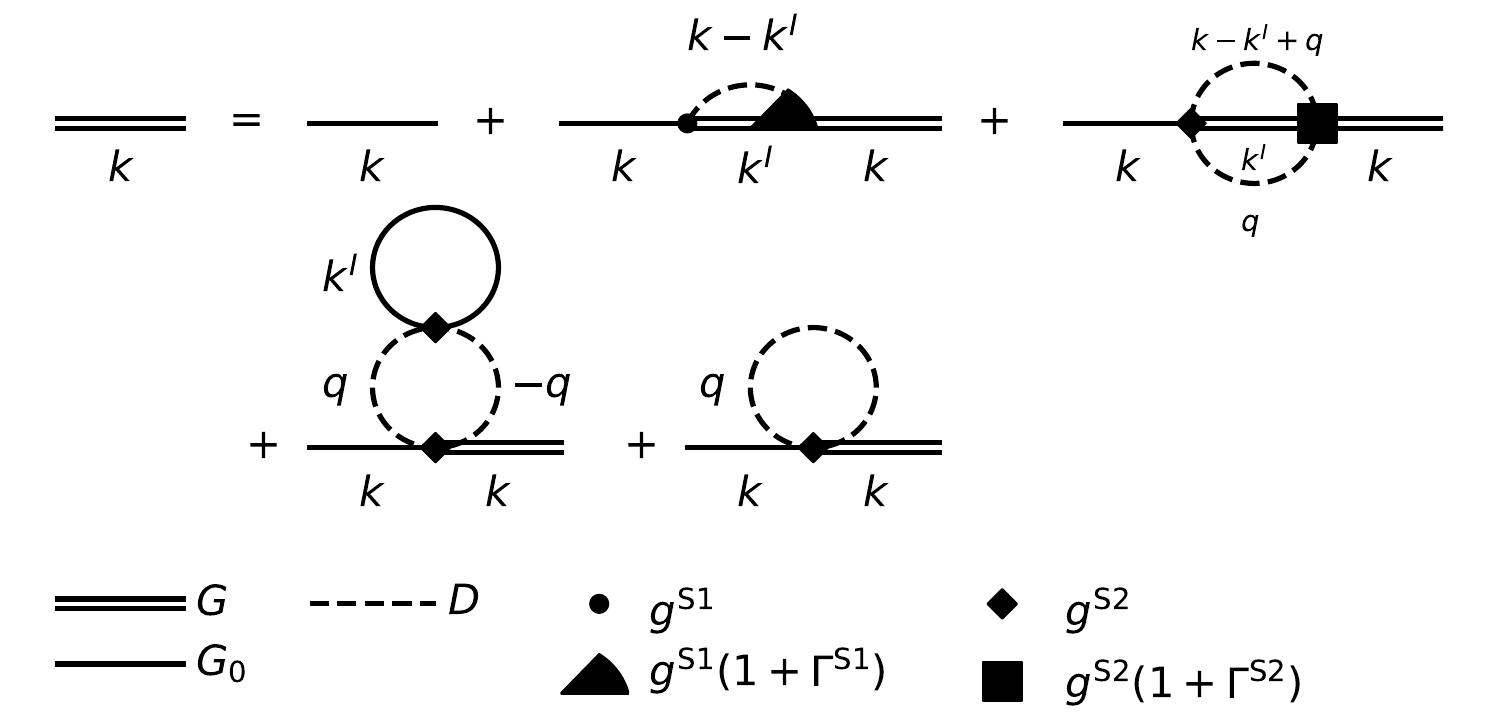}
    \caption{Feynman diagrams illustrate the Dyson equation on the form $G = G_0 + G_0 \Sigma G$. The self-energy has four contributions, named in the order $\Sigma = \Sigma^{\text{S}1}+\Sigma^{\text{S}2} + \Sigma^{\text{EL}} + \Sigma^{\text{ML}}$. The single-magnon sunset diagram $\Sigma^{\text{S}1}$ involves a spin flip at each interaction vertex. The double-magnon diagrams $\Sigma^{\text{S}2}$, $\Sigma^{\text{EL}}$, and $\Sigma^{\text{ML}}$ do not involve spin flips. $\Sigma^{\text{S}2}$ comes from a sunset diagram containing two magnons (S$2$), while $\Sigma^{\text{EL}}$ and $\Sigma^{\text{ML}}$ are tadpolelike with an electron-loop (EL) and a magnon-loop (ML), respectively. Vertex corrections are illustrated in the sunset diagrams and quantified by $\Gamma^{\text{S}1}$ and $\Gamma^{\text{S}2}$ for the single-magnon and double-magnon sunset diagram, respectively. The interaction strengths $g^{\text{S}1}$ and $g^{\text{S}2}$ depend on the momenta and magnon modes at the vertices.}
    \label{fig:Dyson}
\end{figure}

The self-energy $\Sigma$ is extracted from the Dyson equation $G = G_0 + G_0 \Sigma G$ using an $S$-matrix expansion of the renormalized Green's function $G$. $G_0$ is the bare Green's function, i.e., ignoring interactions. 
The FT to imaginary frequency is $G(\boldsymbol{k}, i\omega_n)  = \int_0^\beta d\tau e^{i\omega_n\tau} G(\boldsymbol{k},\tau)$,
where $\beta = 1/T$ is the inverse temperature. The Matsubara frequencies are $i\omega_n = i(2n+1) \pi T$ for fermions, and $i\omega_\nu = i2\nu \pi T$ for bosons.
The bare Green's function is
\begin{align}
    G_0(\boldsymbol{k},i\omega_n) =& \operatorname{diag}\big(i\omega_n-\epsilon_{\boldsymbol{k},\uparrow},i\omega_n-\epsilon_{\boldsymbol{k},\downarrow}, \nonumber\\
    &i\omega_n+\epsilon_{-\boldsymbol{k},\uparrow}, {i\omega_n+\epsilon_{-\boldsymbol{k},\downarrow}} \big)^{-1}.
\end{align}
We split the self-energy into four contributions $\Sigma = \Sigma^{\text{S}1}+\Sigma^{\text{S}2} + \Sigma^{\text{EL}} + \Sigma^{\text{ML}}$. These are illustrated in Fig.~\ref{fig:Dyson}, and detailed expressions are derived in Appendix \ref{app:selfenergy}. We explain our reasons for ignoring the electron-loop (EL) diagram $\Sigma^{\text{EL}}$ and the magnon-loop (ML) diagram $\Sigma^{\text{ML}}$ in Appendix \ref{app:SEAM}. These tadpolelike diagrams are nonzero, unlike the single-magnon tadpole diagram which is not permitted since single-magnon EMC flips the electron spin. $\Sigma^{\text{S}1}$ originates with the single-magnon sunset diagram (S$1$). Apart from some details of the matrix structure, its derivation is covered in most textbooks on many-particle perturbation theory \cite{abrikosov, BruusFlensberg}. The single-magnon interaction Hamiltonian can be written 
\begin{equation}
    H_{\text{int}}^{(1)} = \sum_{\boldsymbol{k}\boldsymbol{q}\alpha\beta\gamma} g_{\boldsymbol{k}+\boldsymbol{q}, \boldsymbol{k}}^{\alpha\beta\gamma} B_{\boldsymbol{q}\gamma} \psi_{\boldsymbol{k}+\boldsymbol{q},\alpha}^\dagger \psi_{\boldsymbol{k}\beta},
\end{equation}
where $\boldsymbol{B}_{\boldsymbol{q}}$ is a vector of magnon operators, e.g., $\boldsymbol{B}_{\boldsymbol{q}} = (\alpha_{\boldsymbol{q}}, \alpha_{-\boldsymbol{q}}^\dagger, \beta_{\boldsymbol{q}}, \beta_{-\boldsymbol{q}}^\dagger)$ in the case of two magnon modes. 
These operators describe the long-lived magnon modes. Any transformation coefficients in the electron and magnon sectors are contained in the coupling $g_{\boldsymbol{k}+\boldsymbol{q}, \boldsymbol{k}}^{\alpha\beta\gamma}$. Then, the self-energy is
\begin{align}
    \Sigma_{\alpha_1\beta_2}^{\text{S}1}(k) =& -  \sum_{k'}\sum_{\beta_1, \gamma_1, \alpha_2, \gamma_2} g_{\boldsymbol{k}, \boldsymbol{k}'}^{\alpha_1\beta_1\gamma_1}g_{\boldsymbol{k}', \boldsymbol{k}}^{\alpha_2\beta_2\gamma_2} \nonumber \\
    & \times D_{\gamma_1\gamma_2}(k-k') G_{\beta_1\alpha_2}(k'). \label{eq:SES1gen}
\end{align}
We let $k = \boldsymbol{k}, i\omega_n$ and $\sum_k = T\sum_{\boldsymbol{k},i\omega_n}$. $D_{\gamma_1\gamma_2}(q)$ is the FT of the magnon propagator $D_{\gamma_1\gamma_2}(\boldsymbol{q}, \tau) = \langle -T_\tau B_{\boldsymbol{q}\gamma_1}(\tau)B_{-\boldsymbol{q},\gamma_2}(0)\rangle$.
Furthermore, the double-magnon interaction Hamiltonian can be written
\begin{equation}
    H_{\text{int}}^{(2)} = \sum_{\boldsymbol{k}\boldsymbol{q}\boldsymbol{q}'\alpha\beta\gamma\gamma'} g_{\boldsymbol{k},\boldsymbol{q},\boldsymbol{q}'}^{\alpha\beta\gamma\gamma'} B_{-\boldsymbol{q},\gamma}B_{\boldsymbol{q}'\gamma'} \psi_{\boldsymbol{k}+\boldsymbol{q}'-\boldsymbol{q},\alpha}^\dagger \psi_{\boldsymbol{k}\beta}.
\end{equation}
The self-energy in the double-magnon sunset diagram (S$2$), derived in Appendix \ref{app:selfenergy}, is
\begin{align}
    \Sigma^{\text{S}2}_{\alpha_1\beta_2}(k) =& \sum_{k',q}\sum_{\substack{ \beta_1 \gamma_1 \gamma'_1 \\ \alpha_2 \gamma_2 \gamma'_2}}  G_{\beta_1\alpha_2}(k') \big[ g_{\boldsymbol{k}',-\boldsymbol{k}+\boldsymbol{k}'-\boldsymbol{q}, -\boldsymbol{q}}^{\alpha_1\beta_1\gamma_1\gamma'_1}   \nonumber \\
    &\times g_{\boldsymbol{k},\boldsymbol{k}-\boldsymbol{k}'+\boldsymbol{q}, \boldsymbol{q}}^{\alpha_2\beta_2\gamma_2\gamma'_2} D_{\gamma_1\gamma_2}(k-k'+q) D_{\gamma'_1\gamma'_2}(-q) \nonumber \\
    &+g_{\boldsymbol{k}',\boldsymbol{q}, \boldsymbol{k}-\boldsymbol{k}'+\boldsymbol{q}}^{\alpha_1\beta_1\gamma_1\gamma'_1}  g_{\boldsymbol{k},\boldsymbol{k}-\boldsymbol{k}'+\boldsymbol{q}, \boldsymbol{q}}^{\alpha_2\beta_2\gamma_2\gamma'_2} \nonumber \\
    &\times   D_{\gamma'_1\gamma_2}(k-k'+q)D_{\gamma_1\gamma'_2}(-q) \big].
    \label{eq:SES2gen}
\end{align}
Detailed expressions for $g_{\boldsymbol{k}+\boldsymbol{q}, \boldsymbol{k}}^{\alpha\beta\gamma}$ and $g_{\boldsymbol{k},\boldsymbol{q},\boldsymbol{q}'}^{\alpha\beta\gamma\gamma'}$ for the case of an AM, are given in Appendix \ref{app:SEAM}.

Since the renormalized Green's function $G$ contains the self-energy, these are coupled self-consistent equations, thereby counting a great number of diagrams. Nevertheless,  the description excludes some diagrams, illustrated as vertex corrections in Fig.~\ref{fig:Dyson}. Migdal's theorem \cite{migdal1958interaction} applies to electron-phonon coupling, and states that vertex corrections are negligible when the electron bandwidth is much larger than the phonon bandwidth \cite{Chubukov2020FEspecialized}. A similar requirement is found in 2D \cite{Migdal2D}, but its applicability to EMC is questionable since processes with small momentum transfer can be strong. Reference \cite{ThingstadEliashberg} discusses the lowest-order vertex correction due to single-magnon processes, which is a fourth-order diagram. The strength of processes involving zero-momentum magnons means that this diagram is not necessarily negligible. In the case of a spin split FS, this vertex correction should be small since zero-momentum single-magnon processes are energetically suppressed. The inclusion of double-magnon EMC, however, gives rise to vertex corrections of first order in interactions. From rough estimates, we find that these corrections should be an order $J_{\text{sd}}/t$ weaker than the included diagrams. An exploration of such corrections is outside the scope of the present paper.

We now assume that the Green's function is on the form
\begin{equation}
\label{eq:Gassumption}
    G = \begin{pmatrix}
        G_{11} & 0 & G_{13} & 0 \\
        0 & G_{22} & 0 & G_{24} \\
        G_{31} & 0 & G_{33} & 0 \\
        0 & G_{42} & 0 & G_{44}
    \end{pmatrix}.
\end{equation}
The element $G_{12}$ would entail a single spin flip, but spin flips will always come in pairs in any self-energy diagrams, so it must be zero. $G_{14}$ would be an anomalous Green's function describing Cooper pairs consisting of electrons with opposite spin. 
For FMs and AMs, only spin-polarized Cooper pairs can have zero net momentum. The anomalous Green's functions on the off-diagonal in Eq.~\eqref{eq:Gassumption} describe such pairing. From the Dyson equation on the form $G^{-1} = G_0^{-1} - \Sigma$, and the fact that $G_0$ is diagonal, it is clear that the self-energy must take the same form as the Green's function. We parametrize the self-energy in terms of Pauli matrix outer products as 
\begin{align}
    \Sigma =& (1-Z)i\omega_n \tau_0 \sigma_0 +\eta \tau_0 \sigma_3 + \chi \tau_3 \sigma_0 -\Sigma_h \tau_3 \sigma_3 \nonumber\\
    &+ \phi_{10} \tau_1\sigma_0 + \phi_{13} \tau_1\sigma_3 + \phi_{20} \tau_2 \sigma_0 + \phi_{23} \tau_2 \sigma_3. \label{eq:Sparam}
\end{align}
Here, $\tau_i$ covers the particle-hole degree of freedom, and $\sigma_i$ covers the spin degree of freedom. $\tau_0$ and $\sigma_0$ are the $2\times2$ identity matrix and $\tau_i \sigma_j$ is a shorthand for the outer product $\tau_i \otimes \sigma_j$. 
The Eliashberg functions, like the self-energy, depend on momentum and frequency, which is suppressed in the notation. Inserting this expression in the Dyson equation, we recognize two disconnected $2\times2$ blocks, one for each spin. Their determinants are given by
\begin{align}
    \Theta_\sigma =& (i\omega_n Z-\sigma\eta)^2 -(\epsilon_{\boldsymbol{k},\sigma}+\chi-\sigma\Sigma_h)^2 \nonumber\\
    &-(\phi_{10}+\sigma\phi_{13})^2 - (\phi_{20} + \sigma\phi_{23})^2, 
\end{align}
with $\sigma = 1$ for spin up and $\sigma = -1$ for spin down.
We then obtain
\begin{widetext}
\begin{equation}
\label{eq:G4x4}
    G = \begin{pmatrix}
        \frac{i\omega_n Z +\epsilon_{\boldsymbol{k},\uparrow}+\chi-\eta-\Sigma_h}{\Theta_\uparrow} & 0 & \frac{\phi_{10}+\phi_{13} - i\phi_{20} - i\phi_{23}}{\Theta_\uparrow} & 0 \\
        0 & \frac{i\omega_n Z +\epsilon_{\boldsymbol{k},\downarrow}+\chi+\eta+\Sigma_h }{\Theta_\downarrow}  & 0 & \frac{\phi_{10}-\phi_{13} - i\phi_{20} +  i\phi_{23}}{\Theta_\downarrow} \\
        \frac{\phi_{10}+\phi_{13} + i\phi_{20} + i\phi_{23}}{\Theta_\uparrow} & 0 & \frac{i\omega_n Z -\epsilon_{\boldsymbol{k},\uparrow}-\chi-\eta+\Sigma_h}{\Theta_\uparrow} & 0 \\
        0 & \frac{\phi_{10}-\phi_{13} + i\phi_{20} - i\phi_{23}}{\Theta_\downarrow} & 0 & \frac{i\omega_n Z -\epsilon_{\boldsymbol{k},\downarrow}-\chi+\eta-\Sigma_h}{\Theta_\downarrow}
    \end{pmatrix}.
\end{equation} 
\end{widetext}
Symmetry relations between the electron correlations in the Green's function results in the symmetries $f(\boldsymbol{k}, i\omega_n) = f(\boldsymbol{k}, -i\omega_n)^*$, where $f$ represents all the Eliashberg functions, $Z(k) = Z(-k)$, $\chi(k) = \chi(-k)$, $\eta(k) = -\eta(-k)$, $\Sigma_h(k) = \Sigma_h(-k)$, and $\phi_{ij}(k) = -\phi_{ij}(-k)$ \cite{ThingstadEliashberg, Maeland2023CC}. 

Using the assumption in Eq.~\eqref{eq:Gassumption} permits a simplification of the self-energy expressions. We find that $\Sigma^{\text{S}1}$ is diagonal, with
\begin{align}
    \Sigma_{11}^{\text{S}1}(k) =& \sum_{k'} \Lambda_{22}^{\text{S}1}(k,k')G_{22}(k'), \label{eq:S1lamstart}\\
    \Sigma_{22}^{\text{S}1}(k) =& \sum_{k'} \Lambda_{11}^{\text{S}1}(k,k')G_{11}(k'), \\
    \Sigma_{33}^{\text{S}1}(k) =& \sum_{k'} \Lambda_{44}^{\text{S}1}(k,k')G_{44}(k'), \\
    \Sigma_{44}^{\text{S}1}(k) =& \sum_{k'} \Lambda_{33}^{\text{S}1}(k,k')G_{33}(k'), \label{eq:S1Lamend}
\end{align}
while the double-magnon sunset diagram gives
\begin{equation}
    \Sigma^{\text{S}2}_{\alpha\beta}(k) = \sum_{k'} \Lambda_{\alpha\beta}^{\text{S}2}(k,k')  G_{\alpha\beta}(k'). \label{eq:S2lam}
\end{equation}
The definitions of $\Lambda_{\alpha\alpha}^{\text{S}1}(k,k')$ and $\Lambda_{\alpha\beta}^{\text{S}2}(k,k') = \sum_q \Lambda_{\alpha\beta}^{\text{S}2}(k,k',q)$ are given in Appendix \ref{app:SEAM} for the AM case. These functions are introduced to compactify the expressions, and contain the parts of the self-energy which are not an electron Green's function. In general, they consist of terms with a product of interaction strengths at the vertices, and one (two) boson propagator(s) for $\Lambda_{\alpha\alpha}^{\text{S}1}$ ($\Lambda_{\alpha\beta}^{\text{S}2}$). 

Equation \eqref{eq:G4x4} for the Green's function can be inserted in the expressions for $\Sigma^{\text{S}1}$ and $\Sigma^{\text{S}2}$ and compared to the parametrization of $\Sigma$ in Eq.~\eqref{eq:Sparam} to derive equations for the Eliashberg functions.
The equations for the pairing functions $\phi_{ij}$ are derived from equations for $\Sigma_{13}(k), \Sigma_{24}(k), \Sigma_{31}(k)$, and $\Sigma_{42}(k)$.
We then use $\phi_{10}(k) = (\Sigma_{13}(k)+\Sigma_{24}(k)+\Sigma_{31}(k)+\Sigma_{42}(k))/4$ and similar relations to get coupled equations for $\phi_{10}(k)$, $i\phi_{20}(k)$, $\phi_{13}(k)$, and $i\phi_{23}(k)$. By interchanging $\phi_{10} \leftrightarrow i\phi_{20}$ and $\phi_{13} \leftrightarrow i\phi_{23}$, we see that the equations for $i\phi_{20}$ and $i\phi_{23}$ correspond to the equations for $\phi_{10}$ and $\phi_{13}$, respectively. This reflects the free complex phase choice of the gaps. We add the equations to get equations for generally complex $\phi_{0} = \phi_{10}+i\phi_{20}$ and $\phi_3 = \phi_{13}+i\phi_{23}$. Finally, it is more convenient to study the spin-polarized gaps rather than a linear combination of them. From the elements of $G$ in Eq.~\eqref{eq:G4x4} we identify 
$\phi_\uparrow = \phi_0 + \phi_3$ and $\phi_\downarrow = \phi_0-\phi_3$. 
We can add and subtract the equations for $\phi_0$ and $\phi_3$ to get equations for $\phi_\uparrow$ and $\phi_\downarrow$.
The Eliashberg equations for the superconducting pairings are
\begin{align}
    \phi_\uparrow(k) =& \sum_{k'} \frac{\Lambda_{13}^{\text{S}2}(k,k')}{\Theta_\uparrow(k')}\phi_\uparrow(k') , \label{eq:phiup} \\
    \phi_\downarrow(k) =& \sum_{k'}\frac{\Lambda_{24}^{\text{S}2}(k,k')}{\Theta_\downarrow(k')}  \phi_\downarrow(k') \label{eq:phidown}.
\end{align}
For the AM model in Sec.~\ref{sec:AM}, we find $\Lambda_{11}^{\text{S}2}(k,k') = \Lambda_{33}^{\text{S}2}(k,k')$, $\Lambda_{22}^{\text{S}2}(k,k') = \Lambda_{44}^{\text{S}2}(k,k')$, $\Lambda_{13}^{\text{S}2}(k,k') = \Lambda_{31}^{\text{S}2}(k,k')$, and $\Lambda_{24}^{\text{S}2}(k,k') = \Lambda_{42}^{\text{S}2}(k,k')$. 
These relations arise from real transformation elements for the electrons, stemming from the fact that the Hamiltonian on matrix form, $\mathcal{H}_{\boldsymbol{k},\sigma}$ in Eq.~\eqref{eq:Hk}, is real and symmetric. Such a property should generally apply to inversion symmetric systems and has been used to simplify the equations.

The full Eliashberg equations for the diagonal terms are too lengthy to write down here. Instead, we proceed directly to derive linearized Eliashberg functions and apply FS averages to handle the momentum dependence. The first step is to gain a better understanding of the effect of the Eliashberg functions $Z, \eta, \Sigma_h,$ and $\chi$. The poles of $G$ provide information about renormalized bands through self-consistent equations, which should be analytically continued to the real frequency axis \cite{Marsiglio1988AnalyticCont, Vidberg1977AnalyticCont}. The poles are the zeros of $\Theta_\sigma(k)$ which are
\begin{align}
    i\omega_n =& \sigma\frac{\eta}{Z} \pm \Bigg[\pqty{\frac{\epsilon_{\boldsymbol{k},\sigma}+\chi-\sigma\Sigma_h}{Z}}^2 + \abs{\frac{\phi_{\sigma}}{Z}}^2\Bigg]^{\frac12} ,
\end{align}
where we used that $Z(k)$ is even in frequency and, therefore, real \cite{EliashbergRevMarsiglio2020}.
We can compare the right hand side to the BCS spectrum $E_{\boldsymbol{k},\sigma} = \pm \sqrt{\epsilon_{\boldsymbol{k},\sigma}^2 + |\Delta_{\boldsymbol{k},\sigma}|^2}$. Thus, we see that $\eta$ shifts the bands $E_{\boldsymbol{k},\sigma}$ depending on spin, which, in this spin decoupled case, is not the same as a magnetic field. A magnetic field would enter in $\epsilon_{\boldsymbol{k},\sigma}+\chi-\sigma\Sigma_h \to \epsilon_{\boldsymbol{k},\sigma}+\chi-\sigma h-\sigma\Sigma_h$ inside the square root. So, $\Sigma_h$ is a renormalization of the magnetic field \cite{Maeland2023CC}, though without a magnetic field, it is more natural to think of it as a renormalization of the spin splitting. $\chi$ is a spin independent shift of the electron bands. We can identify the spin-polarized superconducting gaps as
$\Delta_{\sigma}(k) = \phi_\sigma(k)/Z(k)$. From the symmetries listed after Eq.~\eqref{eq:G4x4}, it is clear that $\phi_\sigma(k) = -\phi_\sigma(-k)$ and so they represent spin triplet gaps that are even in frequency and odd in momentum, or odd in frequency and even in momentum \cite{Linder2019Oddw, Sigrist}.

The functions $\chi(k)$ and $\Sigma_h(k)$ shift the electron bands. To apply FS averages, the position of the FS must be fixed. Therefore the effects of $\chi$ and $\Sigma_h$ must be thought of as absorbed in our definition of the chemical potential $\mu$ and $J_{\text{sd}}S$. See also our argument for regarding  $\Sigma^{\text{ML}}$ and $\Sigma^{\text{EL}}$ as included in our choice of $J_{\text{sd}}S$ in Appendix \ref{app:SEAM}. $\eta$ will not shift the position of the FS and hence can be included in FS averaged equations. However, when solving the equations, we found that $\eta(k)=0$, so, for brevity, we omit it here.

\subsection{Fermi surface average}
A spin-split system has a spin dependent DOS, $N_\sigma(\epsilon) = \sum_{\boldsymbol{k}}\delta(\epsilon-\epsilon_{\boldsymbol{k},\sigma})$. We define a spin dependent FS average as
\begin{equation}
    \langle f(\boldsymbol{k}) \rangle_{\text{FS}_\sigma} = \frac{1}{N_{F,\sigma}}\sum_{\boldsymbol{k}} \delta(\epsilon_{\boldsymbol{k},\sigma}) f(\boldsymbol{k}),
\end{equation}
where $N_{F,\sigma} = N_\sigma(\epsilon=0)$ is the DOS for spin $\sigma$ on the FS for spin $\sigma$, $\text{FS}_\sigma$. We want to find an FS averaged version of equations on the form
\begin{equation}
    f(k) = \sum_{k'} \Lambda(k,k')\frac{h(k')}{\Theta_{\sigma'}(i\omega_{n'},\epsilon_{\boldsymbol{k}',\sigma'})},
\end{equation}
where $\Theta_{\sigma'}(k')$ depends on $\boldsymbol{k}'$ only through $\epsilon_{\boldsymbol{k}',\sigma'}$. We assume $f(k) = f(i\omega_n)\psi(\boldsymbol{k})$ and $h(k) = h(i\omega_n)\psi(\boldsymbol{k})$. If we multiply both sides by $\sum_{\boldsymbol{k}} \delta(\epsilon_{\boldsymbol{k},\sigma})\psi(\boldsymbol{k})/N_{F,\sigma}$ we recognize $\langle \psi^2(\boldsymbol{k}) \rangle_{\text{FS}_\sigma}$ on the left hand side.
Then, we convert the sum over $\boldsymbol{k}'$ to an integral, and split it into a part parallel and a part perpendicular to a constant energy contour of $\epsilon_{\boldsymbol{k}',\sigma}$. The sum over $\boldsymbol{k}'$ is dominated by $\boldsymbol{k}'$ close to $\text{FS}_{\sigma'}$, assuming the electron bandwidth is the largest energy scale in the system. We neglect the dependence of the parallel integral on the perpendicular momentum, and restrict the parallel integral to the FS for spin $\sigma'$. The end result is
\begin{align}
    f(i\omega_n)  =& \frac{1}{N_{F,\sigma}\langle \psi^2(\boldsymbol{k}) \rangle_{\text{FS}_\sigma}} \frac{1}{\beta}\sum_{i\omega_{n'}} h(i\omega_{n'}) \int  \frac{d\epsilon}{\Theta_{\sigma'}(i\omega_{n'},\epsilon)} \nonumber \\
    &\times \sum_{\boldsymbol{k},\boldsymbol{k}'}  \delta(\epsilon_{\boldsymbol{k},\sigma})\delta(\epsilon_{\boldsymbol{k}',\sigma'})\psi(\boldsymbol{k}) \Lambda(k,k')\psi(\boldsymbol{k}')  \label{eq:genFSavg} ,
\end{align}
where the parallel momentum integral has been converted back to a sum. 
We have approximated the length of the constant energy contour to be independent of $\epsilon$ and set it to the length of the FS. This is equivalent to approximating $N_{\sigma'}(\epsilon) \approx N_{F,\sigma'}$. This is generally not a very precise approximation, but since the behavior close to the FS dominates the physics, it should still give reasonable results. Note that this approximation means we treat the system as particle-hole symmetric with a constant DOS.

Assuming that the electron bandwidth is much larger than the boson bandwidth, a very reasonable approximation for magnons,  we extend the integration limit for $\epsilon$ to infinity. We have $\Theta_\sigma(k) = [i\omega_n Z(k)]^2 -(\epsilon_{\boldsymbol{k},\sigma})^2 -|\phi_{\sigma}(k)|^2$. We now assume $Z(k) = Z(i\omega_n)$ and $\phi_{\sigma}(k) = \phi_\sigma (i\omega_n) \psi_\sigma (\boldsymbol{k})$. If we linearize the equations for a temperature close to the critical temperature $T_c$ for superconductivity, we can neglect the pairing amplitude $\phi_\sigma (k)$ in the denominator. This gives $\Theta_\sigma(k) \approx [i\omega_n Z(i\omega_n)]^2 -(\epsilon_{\boldsymbol{k},\sigma})^2$ such that the momentum dependence is solely in the form $\epsilon_{\boldsymbol{k},\sigma}$. Then, using
\begin{align}
    \int_{-\infty}^{\infty} d\epsilon \frac{1}{[i\omega_n Z(i\omega_n)]^2-\epsilon^2} = \frac{-\pi}{\sqrt{[\omega_n Z(i\omega_n)]^2}},
\end{align}
we can derive the FS averaged, linearized Eliashberg equations. 

Some care must be exercised in performing the FS average in the equation for $Z(i\omega_n)$, since this quantity is related to both spin up and down. The key is first to FS average the expressions for $\Sigma_{ii}(k)$. For instance, $\Sigma_{11}(k)$ is related to spin up, so we FS average $\boldsymbol{k}$ to $\text{FS}_\uparrow$. We have $\Sigma_{ii}(k) = \Sigma_{ii}(i\omega_n)$ since we have assumed $Z(k) = Z(i\omega_n)$. Using the general FS average in Eq.~\eqref{eq:genFSavg}, we get
\begin{align}
    \Sigma_{11}(i\omega_n) =& \frac{-\pi}{N_{F,\uparrow}}\frac{1}{\beta}\sum_{i\omega_{n'}} i\operatorname{sgn}[\omega_{n'}Z(i\omega_{n'})]\nonumber \\
    &\times \bigg[\sum_{\boldsymbol{k}\boldsymbol{k}'}\delta(\epsilon_{\boldsymbol{k},\uparrow})\delta(\epsilon_{\boldsymbol{k}',\downarrow})\Lambda_{22}^{\text{S}1}(k,k') \nonumber \\
    &+\sum_{\boldsymbol{k}\boldsymbol{k}'}\delta(\epsilon_{\boldsymbol{k},\uparrow})\delta(\epsilon_{\boldsymbol{k}',\uparrow})\Lambda_{11}^{\text{S}2}(k,k') \bigg].
\end{align}
In the second line, related to the single-magnon sunset diagram, the two momenta belong to FSs for different spins due to the spin flip. In the third line, related to double-magnon processes, both momenta belong to the same spin since there is no spin flip. For $\Sigma_{22}(k)$ and $\Sigma_{44}(k)$ we let $\boldsymbol{k} \in \text{FS}_\downarrow$, while for $\Sigma_{33}(k)$ we let $\boldsymbol{k} \in \text{FS}_\uparrow$. This gives analogous FS averaged expressions to that for $ \Sigma_{11}(i\omega_n)$. Then, from $[1-Z(i\omega_n)]i\omega_n = [\Sigma_{11}(i\omega_n)+\Sigma_{22}(i\omega_n)+\Sigma_{33}(i\omega_n)+\Sigma_{44}(i\omega_n)]/4$ we arrive at the equation for $Z(i\omega_n)$. The FS averaged equations for $\phi_\sigma(i\omega_n)$ are found straightforwardly from the general expression for the FS average and Eqs.~\eqref{eq:phiup} and \eqref{eq:phidown}.

It is convenient to define the following dimensionless couplings
\begin{align}
    \lambda_{1,\uparrow\downarrow}(i\omega_n- i\omega_{n'}) =& \frac{1}{2N_{F,\uparrow}} \sum_{\boldsymbol{k}\boldsymbol{k}'}\delta(\epsilon_{\boldsymbol{k},\uparrow})\delta(\epsilon_{\boldsymbol{k}',\downarrow})\nonumber\\
    &\times[\Lambda_{22}^{\text{S}1}(k,k')+\Lambda_{44}^{\text{S}1}(k,k')], \label{eq:lambdasstart} \\
    \lambda_{1,\downarrow\uparrow}(i\omega_n- i\omega_{n'}) =& \frac{1}{2N_{F,\downarrow}} \sum_{\boldsymbol{k}\boldsymbol{k}'}\delta(\epsilon_{\boldsymbol{k},\downarrow})\delta(\epsilon_{\boldsymbol{k}',\uparrow})\nonumber\\
    &\times[\Lambda_{11}^{\text{S}1}(k,k') +\Lambda_{33}^{\text{S}1}(k,k') ], \\
    \lambda_{1,\uparrow\uparrow}(i\omega_n- i\omega_{n'}) =& \frac{1}{N_{F,\uparrow}} \sum_{\boldsymbol{k}\boldsymbol{k}'}\delta(\epsilon_{\boldsymbol{k},\uparrow})\delta(\epsilon_{\boldsymbol{k}',\uparrow})\Lambda_{11}^{\text{S}2}(k,k'),\\
    \lambda_{1,\downarrow\downarrow}(i\omega_n- i\omega_{n'}) =& \frac{1}{N_{F,\downarrow}} \sum_{\boldsymbol{k}\boldsymbol{k}'}\delta(\epsilon_{\boldsymbol{k},\downarrow})\delta(\epsilon_{\boldsymbol{k}',\downarrow})\Lambda_{22}^{\text{S}2}(k,k'),\\
    \lambda_{2,\uparrow}(i\omega_n- i\omega_{n'}) =&  \frac{-1}{N_{F,\uparrow} \langle \psi_\uparrow^2(\boldsymbol{k}) \rangle_{\text{FS}_\uparrow}} \sum_{\boldsymbol{k}, \boldsymbol{k}'} \delta(\epsilon_{\boldsymbol{k},\uparrow}) \delta(\epsilon_{\boldsymbol{k}',\uparrow}) \label{eq:lam2} \nonumber\\
    &\times \psi_\uparrow(\boldsymbol{k}) \Lambda_{13}^{\text{S}2}(k,k') \psi_\uparrow(\boldsymbol{k}') ,\\
    \lambda_{2,\downarrow}(i\omega_n- i\omega_{n'}) =&  \frac{-1}{N_{F,\downarrow} \langle \psi_\downarrow^2(\boldsymbol{k}) \rangle_{\text{FS}_\downarrow}} \sum_{\boldsymbol{k}, \boldsymbol{k}'} \delta(\epsilon_{\boldsymbol{k},\downarrow}) \delta(\epsilon_{\boldsymbol{k}',\downarrow}) \nonumber\\
    &\times \psi_\downarrow(\boldsymbol{k}) \Lambda_{24}^{\text{S}2}(k,k') \psi_\downarrow(\boldsymbol{k}') . \label{eq:lambdas}
\end{align}
Note that they only depend on frequency through the difference $i\omega_n- i\omega_{n'} = i\omega_{\nu}$, which is a bosonic Matsubara freqeuency.
For an AM, we find $N_{F,\uparrow} = N_{F,\downarrow} = N_F$, $\lambda_{1,\uparrow\downarrow}(i\omega_{\nu}) = \lambda_{1,\downarrow\uparrow}(i\omega_{\nu}) \equiv \lambda_{1}^{\text{S}1}(i\omega_{\nu})$, $ \lambda_{1,\uparrow\uparrow}(i\omega_{\nu}) =  \lambda_{1,\downarrow\downarrow}(i\omega_{\nu}) \equiv \lambda_1(i\omega_{\nu})$, and $\lambda_{2,\uparrow}(i\omega_{\nu}) = \lambda_{2,\downarrow}(i\omega_{\nu}) \equiv \lambda_{2}(i\omega_{\nu})$ due to the $d$-wave nature of the spin splitting. Additionally, $\lambda_{1}^{\text{S}1}(i\omega_{\nu}), \lambda_1(i\omega_{\nu}), \lambda_{2}(i\omega_{\nu})$ are real and even in frequency. $\lambda_{1}^{\text{S}1}(i\omega_{\nu})$ originates with the single-magnon sunset diagram, while $\lambda_{1}(i\omega_{\nu})$ and $\lambda_{2}(i\omega_{\nu})$ originate with the double-magnon sunset diagram. In a half-metallic FM only $N_{F,\uparrow}$, $\lambda_{1,\uparrow\uparrow}(i\omega_n- i\omega_{n'})$, and $\lambda_{2,\uparrow}(i\omega_n- i\omega_{n'})$ are nonzero. Assuming we are in one of those two situations for brevity, the linearized, FS averaged Eliashberg equations are
\begin{align}
    Z(i\omega_n) =& 1 + \frac{\pi}{\omega_n}\frac{1}{\beta}\sum_{i\omega_{n'}}[\lambda_{1}^{\text{S}1}(i\omega_n- i\omega_{n'})\nonumber\\
    &+\lambda_{1}(i\omega_n- i\omega_{n'})]\operatorname{sgn}(\omega_{n'}), \label{eq:ZFS} \\
    \phi_\uparrow(i\omega_n) =& \frac{\pi}{\beta} \sum_{i\omega_{n'}} \lambda_{2}(i\omega_n- i\omega_{n'})\frac{\phi_\uparrow(i\omega_{n'})}{|\omega_{n'} Z(i\omega_{n'})|} , \label{eq:phiupFS}\\
    \phi_\downarrow(i\omega_n) =& \frac{\pi}{\beta} \sum_{i\omega_{n'}}\lambda_{2}(i\omega_n- i\omega_{n'}) \frac{\phi_\downarrow(i\omega_{n'})}{|\omega_{n'} Z(i\omega_{n'})|}. \label{eq:phidownFS}  
\end{align}
From this, note that both $\lambda_{1}^{\text{S}1}(i\omega_{\nu})$ and $\lambda_{1}(i\omega_{\nu})$ contribute to the mass renormalization. Meanwhile, $\lambda_{2}(i\omega_{\nu})$ contributes to the superconducting pairing. Also, note that $Z(i\omega_{n})$ enters in the equations for $\phi_\sigma(i\omega_n)$, showing that the mass renormalization affects the superconducting pairing.

\subsection{Critical temperature estimates}
\subsubsection{Bardeen-Cooper-Schrieffer}
In BCS theory, the estimate of the critical temperature is \cite{BCS, SFsuperconductivity}
\begin{equation}
    T_c^{\text{BCS}} \approx 1.13 \omega_c e^{-1/\lambda},
\end{equation}
where $\omega_c$ is the maximum boson frequency and $\lambda$ is the dimensionless coupling.

\subsubsection{Eliashberg}
Since the dimensionless couplings are even in frequency, the sign function $\operatorname{sgn}(\omega_{n'})$ allows us to compute the sum over frequencies exactly in Eq.~\eqref{eq:ZFS}. Let us take $n\geq0$ as an example since the derivation for $n < 0$ is analogous, and we can use that $Z(i\omega_n) = Z(-i\omega_n)$. Then, the contributions at $n' = -1$ and $n'=2n+1$ cancel due to the sign function and the fact that $i\omega_n- i\omega_{-1} = -(i\omega_n- i\omega_{2n+1})$. We are left with the contribution at $n'=n$ and two times the contributions from $n'=0,\dots, n-1$. As a result,
\begin{align}
    Z(i\omega_n) =& 1 + \frac{1}{2n+1}\bigg(\lambda_{1}^{\text{S}1}(0)+\lambda_{1}(0)\nonumber\\
    &+2\sum_{\nu=1}^{n}[\lambda_{1}^{\text{S}1}(i\omega_\nu)+\lambda_{1}(i\omega_\nu)]\bigg).
\end{align}
Note that $Z(0) = Z(i\omega_{n=0}) = 1+\lambda_{1}^{\text{S}1}(0)+\lambda_{1}(0)$. 

The equations for $\phi_\uparrow$ and $\phi_\downarrow$ are the same, so we expect the same $T_c$ for both the spin up and spin down gaps. We introduce a symmetric cutoff $|\omega_n|<M$ and symmetrize to get
\begin{align}
     \Tilde{\phi}_n =& \sum_{n'=0}^N M_{nn'}^s(T)\Tilde{\phi}_{n'}. \label{eq:gapeqEliash}
\end{align}
with $\omega_N \approx M$, $\Tilde{\phi}_n = \phi_\sigma(i\omega_n)/\sqrt{|2n+1|Z(i\omega_n)}$, 
\begin{equation}
M_{nn'}^s(T) = \frac{\lambda_{2}(i\omega_n- i\omega_{n'})+\zeta_s \lambda_{2}(i\omega_n + i\omega_{n'})}{\sqrt{|2n+1|Z(i\omega_n)|2n'+1| Z(i\omega_{n'})}}, 
\end{equation}
$\zeta_e = 1$ for even-frequency gap, and $\zeta_o = -1$ for odd-frequency gap. We find the largest eigenvalue $\rho_s(T)$ of $M^s(T)$ and find the critical temperature $T_c^E$ by solving $\rho_s(T_c^E) = 1$, where Eq.~\eqref{eq:gapeqEliash} is satisfied. Then, we can recover $\phi_\sigma(i\omega_n)$ from the eigenvector, which tells us the frequency dependence of the gap just below $T_c$. The functions $\psi_\sigma(\boldsymbol{k})$ cover the momentum dependence.

\subsubsection{Allen-Dynes}
In a seminal work \cite{Allen1975Aug}, Allen and Dynes (AD) suggested a formula that approximates the critical temperature obtained from solving the Eliashberg equations. 
It was originally introduced to fit well with Eliashberg solutions and experimental measurements in phonon-mediated, even-frequency, $s$-wave superconductors \cite{Allen1975Aug}. Generalizing to more unconventional pairing mechanisms and pairing symmetries, we write \cite{ThingstadEliashberg}
\begin{equation}
    T_c^{\text{AD}} = \frac{\omega_{\text{log}}}{1.2} \exp(-\frac{1.04 Z(0)}{\lambda_2(0)}), \label{eq:AD}
\end{equation}
where $\omega_{\text{log}}$ is a logarithmic average of the boson spectrum defined in Appendix \ref{app:omlogdef} for the AM case. When low-energy bosons dominate the coupling, as in EMC, one typically has $\omega_{\text{log}} \ll \omega_c$, which reduces the estimate of $T_c$ compared to BCS theory \cite{ThingstadEliashberg}. The fact that $Z(0) > 1$ in the exponential also reduces the estimate of $T_c$ compared to BCS theory, indicating that many-body renormalization effects act in a pair-breaking manner. We emphasize that the AD estimate in Eq.~\eqref{eq:AD} is primarily used to get an understanding of the resultant $T_c$, the most reliable estimate of $T_c$ comes from solving the linearized Eliashberg equations, giving $T_c^E$.

\section{Altermagnet} \label{sec:AM}

\begin{figure}
    \centering
    \includegraphics[width = \linewidth]{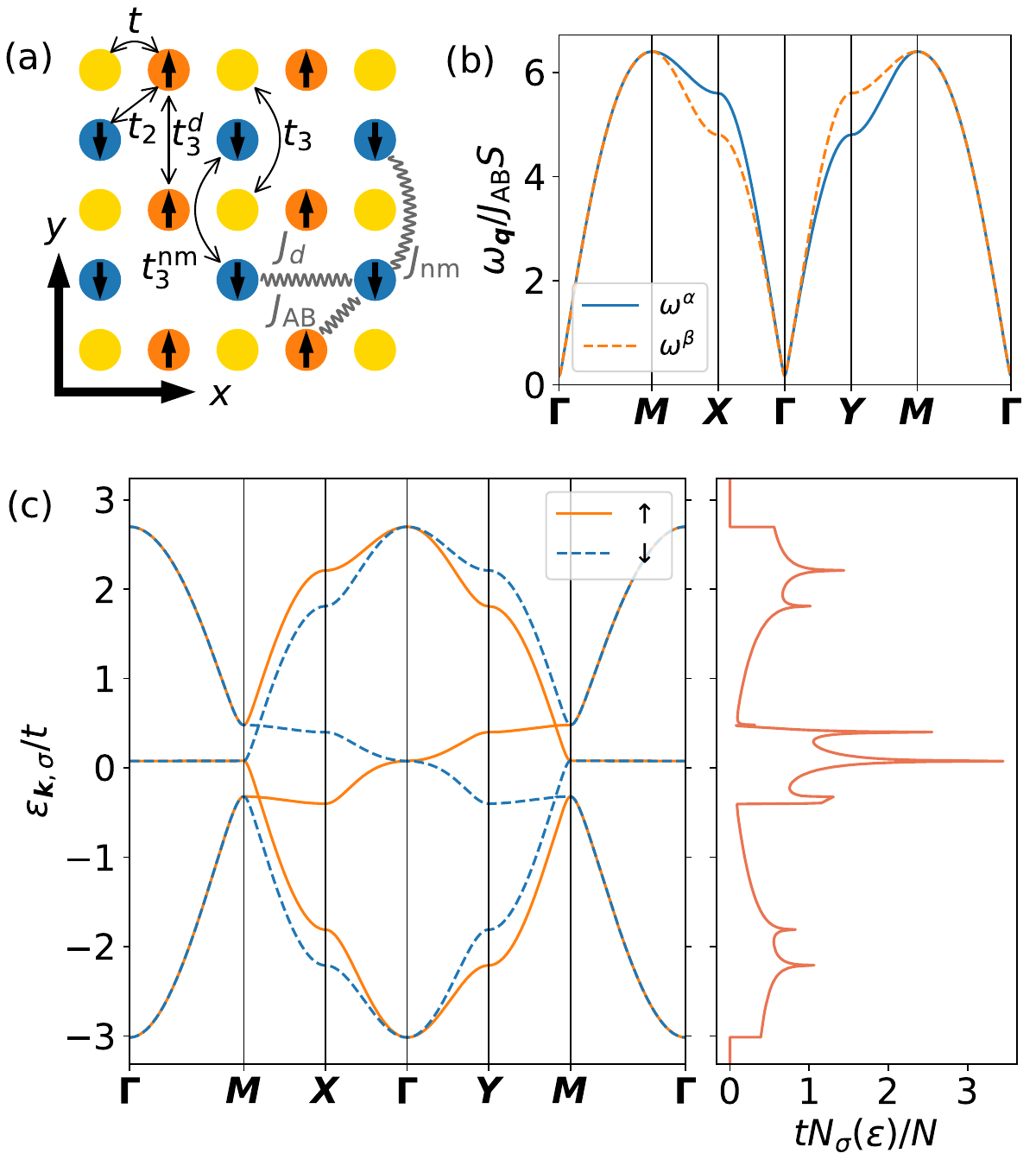}
    \caption{(a) The 2D Lieb lattice for our microscopic model of an AM. The gold sites are nonmagnetic, and the orange (blue) sites have a localized spin with spin up (down). Arrows illustrate the hopping parameters, while wavy lines illustrate the exchange interactions. (b) The magnon bands are plotted along a path in the 1BZ. The high-symmetry points in the 1BZ are illustrated in Fig.~\ref{fig:FS}. (c) The electron bands show a $d$-wave spin splitting resulting in a spin-independent density of states $N_\sigma(\epsilon)$, which is plotted in the right panel. Parameters: $t_2/t = 0.04$, $t_3/t =  0.02$, $\mu/t = 0$, $J_{\text{sd}}S/t = 0.4$, $J_d/J_{\text{AB}} = -0.2$, $J_{\text{nm}}/J_{\text{AB}} = -0.4$, and $K/J_{\text{AB}} = 0.002$.}
    \label{fig:AM}
\end{figure}

As a specific model of a 2D AM, we extend the microscopic model used in Ref.~\cite{Brekke2023Aug} to include up to third-nearest-neighbor hopping
\begin{align}
    H_e =& -\sum_{i,j,\sigma}  t_{i,j}c_{i,\sigma}^{\dagger} c_{j,\sigma} -\mu \sum_{i, \sigma}  c_{i,\sigma}^{\dagger} c_{i,\sigma} + H_{\text{sd}}, 
    \label{eq:He}
\end{align}
where $\mu$ is the chemical potential and $H_{\text{sd}}$ is defined in Eq.~\eqref{eq:Jsd}. 
We consider a Lieb lattice \cite{Lieb1989Mar} illustrated in Fig.~\ref{fig:AM}(a). This involves three sublattices, where sublattice $A$ ($B$) contains localized spins with spin up (down) and sublattice ``nm'' is nonmagnetic. 
The hopping parameters are $t_{i,j} = t$ for nearest-neighbor hopping, $t_2$ for next-nearest-neighbor hopping, $t_3^d$ for direct third-nearest-neighbor hopping via vacuum, $t_3^{\text{nm}}$ for third-nearest-neighbor hopping via the nonmagnetic sites, and $t_3$ for third-nearest neighbor hopping between the nonmagnetic sites. Figure \ref{fig:AM}(a) illustrates these hopping parameters. For simplicity we set $t_3^d = t_3^{\text{nm}} = t_3$ throughout this paper. 

The Hamiltonian \cite{Brekke2023Aug, Cui2023AMmagnonmodel}
\begin{align}
    H_m =& \sum_{\langle i,j\rangle} J_{\text{AB}}\boldsymbol{S}^A_{i} \cdot \boldsymbol{S}^B_{j} - \sum_{i} K \big[(S^A_{i,z})^2 + (S^B_{i,z})^2\big]  \nonumber\\ 
    &+ \sum_{\langle i_x,j_x\rangle} \big(J_{\text{nm}}\boldsymbol{S}^A_{i} \cdot \boldsymbol{S}^A_{j} + J_{d}\boldsymbol{S}^B_{i} \cdot \boldsymbol{S}^B_{j}\big) \nonumber\\
    &+ \sum_{\langle i_y,j_y\rangle} \big(J_{d}\boldsymbol{S}^A_{i} \cdot \boldsymbol{S}^A_{j} + J_{\text{nm}}\boldsymbol{S}^B_{i} \cdot \boldsymbol{S}^B_{j}\big),
    \label{eq:Hm}
\end{align}
describes the localized spins on sublattices $A$ and $B$. Here $\langle i_{x(y)},j_{x(y)}\rangle$ indicates that the sum is taken over nearest neighbors on the same sublattice in the $x$ $(y)$ direction, while $\langle i,j\rangle$ denotes nearest neighbors among the magnetic sites. $J_{\text{AB}}>0$ is the antiferromagnetic exchange between sublattices, $J_d$ is direct exchange within sublattices via vacuum, and $J_{\text{nm}}$ is intrasublattice exchange via a nonmagnetic site. Figure \ref{fig:AM}(a) illustrates the different exchange interactions. $K$ is the strength of the easy-axis anisotropy along the $z$ axis.

\subsection{Electrons}
Compared to Ref.~\cite{Brekke2023Aug}, we have changed the sign convention on the hopping terms. Otherwise, we proceed in the same manner in the following. An FT of the electron Hamiltonian to momentum space gives a description in terms of three types of electron operators $c_{\ell, \boldsymbol{k},\sigma}$, with $\ell \in \{ A, \text{nm}, B\}$, one for each sublattice. The Hamiltonian is then written on matrix form as $H_e = \sum_{\boldsymbol{k}, \sigma, \ell, \ell'}c_{\ell,\boldsymbol{k},\sigma}^\dagger [\mathcal{H}_{\boldsymbol{k}, \sigma}]_{\ell, \ell'}c_{\ell',\boldsymbol{k},\sigma}$, with $\mathcal{H}_{\boldsymbol{k},\sigma}$ defined in Eq.~\eqref{eq:Hk}. Diagonalization by a unitary matrix $U_{\boldsymbol{k},\sigma}$ gives
\begin{equation}
    H_e = \sum_{n, \boldsymbol{k}, \sigma} \epsilon_{n, \boldsymbol{k}, \sigma} d^\dagger_{n, \boldsymbol{k}, \sigma}d_{n, \boldsymbol{k}, \sigma},
\end{equation}
where the operators destroying and creating eigenstates are given by $d_{n,\boldsymbol{k},\sigma} =  \sum_{\ell} q^*_{n, \ell, \boldsymbol{k}, \sigma} c_{\ell, \boldsymbol{k}, \sigma}$ and the corresponding adjoint operators.

The three spin-split bands $\epsilon_{n,\boldsymbol{k},\sigma}$ are shown in Fig.~\ref{fig:AM}(c), displaying the $d$-wave spin splitting characteristic for AMs \cite{Smejkal2022Dec}. The second- and third-nearest-neighbor hopping terms break the accidental particle-hole symmetry considered in Ref.~\cite{Brekke2023Aug}. The FS is most similar to the case in Ref.~\cite{Brekke2023Aug} for $\mu > 0$ now. The middle bands have energies $\pm J_{\text{sd}}S$ at the $\boldsymbol{X}$ and $\boldsymbol{Y}$ points. 
The third-nearest-neighbor hopping terms add dispersion on the $\boldsymbol{M}\boldsymbol{X}$ and $\boldsymbol{M}\boldsymbol{Y}$ lines giving a more realistic model for an AM. Our motivation for including them is also to justify an FS average in the most interesting region of chemical potentials, $\mu \lesssim J_{\text{sd}}S$. 
Here, Ref.~\cite{Brekke2023Aug} predicts the strongest coupling due to a combination of a large DOS, and electrons being localized on mainly one sublattice. Hence, this is the most interesting region to investigate using strong-coupling Eliashberg theory. Figure \ref{fig:AM}(c) also shows the spin-dependent DOS $N_\sigma(\epsilon) = \sum_{n,\boldsymbol{k}} \delta(\epsilon-\epsilon_{n,\boldsymbol{k},\sigma})$, which is the same for both spins due to the $d$-wave spin splitting.

\subsection{Magnons}
The magnetic Hamiltonian is treated using an HP transformation and an FT. This gives a description in terms of magnon operators $a_{\boldsymbol{q}}$ and $b_{\boldsymbol{q}}$ related to spin fluctuations on sublattice $A$ and $B$, respectively. The Hamiltonian describing the magnons to second order in magnon operators, is diagonalized by a Bogoliubov transformation $a_{\boldsymbol{q}} = u_{\boldsymbol{q}}\alpha_{\boldsymbol{q}} + v_{\boldsymbol{q}}\beta_{-\boldsymbol{q}}^\dagger$, $b_{-\boldsymbol{q}}^{\dagger} = v_{\boldsymbol{q}}^* \alpha_{\boldsymbol{q}}  +u^*_{\boldsymbol{q}}\beta_{-\boldsymbol{q}}^\dagger$, yielding 
\begin{equation}
    H_m = \sum_{\boldsymbol{q}} \big(\omega^{\alpha}_{\boldsymbol{q}} \alpha_{\boldsymbol{q}}^\dagger \alpha_{\boldsymbol{q}} + \omega^{\beta}_{\boldsymbol{q}}\beta_{\boldsymbol{q}}^\dagger \beta_{\boldsymbol{q}} \big).
\end{equation}
By defining
\begin{align}
    A_{\boldsymbol{q}} =& 2S(J_d \cos2q_y + J_{\text{nm}}\cos2q_x \nonumber\\
    &-J_d-J_{\text{nm}}+2J_{\text{AB}}+K), \\
    B_{\boldsymbol{q}} =& 4SJ_{\text{AB}} \cos q_x \cos q_y, \\ 
    C_{\boldsymbol{q}} =& 2S(J_d \cos2q_x + J_{\text{nm}}\cos2q_y \nonumber\\
    &-J_d-J_{\text{nm}}+2J_{\text{AB}}+K),
\end{align}
we can write
\begin{align}
    u_{\boldsymbol{q}} = \frac{i}{\sqrt{2}}\sqrt{\frac{A_{\boldsymbol{q}}+C_{\boldsymbol{q}}}{\sqrt{(A_{\boldsymbol{q}}+C_{\boldsymbol{q}})^2-4B_{\boldsymbol{q}}^2}}+1}, \\
    v_{\boldsymbol{q}} = \frac{i}{\sqrt{2}}\sqrt{\frac{A_{\boldsymbol{q}}+C_{\boldsymbol{q}}}{\sqrt{(A_{\boldsymbol{q}}+C_{\boldsymbol{q}})^2-4B_{\boldsymbol{q}}^2}}-1}, \\
    \omega^{\alpha}_{\boldsymbol{q}} = \frac{A_{\boldsymbol{q}}-C_{\boldsymbol{q}}}{2} + \frac{1}{2}\sqrt{(A_{\boldsymbol{q}}+C_{\boldsymbol{q}})^2-4B_{\boldsymbol{q}}^2}, \\
    \omega^{\beta}_{\boldsymbol{q}} = \frac{C_{\boldsymbol{q}}-A_{\boldsymbol{q}}}{2} + \frac{1}{2}\sqrt{(A_{\boldsymbol{q}}+C_{\boldsymbol{q}})^2-4B_{\boldsymbol{q}}^2}.
\end{align}
The magnon bands are illustrated in Fig.~\ref{fig:AM}(b) and display a $d$-wave spin splitting if $J_d \neq J_{\text{nm}}$. The magnon gap is $\omega_0 = \omega_{\boldsymbol{q}=0}^{\alpha/\beta} = 2S\sqrt{K(K+4J_{\text{AB}})}$ and is caused by the easy-axis anisotropy breaking a continuous symmetry. Note the exchange-enhancement of the magnon gap, common to antiferromagnets.

\subsection{Electron-magnon coupling}
The classical part of the spin operator in $-J_{\text{sd}}\sum_{i, \sigma, \sigma'} \boldsymbol{S}_i\cdot c_{i,\sigma}^{\dagger}\boldsymbol{\sigma}_{\sigma \sigma'} c_{i,\sigma'}$ resulted in the spin splitting of the electron bands. The quantum part of the spin operator, described in terms of magnons by using the HP transformation, results in EMC. Throughout this paper, we will place the FS in the middle band and, therefore, only consider scatterings involving that band. Hence, we drop the band index on the electrons. We include both single- and double-magnon processes, $H_{\text{em}} = H^{(1)}_{\text{em}}+ H^{(2)}_{\text{em}}$ with
\begin{align}
    H^{(1)}_{\text{em}} &= -J_{\text{sd}}\frac{ \sqrt{2S}}{\sqrt{N}} \sum_{\boldsymbol{k}, \boldsymbol{q}} \Big(\big[(\Omega^A_{\boldsymbol{k}+\boldsymbol{q},\boldsymbol{k},\downarrow, \uparrow}u_{\boldsymbol{q}}+\Omega^B_{\boldsymbol{k}+\boldsymbol{q},\boldsymbol{k},\downarrow, \uparrow}v^*_{\boldsymbol{q}}) \alpha_{\boldsymbol{q}} \nonumber\\
    &+ (\Omega^A_{\boldsymbol{k}+\boldsymbol{q},\boldsymbol{k},\downarrow, \uparrow}v_{\boldsymbol{q}} +\Omega^B_{\boldsymbol{k}+\boldsymbol{q},\boldsymbol{k},\downarrow, \uparrow}u_{\boldsymbol{q}}^*)\beta^{\dagger}_{-\boldsymbol{q}}\big] d^\dagger_{\boldsymbol{k}+\boldsymbol{q},\downarrow}d_{\boldsymbol{k},\uparrow} \nonumber\\ 
    &+ \big[(\Omega^A_{\boldsymbol{k}+\boldsymbol{q},\boldsymbol{k},\uparrow, \downarrow}u^*_{\boldsymbol{q}}+ \Omega^B_{\boldsymbol{k}+\boldsymbol{q},\boldsymbol{k},\uparrow, \downarrow}v_{\boldsymbol{q}})\alpha^\dagger_{-\boldsymbol{q}} \nonumber\\
    &+ (\Omega^A_{\boldsymbol{k}+\boldsymbol{q},\boldsymbol{k},\uparrow ,\downarrow}v^*_{\boldsymbol{q}}+\Omega^B_{\boldsymbol{k}+\boldsymbol{q},\boldsymbol{k},\uparrow, \downarrow}u_{\boldsymbol{q}}) \beta_{\boldsymbol{q}}\big]  d^\dagger_{\boldsymbol{k}+\boldsymbol{q},\uparrow}d_{\boldsymbol{k},\downarrow}\Big), \label{eq:H1em}
\end{align}
\begin{align}
    H^{(2)}_{\text{em}} &= - \sum_{\boldsymbol{k}, \boldsymbol{q}, \boldsymbol{q}', \sigma}
    \frac{J_{\text{sd}}\sigma}{N} \Big[ \Omega^B_{\boldsymbol{k}+\boldsymbol{q}'-\boldsymbol{q},\boldsymbol{k},\sigma,\sigma} \big(v^*_{\boldsymbol{q}}v_{\boldsymbol{q}'} \alpha_{-\boldsymbol{q}} \alpha^\dagger_{-\boldsymbol{q}'} \nonumber\\
    &+ v^*_{\boldsymbol{q}} u_{\boldsymbol{q}'} \alpha_{-\boldsymbol{q}}\beta_{\boldsymbol{q}'} + u^*_{\boldsymbol{q}}v_{\boldsymbol{q}'}\beta_{\boldsymbol{q}}^\dagger \alpha^\dagger_{-\boldsymbol{q}'} + u^*_{\boldsymbol{q}} u_{\boldsymbol{q}'}\beta^\dagger_{\boldsymbol{q}}\beta_{\boldsymbol{q}'}\big) \nonumber\\ 
    &-\Omega^A_{\boldsymbol{k}+\boldsymbol{q}'-\boldsymbol{q},\boldsymbol{k},\sigma, \sigma}\big(u^*_{\boldsymbol{q}} u_{\boldsymbol{q}'} \alpha^\dagger_{\boldsymbol{q}} \alpha_{\boldsymbol{q}'} + u^*_{\boldsymbol{q}} v_{\boldsymbol{q}'} \alpha^\dagger_{\boldsymbol{q}}\beta^\dagger_{-{\boldsymbol{q}'}} \nonumber\\
    &+ v^*_{\boldsymbol{q}}u_{\boldsymbol{q}'} \beta_{-\boldsymbol{q}}\alpha_{\boldsymbol{q}'} + v^*_{\boldsymbol{q}} v_{\boldsymbol{q}'}\beta_{-\boldsymbol{q}} \beta^\dagger_{-\boldsymbol{q}'} \big)\Big]d^\dagger_{\boldsymbol{k}+\boldsymbol{q}'-\boldsymbol{q},\sigma}d_{\boldsymbol{k},\sigma}. \label{eq:H2em}
\end{align}
Here, the electron and magnon operators have been transformed to their diagonal bases, and $\Omega^\ell_{\boldsymbol{k}',\boldsymbol{k},\sigma', \sigma} \equiv q^*_{\ell, \boldsymbol{k}',\sigma'} q_{\ell, \boldsymbol{k},\sigma}$. From $ H^{(1)}_{\text{em}}$, we see that the $\alpha$ magnon mode carries spin $-1$ along the $z$ axis while the $\beta$ magnon mode carries spin $1$ along the $z$ axis leading to a specific spin-flip structure of the interactions. In $H^{(2)}_{\text{em}}$, we see that all combinations of magnon operators carry in total 0 spin along the $z$ axis, leading to no spin flip of the electron.

In BCS theory, a Schrieffer-Wolff transformation \cite{Schrieffer1966Wolff} is applied to derive an effective electron-electron interaction mediated by the magnons. 
The derivation of the linearized BCS gap equation was done in Ref.~\cite{Brekke2023Aug} and is not repeated here. We perform new calculations to find the BCS prediction of $T_c$ when including third-nearest-neighbor hopping and a change of parameters, using Eq.~\eqref{eq:linBCS}.

\subsection{Superconductivity}

\begin{figure}
    \centering
    \includegraphics[width = \linewidth]{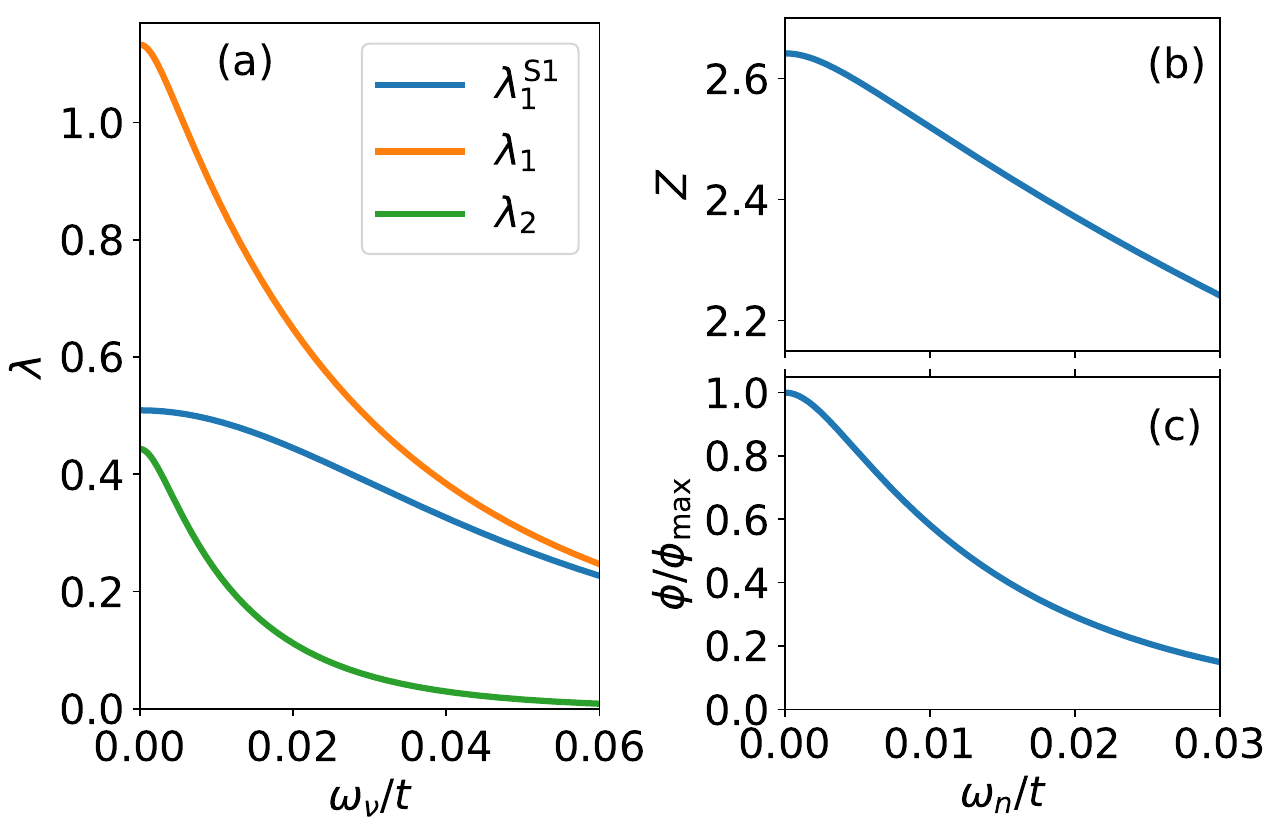}
    \caption{(a) The dimensionless couplings $\lambda_{1}^{\text{S}1}(i\omega_{\nu}), \lambda_1(i\omega_{\nu}),$ and $\lambda_{2}(i\omega_{\nu})$ as functions of bosonic Matsubara frequencies up to $\omega_\nu \approx 2M$. This includes 534 discrete Matsubara frequencies at the chosen temperature, so the curves look smooth, and markers are omitted. (b) $Z$ and (c) $\phi/\phi_{\text{max}}$ as a function of fermionic Matsubara frequencies up to $\omega_n \approx M$. All functions in this plot are real and even in frequency. At larger frequencies $Z \to 1$ and $\lambda_{1}^{\text{S}1}, \lambda_1, \lambda_{2}, \phi \to 0$. We plot a normalized version of $\phi = \phi_\sigma$ since its amplitude is arbitrary when solving linearized equations. Parameters: $t_2/t = 0.04$, $t_3/t =  0.02$, $\mu/t = 0.39$, $J_{\text{sd}}S/t = 0.4$, $S=3/2$, $J_{\text{AB}}S/t = 0.01$, $J_d/J_{\text{AB}} = -0.2$, $J_{\text{nm}} = -0.4$, $K/J_{\text{AB}} = 0.002$, $T/t = 1.8\times 10^{-5} \approx T_c^E/t$, $M/t = 0.03$, $N = 80^2$ points in sum over $\boldsymbol{q}$, and 54 points on the FS for each spin. $\psi_\sigma(\boldsymbol{k})$ is set to the normalized solution of the linearized BCS gap equation, shown in Fig.~\ref{fig:TcAM}(c).}
    \label{fig:lamZphi}
\end{figure}

The frequency sum in $\Lambda_{\alpha\beta}^{\text{S}2}(k,k')$ can be computed analytically. This gives some terms that survive at zero temperature and additional terms that contain factors of $n_B(\omega_{\boldsymbol{q}}^{\alpha/\beta})$. Still, the free sum over momentum $\boldsymbol{q}$ means that computing the dimensionless couplings in Eqs.~\eqref{eq:lambdasstart}-\eqref{eq:lambdas} puts a heavy demand on computational resources. As a simplification, we assume the temperature is much smaller than the magnon gap, $T \ll \omega_0$, such that $n_B(\omega_{\boldsymbol{q}}^{\alpha/\beta}) \approx 0$, and ignore the contributions coming from thermal magnons. Then, the only temperature dependence of the dimensionless couplings is through the positions of the bosonic Matsubara frequencies. At a sufficiently low temperature, we have enough discrete points such that $\lambda_{1}^{\text{S}1}(i\omega_{\nu}), \lambda_1(i\omega_{\nu}),$ and $\lambda_{2}(i\omega_{\nu})$ are smooth curves. From this, we can interpolate to any temperature satisfying $T \ll \omega_0$ using quadratic interpolation. The momentum sums are converted to integrals and calculated using 
\begin{equation}
    \int d\boldsymbol{k} f(\boldsymbol{k}) \delta(\epsilon_{\boldsymbol{k},\sigma}) = \int_\Gamma d\boldsymbol{k}\cdot \boldsymbol{e_{k}}  \frac{f(\boldsymbol{k})}{|\nabla_{\boldsymbol{k}}\epsilon_{\boldsymbol{k},\sigma}|},
\end{equation}
where $\Gamma$ is the line where $\epsilon_{\boldsymbol{k},\sigma} = 0$ and $\boldsymbol{e_{k}}$ is a unit vector along that line. A sum over equidistant points on the FS approximates the line integral. 

Figure \ref{fig:lamZphi}(a) shows the dimensionless couplings $\lambda_{1}^{\text{S}1}(i\omega_{\nu}), \lambda_1(i\omega_{\nu}),$ and $\lambda_{2}(i\omega_{\nu})$. While the single-magnon sunset diagram does not contribute to superconductivity, it has a significant effect on renormalizations as seen in $\lambda_{1}^{\text{S}1}(i\omega_{\nu})$. $\lambda_1(i\omega_{\nu})$ and $\lambda_{2}(i\omega_{\nu})$ are the contribution from the double-magnon diagram to renormalization and superconducting pairing, respectively. Note that $\lambda_2(i\omega_{\nu}) < \lambda_1(i\omega_{\nu})$, which, according to the AD estimate in Eq.~\eqref{eq:AD}, will have a negative effect on $T_c$. If the gap had been $s$-wave, $\psi(\boldsymbol{k})=1$, we would have $\lambda_1(i\omega_{\nu}) = \lambda_2(i\omega_{\nu})$ ($\Lambda_{13}^{\text{S}2}(k,k') = -\Lambda_{11}^{\text{S}2}(k,k')$ since the electron transformation coefficients are real). The fact that $\psi_\sigma(\boldsymbol{k})$ is odd in momentum for even-frequency, spin-polarized pairing seems to have a detrimental effect on $T_c$. However, Ref.~\cite{ThingstadEliashberg} found that $\lambda_1(i\omega_{\nu}) \approx \lambda_2(i\omega_{\nu})$ for both even- and odd-momentum gaps when considering single-magnon processes. 

Reference \cite{ThingstadEliashberg} considers an AFM/NM/AFM trilayer. As discussed in Sec.~\ref{sec:EMCcolinear}, single-magnon processes are strongly dominated by $\boldsymbol{k} \approx \boldsymbol{k}'$ processes, especially when the magnon gap is small. Double-magnon processes have a similar effect, but the effect is much less significant due to the free sum over $\boldsymbol{q}$. As a result, $\lambda_2$ in Ref.~\cite{ThingstadEliashberg} should be dominated by the $\boldsymbol{k} \approx \boldsymbol{k}'$ parts of the sum where $\psi(\boldsymbol{k})$ and $\psi(\boldsymbol{k}')$ have the same sign. In the AM case studied here, where double-magnon processes dominate the superconducting pairing, the $\boldsymbol{k}' \approx -\boldsymbol{k}$ parts of the sum are not negligible. Since $\psi_\sigma(\boldsymbol{k})$ and $\psi_\sigma(\boldsymbol{k}')$ have opposite sign here, these parts contribute with the opposite sign compared to when $\boldsymbol{k} \approx \boldsymbol{k}'$. As a result, we obtain $\lambda_2(i\omega_{\nu}) < \lambda_1(i\omega_{\nu})$.

In order to increase $\lambda_2$ compared to $\lambda_1$, one would need to make the low-energy magnons more dominant. This can be achieved by lowering the magnon gap, i.e., decreasing $K/J_{\text{AB}}$ or $J_{\text{AB}}S/t$, or decreasing both. However, lowering the magnon gap too far can destabilize the magnet even at very low temperatures or destroy the Fermi liquid nature of the quasiparticles close to the FS \cite{KargarianFMTI, Maeland2021Sep}. Hence, a compromise must be made. 
It is also of interest to minimize $\lambda_{1}^{\text{S}1}(i\omega_{\nu})$. Since this involves scattering between FSs of different spin, the key is to separate the FSs as much as possible, which happens to be the situation when $\mu \to J_{\text{sd}}S$. Hence, by lowering $\mu$ compared to Fig.~\ref{fig:lamZphi}, $\lambda_{1}^{\text{S}1}(i\omega_{\nu})$ increases since the reciprocal distance between the FSs decreases. Moreover, $\lambda_{1}^{\text{S}1}(i\omega_{\nu})$ is reduced if the intermediate and large momentum magnons have a higher energy since only they contribute when the FSs are well separated. Increasing the magnon bandwidth affects the validity of FS averages and the Migdal theorem \cite{migdal1958interaction, Migdal2D}, so again, a compromise must be made. Making $J_d$ and $J_{\text{nm}}$ antiferromagnetic such that they frustrate the magnetic state, would make magnons away from zero momentum more relevant since $|u_{\boldsymbol{q}}|$ and $|v_{\boldsymbol{q}}|$ would increase for $\boldsymbol{q}\neq 0$ \cite{Erlandsen2020TIAFM}. In this system, this is disadvantageous since it would increase $\lambda_{1}^{\text{S}1}(i\omega_{\nu})$, so we consider $J_d$ and $J_{\text{nm}}$ to be ferromagnetic.

Even-momentum gaps do not necessarily suffer from $\lambda_2(i\omega_{\nu}) < \lambda_1(i\omega_{\nu})$. In fact, for $d_{x^2-y^2}$-wave pairing we get $\lambda_2(i\omega_{\nu}) > \lambda_1(i\omega_{\nu})$ though only by a small amount. For $s$-wave pairing, $\lambda_2(i\omega_{\nu}) = \lambda_1(i\omega_{\nu})$. Hence, one may wonder if odd-frequency pairing \cite{Linder2019Oddw} is preferred in this system. However, there is no solution to the eigenvalue problem in Eq.~\eqref{eq:gapeqEliash} for odd-frequency gaps at any temperature. The largest eigenvalue $\rho_o(T) \ll 1$ for any temperature, and in fact decreases when the temperature is decreased. We used $\psi_\sigma(\boldsymbol{k}) = \cos(2\theta_{\boldsymbol{k}})$ with $\theta_{\boldsymbol{k}}$ equal to the angle between $\boldsymbol{k}$ and the $k_x$ axis to represent a $d_{x^2-y^2}$-wave gap. We also tried $\psi_\sigma(\boldsymbol{k}) = \sin(2\theta_{\boldsymbol{k}})$ corresponding to $d_{xy}$-wave pairing. In that case, $\lambda_2(i\omega_{\nu}) < \lambda_1(i\omega_{\nu})$, and $\lambda_2(i\omega_{\nu})$ shows a sign change at intermediate frequency. This could be advantageous to odd-frequency pairing. However, we again find $\rho_o(T) \ll 1$ for any temperature, possibly since the magnitude of $\lambda_2(i\omega_{\nu})$ is very low when it is negative. Hence, we exclude odd-frequency pairing as a competing order in this system.

The unlikelihood of odd-frequency pairing mediated by EMC appears to be a general feature for AMs.
It is worth considering phonon-mediated superconductivity in AMs, although phonons typically do not give odd-frequency superconductivity \cite{Linder2019Oddw}. However, applying a magnetic field that is weak enough not to affect the compensated magnetic ordering in the AM state, might induce an odd-frequency component in a phonon-mediated superconducting state \cite{Maeland2023CC, Aperis2015magneticField}. Additionally, a phonon-mediated interaction between electrons typically yields $s$-wave superconducting gaps. However, phonons may also yield $d$-wave superconductivity if included in combination with spin-fluctuations, or provided that vertex corrections in the electron self-energy are accounted for \cite{Schnell2006dwavephonon, Schrodi2021dwavephonon}.

\begin{figure*}
    \centering
    \includegraphics[width = \linewidth]{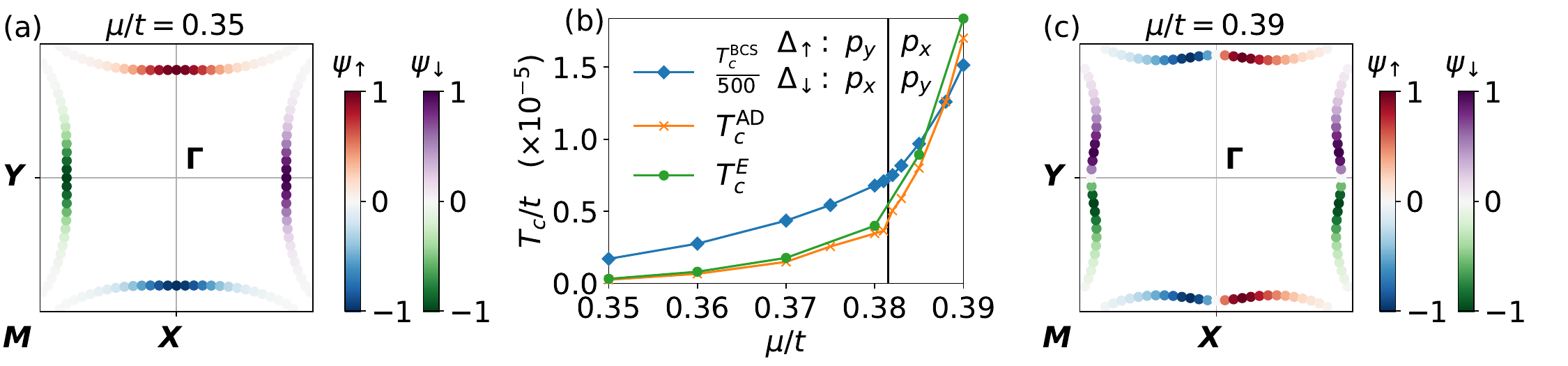}
    \caption{(b) Critical temperature from BCS theory $T_c^{\text{BCS}}$ (divided by 500, blue line, diamonds), Allen-Dynes estimate $T_c^{\text{AD}}$ (orange line, crosses), and from solving the linearized Eliashberg equations $T_c^{E}$ (green line, circles at calculated points). Parameters: $t_2/t = 0.04$, $t_3/t =  0.02$, $J_{\text{sd}}S/t = 0.4$, $S=3/2$, $J_{\text{AB}}S/t = 0.01$, $J_d/J_{\text{AB}} = -0.2$, $J_{\text{nm}}/J_{\text{AB}} = -0.4$, $K_z/J_{\text{AB}} = 0.002$, $M/t = 0.03$, $N = 80^2$ points in sum over $\boldsymbol{q}$, and 54 or 58 points on the FS for each spin, depending on the chemical potential. $\psi_\sigma(\boldsymbol{k})$ is set to the normalized solution of the linearized BCS gap equation, shown for $\mu/t = 0.35$ in (a) and $\mu/t = 0.39$ in (c). The black vertical line in (b) shows the transition from nodeless $p_y$- to nodal $p_x$-wave symmetry for the spin-up gap. The magnon gap is $\omega_0/t \approx 0.0018$, the maximum magnon frequency is $\omega_c/t \approx 0.064$, and, if $t = 1$~eV, then $T_c^{\text{BCS}} \approx 88$~K and $T_c^E \approx 0.21$~K at $\mu/t = 0.39$.}
    \label{fig:TcAM}
\end{figure*}

\begin{figure}
    \centering
    \includegraphics[width = \linewidth]{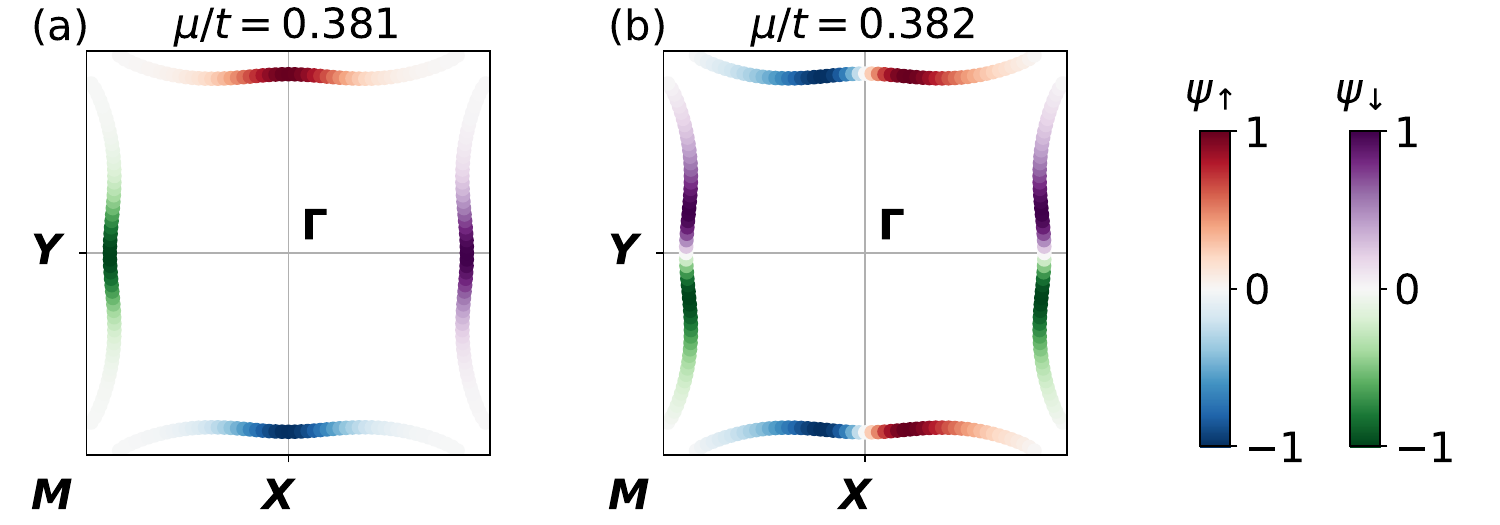}
    \caption{The normalized solutions to the linearized BCS gap equation, $\psi_\sigma(\boldsymbol{k})$, are shown on the FS for (a) $\mu/t = 0.381$ and (b) $\mu/t = 0.382$. The transition from nodeless $p_y$- to nodal $p_x$-wave symmetry for the spin-up gap occurs without any dramatic change in the shape of the FS. Here, there are 110 points on the FS for each spin and otherwise the parameters are the same as in Fig.~\ref{fig:TcAM}.}
    \label{fig:Gap}
\end{figure}

When focusing on even-frequency $p$-wave gaps, solutions exist for $\rho_e(T_c^E) = 1$. Figures \ref{fig:lamZphi}(b) and \ref{fig:lamZphi}(c) show $Z(i\omega_n)$ and $\phi_\sigma(i\omega_n)$ for a temperature close to $T_c^E$, where the largest eigenvalue is $\rho_e \approx 1.0$. The result for $Z(i\omega_n)$ indicates significant mass renormalization effects. For even-frequency gaps, $\rho_e(T)$ increases when we decrease the temperature until it crosses $1$ at $T_c^E$. We set $\psi_\sigma(\boldsymbol{k})$ to normalized versions of the BCS gap solutions $\Delta_{\boldsymbol{k},\sigma}$ in order to maximize $\lambda_2(i\omega_\nu)$. These are shown in Fig.~\ref{fig:TcAM}(a) for $\mu/t = 0.35$ and Fig.~\ref{fig:TcAM}(c) for $\mu/t = 0.39$, and can be chosen to be real. From the linearized BCS gap equation [Eq.~\eqref{eq:linBCS}] we find a transition from nodeless $p_y$-wave symmetry to nodal $p_x$-wave symmetry for $\Delta_{\boldsymbol{k},\uparrow}$ at $0.381 < \mu/t < 0.382$. $\Delta_{\boldsymbol{k},\downarrow}$ has a similar transition from nodeless to nodal $p$-wave. This transition occurs because at larger chemical potential, it is an advantage to have a sign change in $\Delta_{\boldsymbol{k},\uparrow}$ across $k_x = 0$, since the coupling $\Tilde{V}_{\boldsymbol{k}\boldsymbol{k}'}^{\text{FS}_\uparrow}$ [Eq.~\eqref{eq:VkkAM}] becomes positive for $\boldsymbol{k}' \approx (-k_x, k_y)$.
As $\mu$ increases, the positive values at these momenta grow in magnitude compared to the negative values at $\boldsymbol{k}' \approx \boldsymbol{k}$ so that at some value of $\mu$, a nodal gap becomes preferred. This is related to the fact that the FS moves closer to the edge of the 1BZ, as seen when comparing Figs.~\ref{fig:TcAM}(a) and (c). However, the exact shape of the FS is not what determines the transition point. As illustrated in Fig.~\ref{fig:Gap}, there is no dramatic change in the shape of the FS across the transition point.
Instead, we find that the magnon gap plays a significant role for the precise location of the transition from nodeless to nodal gap. With parameters as in Ref.~\cite{Brekke2023Aug}, where the magnon gap is larger, the transition occurs at $0.314 < \mu/t < 0.315$. With smaller magnon gap, $\Tilde{V}_{\boldsymbol{k}\boldsymbol{k}'}^{\text{FS}_\uparrow}$ increases in magnitude for $\boldsymbol{k}\approx \boldsymbol{k}'$ while there are relatively small changes when $\boldsymbol{k}$ and $\boldsymbol{k}'$ are far apart. Hence, the developing sign change at $\boldsymbol{k}' \approx (-k_x, k_y)$ becomes less important at low magnon gaps, which means that the transition moves to higher values of $\mu$.

Figure \ref{fig:TcAM}(b) shows $T_c^E$ as a function of chemical potential. The general BCS prediction, namely that $T_c$ increases significantly close to $\mu \lesssim J_{\text{sd}}S$ \cite{Brekke2023Aug}, remains. However, we find a dramatic suppression of $T_c$ compared to the BCS prediction by a factor of about $500$. This is due to the many-body effects that BCS theory neglects. For comparison, we plot the BCS ($T_c^{\text{BCS}}$) and AD ($T_c^{\text{AD}}$) estimates together with the solution to the eigenvalue problem in Eq.~\eqref{eq:gapeqEliash}, i.e., $T_c^E$. Since the AD estimate was constructed for single-boson-, specifically phonon-mediated superconductors, it must be noted that it gives a surprisingly good guess of $T_c^E$ in a system where double-magnon processes provide the superconducting pairing. Still, it does underestimate $T_c$ slightly, as was known for phonon-mediated superconductors \cite{Allen1975Aug}.

The factor of $500$ decrease in critical temperature compared to BCS is far more than what $1.13\omega_c \to \omega_{\text{log}}/1.2$ causes. We find $\omega_{\text{log}}/t \approx 0.01$ while $\omega_c/t \approx 0.064$. Hence, this gives a factor of 9 decrease, similar to that found for AFM/NM interfaces when comparing BCS to Eliashberg predictions \cite{Sun2023Aug, ThingstadEliashberg}. The major effect is in the exponentials, where $e^{-1/\lambda}$ goes to $e^{-1.04 Z(0)/\lambda_2(0)}$. When we use the normalized BCS gap solutions for the momentum dependence $\psi_\sigma(\boldsymbol{k})$, we get $\lambda_2(0) \approx \lambda$, where $\lambda$ is the largest eigenvalue in the linearized BCS gap equation [Eq.~\eqref{eq:linBCS}] at the same parameters. We have $Z(0) = 1 + \lambda_{1}^{\text{S}1}(0)+\lambda_1(0)$. The fact that there are two sources of mass renormalization and only one source of pairing ($\lambda_2$) is a disadvantage of double-magnon-mediated superconductivity. Furthermore, the fact that $\lambda_2(i\omega_{\nu}) < \lambda_1(i\omega_{\nu})$ leads to a significant decrease in $T_c^E$ compared to $T_c^{\text{BCS}}$. Phonon-mediated $s$-wave superconductors, in contrast, have $\lambda_2(0) \approx \lambda$ and $Z(0) \approx 1 +\lambda$ giving a less dramatic correction to $T_c$ compared to BCS estimates \cite{Allen1975Aug}. 

Compared to the BCS prediction $T_c^{\text{BCS}}$, the Eliashberg solution $T_c^E$ decreases more rapidly when $\mu$ decreases, as seen in Fig.~\ref{fig:TcAM}(b). This is because $\lambda_{1}^{\text{S}1}(i\omega_{\nu})$ increases when $\mu$ decreases, since this brings the FSs for spin up and spin down closer. Hence, the renormalization effects from single-magnon processes become more detrimental to superconductivity. 

Note the jump in $T_c^{\text{AD}}$ when the gap changes symmetry across the black vertical line in Fig.~\ref{fig:TcAM}(b). This is because $\omega_{\text{log}}$ increases by 25\% when going from nodeless to nodal gap when increasing $\mu$. Focusing on the FS for spin up, the electron band is flattest on the $\boldsymbol{\Gamma Y}$ line where $p_x$-wave has a node and $p_y$-wave is maximal. Hence, for nodeless $p_y$-wave, the low-energy-magnon scattering for electrons close to the intersection of the FS and the line $\boldsymbol{\Gamma Y}$ dominate more than any low-energy-magnon scatterings do for the nodal $p_x$-wave gap. Therefore $\omega_{\text{log}}$ is smaller for nodeless $p_y$-wave gap. BCS theory does not catch this effect. Hence, this indicates that the nodal $p_x$-wave gap may survive to lower $\mu$ in Eliashberg theory. Due to the exponential dependence on the dimensionless couplings, this should only be a small effect. Also, the transition point is strongly dependent on material parameters. 
Therefore we do not explore the Eliashberg correction to the transition line.

The magnon gap in Fig.~\ref{fig:TcAM} is $\omega_0/t \approx 0.0018$ while the critical temperature for superconductivity is at most $T_c^E/t \approx  1.8\times 10^{-5} $. Hence, we are at a much lower temperature than the magnon gap, ensuring the validity of the HP transformation and the validity of ignoring any thermal effects in the dimensionless couplings proportional to $n_B(\omega_{\boldsymbol{q}}^{\alpha/\beta})$. 
We found only small changes in the dimensionless couplings when increasing the number of points on the FS or the number of points $N$ in the sum over $\boldsymbol{q}$.
Double-magnon processes are more spread out in momentum compared to single-magnon processes, so fewer points are needed to get accurate sums. The cutoff $M = 0.03t$ is large enough to see $\phi_\sigma(i\omega_n)$ drop to 15 \% of its maximum. Increasing the cutoff gives only a slight increase in the estimate of $T_c^E$. 
From considering a larger cutoff, more points in FS average, or more points in the sum over $\boldsymbol{q}$ than in Fig.~\ref{fig:TcAM}, we expect that the numerical estimate of $T_c^E$ is at least accurate to within 5 \% relative deviation. In that case, we expect that the FS average should introduce more errors than the numerics. 
$\omega_{\text{log}}/t \approx 0.01$ quantifies the typical energy exchange in the interactions. This means that momenta where $|\epsilon_{\boldsymbol{k},\sigma}| < \omega_{\text{log}}$ should be most relevant, supporting the validity of the FS average at the chosen chemical potentials, especially with the added electron dispersion due to the third-nearest-neighbor hoppings. 

\begin{figure}
    \centering
    \includegraphics[width = \linewidth]{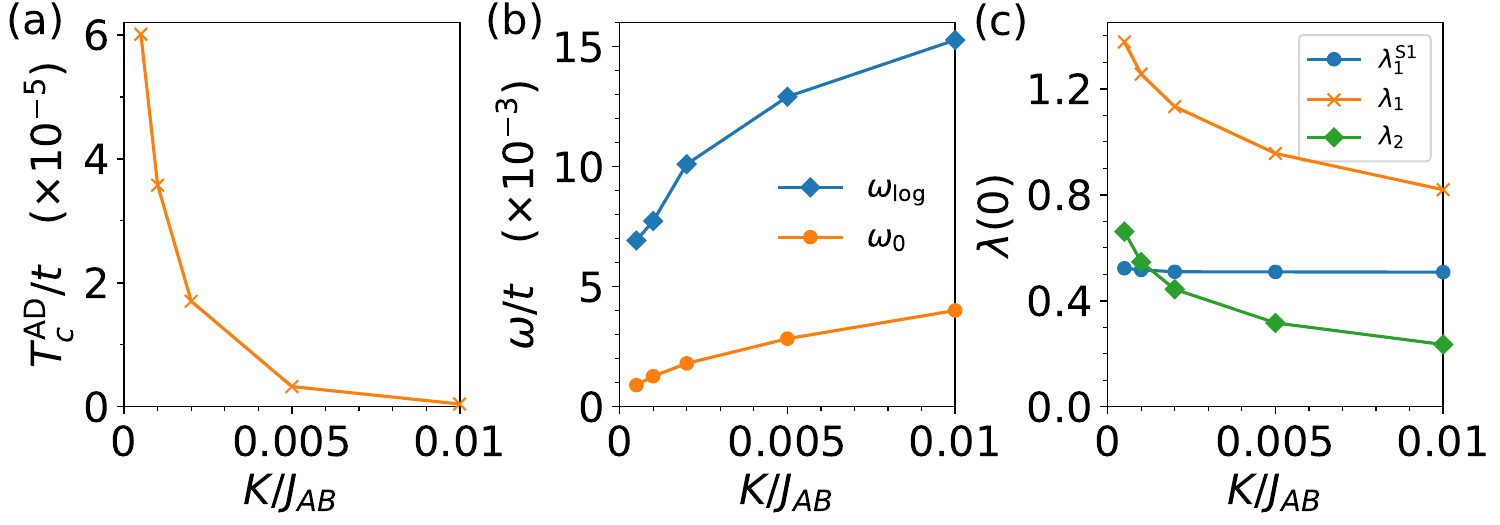}
    \caption{(a) The Allen-Dynes estimate $T_c^{\text{AD}}$ of the critical temperature, (b) $\omega_0$ and $\omega_{\text{log}}$, and (c) the dimensionless couplings at zero frequency, all plotted as a function of the easy-axis anisotropy $K$. Parameters: $t_2/t = 0.04$, $t_3/t =  0.02$, $J_{\text{sd}}S/t = 0.4$, $\mu/t = 0.39$, $S=3/2$, $J_{\text{AB}}S/t = 0.01$, $J_d/J_{\text{AB}} = -0.2$, and $J_{\text{nm}}/J_{\text{AB}} = -0.4$. $\psi_\sigma(\boldsymbol{k})$ is set to the normalized solution of the linearized BCS gap equation. To get accurate results, we used 110 points on the FS and $N = 150^2$ points in the sum over $\boldsymbol{q}$ at the lowest magnon gap.}
    \label{fig:TcK}
\end{figure}

We mentioned that a lower magnon gap should give a larger $\lambda_2(0)$ and $\lambda_2(0)/\lambda_1(0)$. Figure \ref{fig:TcK} explores this and the expected behavior of $T_c$ when lowering the magnon gap $\omega_0$. Since the AD estimate only slightly underestimates $T_c^E$ we do not solve the linearized Eliashberg equations in this figure. As expected, the logarithmic average $\omega_{\text{log}}$ decreases when lowering the magnon gap since low-energy magnons become more important. Still, $T_c^{\text{AD}}$ increases due to the increase of $\lambda_2(0)$ and $\lambda_2(0)/\lambda_1(0)$. On the other hand, $\lambda_1^{\text{S}1}$ varies very little with $K$ since it does not depend on the low-energy magnons. We predict that reaching a large $T_c$ in AMs will require a very small magnon gap.

The impact of many-body effects on $T_c$ for magnon-mediated superconductivity in AMs motivates us to explore an alternative system. In this system, double-magnon processes should still result in spin-polarized Cooper pairs, while single-magnon processes do not contribute to renormalization. This condition necessitates that only one spin component has an FS, a scenario typically observed in a half-metallic FM.

\section{Double-magnon processes in half-metallic ferromagnets} \label{sec:FM}

\begin{figure}
    \centering
    \includegraphics[width = 0.8\linewidth]{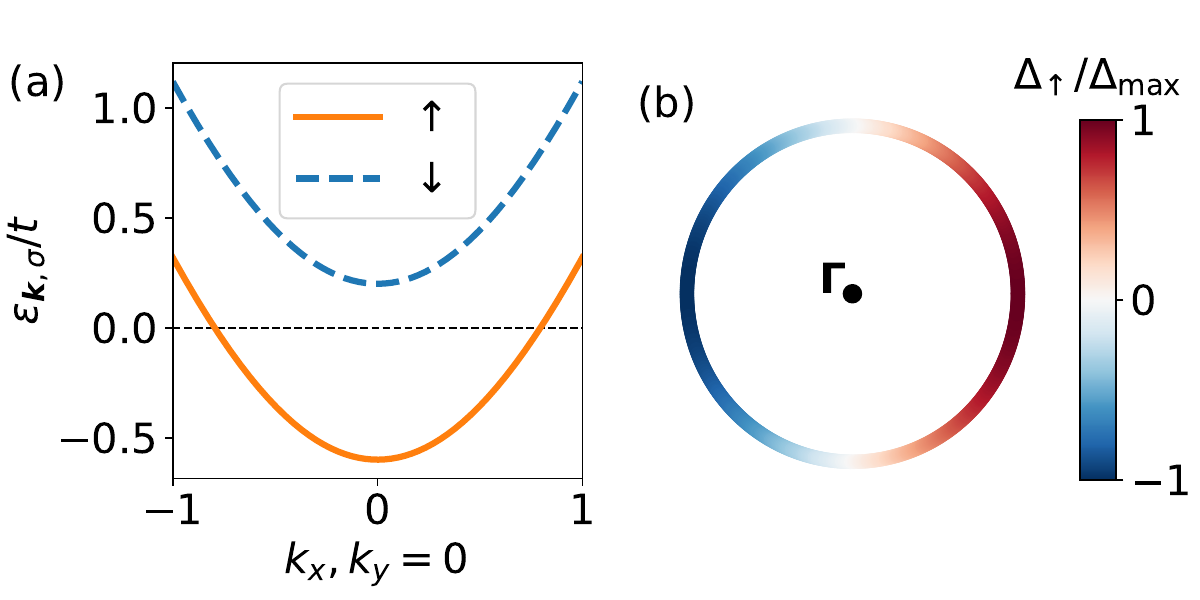}
    \caption{(a) The electron bands in a metallic FM for a case where only spin up crosses the FS. The bands are plotted along $k_x$, with $k_y=0$, giving an approximately circular FS around the $\boldsymbol{\Gamma}$ point. (b) A normalized $p_x$-wave gap for spin up, $\Delta_{\boldsymbol{k},\uparrow}$, shown on the FS. With parameters chosen as $t = 1$~eV, $S=1$, $\mu/t = -3.8$, $J_{\text{sd}}S/t = 0.4$, $JS/t = 0.005$, $J_x/J = 0.9999$, and $J_y/J = 0.3$, we get $\omega_c/t \approx 0.064$, $\omega_0/t \approx 0.0003$, $\lambda \approx 0.138$, and $T_c^{\text{BCS}} \approx 0.6$~K.}
    \label{fig:FM}
\end{figure}

In Appendix \ref{app:FM}, we present the linearized BCS gap equation for a half-metallic FM, describing spin-polarized superconductivity mediated by double-magnon processes. We consider a square lattice and a case where only spin up has an FS, as illustrated in Fig.~\ref{fig:FM}(a). To get double-magnon processes that contribute to superconducting pairing, we must consider an exchange anisotropic Hamiltonian for the magnet, with $J_x, J_y < J_z = J$, $J_x \neq J_y$, and $2\Delta J = J_x-J_y$. The size of $\Delta J$ is related to the strength of SOC in the system \cite{Skomski2003anisotropySOC}.
With details relegated to Appendix \ref{app:FM}, we find that any superconducting pairing is proportional to $\Delta J$, a disadvantage compared to the AM case where no SOC is required to generate the superconducting pairing.

Unlike the Lieb lattice model for an AM, the DOS is relatively low in the tight-binding model on the square lattice, especially in the ranges of chemical potential where only one spin component crosses the FS. A very small magnon gap and a substantial $\Delta J$ are necessary to get $\lambda > 0.1$. The magnon gap decreases when $J_x$ is closer to $J$. For smaller gaps, the magnitude of the Bogoliubov transformation factors $|u_{\boldsymbol{q}}|$ and $|v_{\boldsymbol{q}}|$ increase close to zero momentum, and the low-energy magnons dominate the interactions. 

Figure \ref{fig:FM}(b) illustrates the gap solution for a set of parameters where $\lambda \approx 0.138$ and $T_c^{\text{BCS}} \approx 0.6$~K. The gap $\Delta_{\boldsymbol{k},\uparrow}$ has a $p$-wave symmetry, similar to the prediction for ferromagnetic superconductors \cite{Linder2007FMSC, Jian2009FMSC}. Here, all linear combinations of $p_x$ and $p_y$ are degenerate due to the four-fold symmetry of the square lattice \cite{Sun2023Aug}. Since $\Delta J/J \approx 0.35$ this should entail a requirement for strong SOC to reach an appreciable $T_c$ \cite{Skomski2003anisotropySOC}.

$T_c^{\text{BCS}}$ and $\lambda$ are so small for reasonable parameters that we omit refining the estimate with strong-coupling theory. Still, we would expect that Eliashberg theory leads to a lower $T_c$. Low-energy magnons dominate, so $\omega_{\text{log}}$ should be small compared to $\omega_c$. We know that $\lambda_1^{\text{S}1} = 0$ since spin down has no FS, which is an advantage. We would also expect $\lambda_2 < \lambda_1$, though they should be closer in magnitude since the low-energy magnons dominate more than with the parameters we considered for the AM model. Hence, we expect that the Eliashberg estimate $T_c^E$ is reduced by a smaller factor compared to the BCS estimate $T_c^{\text{BCS}}$ than in the AM case. However, unlike the AM case, the BCS estimate for the critical temperature is small. One could imagine another lattice where a half-metallic FM can be realized with a larger DOS on the FS, but we leave our study here. The critical temperature due to EMC in a half-metallic FM is limited by the strength of SOC, which is a relativistic effect \cite{Skomski2003anisotropySOC}.

\section{Conclusion} \label{sec:conclusion}
We have discussed the role of double-magnon electron-magnon scattering in colinear magnets, focusing on magnon-mediated superconductivity. Single-magnon processes proved to be much stronger than double-magnon processes, provided both are active. As a result, double-magnon processes play little role in superconducting pairing in conventional antiferromagnets. On the other hand, they are the only source of magnon-mediated pairing in altermagnets, unconventional antiferromagnets that have nonrelativistic $d$-wave spin splitting due to broken structural symmetries. A strong-coupling approach to superconductivity revealed that many-body renormalization effects drastically reduce the possible critical temperature in altermagnets due to magnon-mediated superconductivity. This involved a detailed derivation of the self-energy diagrams that result from electron-magnon coupling with two electrons and two magnons at each vertex. At the same time, single-magnon processes also contribute to the many-body effects. We suggested ways in which $T_c$ can be increased compared to our estimates.
In an attempt to limit the many-body effects driven by single-magnon processes, we also considered double-magnon-mediated superconductivity in a ferromagnetic half-metal. Here, we found that the strength of spin-orbit coupling affects the critical temperature for superconductivity.

\section*{Acknowledgments} 
This work was funded by the Research Council of Norway (RCN) through its Centres of Excellence funding scheme Project No.~262633, ``QuSpin," and RCN Project No.~323766, ``Equilibrium and out-of-equilibrium quantum phenomena in superconducting hybrids with antiferromagnets and topological insulators."

\appendix

\section{BCS theory for half-metallic ferromagnet} \label{app:FM}

Electron bands in metallic FMs are spin split. 
We consider a square lattice, with all $\boldsymbol{S}_i = S \hat{z}$ in the classical ground state. The electron Hamiltonian is
\begin{align}
    H_e = -t\sum_{\left<i,j\right>,\sigma}  c_{i,\sigma}^{\dagger} c_{j,\sigma} -\mu \sum_{i, \sigma}  c_{i,\sigma}^{\dagger} c_{i,\sigma} + H_{\text{sd}} .
\end{align}
Applying an FT gives
\begin{equation}
    H_e = \sum_{\boldsymbol{k}\sigma} (\epsilon_{\boldsymbol{k}}-\sigma J_{\text{sd}}S) c_{\boldsymbol{k}\sigma}^\dagger c_{\boldsymbol{k}\sigma} = \sum_{\boldsymbol{k}\sigma} \epsilon_{\boldsymbol{k},\sigma} c_{\boldsymbol{k}\sigma}^\dagger c_{\boldsymbol{k}\sigma}.
\end{equation}
Here, $J_{\text{sd}}S$ acts like a magnetic field, and $\epsilon_{\boldsymbol{k}} = -\mu -2t(\cos k_x + \cos k_y)$.

FM groundstates are typically considered to be classical. Hence, we would expect that only thermal magnons can give double-magnon processes. Anisotropic exchange interactions can make FM states quantum mechanical \cite{Kamra2020MagnonSqueezingRev}. Anisotropic exchange originates with SOC and should, therefore, be a weak effect \cite{Skomski2003anisotropySOC}. In AMs, \textit{quantum} double-magnon processes exist without SOC. Here, \textit{quantum} is used in the sense that terms originate with commutators, not Bose factors (thermal magnons). The reason anisotropic exchange can lead to quantum effects is that a Bogoliubov transformation is required to derive the long-lived magnons. We consider the Hamiltonian
\begin{equation}
    H_m = -\sum_{\langle i,j \rangle, \alpha} J_\alpha S_{i,\alpha} S_{j,\alpha},
\end{equation}
for the localized spins, with $J = J_z, 2\Bar{J} = J_x +J_y$, and $2\Delta J = J_x-J_y$.
Inserting an HP transformation and an FT gives an off-diagonal description in terms of magnon operators $a_{\boldsymbol{q}}$ and $a_{\boldsymbol{q}}^\dagger$ to second order. Then, we introduce a Bogoliubov transformation $A_{\boldsymbol{q}} = u_{\boldsymbol{q}}a_{\boldsymbol{q}}+v_{\boldsymbol{q}}a_{-\boldsymbol{q}}^\dagger, A_{-\boldsymbol{q}}^\dagger = u_{\boldsymbol{q}}a_{-\boldsymbol{q}}^\dagger+v_{\boldsymbol{q}}a_{\boldsymbol{q}}$.
In order to diagonalize the magnon Hamiltonian, we require that
\begin{equation}
    u_{\boldsymbol{q}} = \sqrt{\frac{1}{2}+\frac{|\gamma_1(\boldsymbol{q})|}{2\sqrt{[\gamma_1(\boldsymbol{q})]^2-[\gamma_2(\boldsymbol{q})]^2}}}, 
\end{equation}
\begin{equation}
    v_{\boldsymbol{q}} = \text{sgn}[\gamma_2(\boldsymbol{q})]\sqrt{\frac{1}{2}-\frac{|\gamma_1(\boldsymbol{q})|}{2\sqrt{[\gamma_1(\boldsymbol{q})]^2-[\gamma_2(\boldsymbol{q})]^2}}},
\end{equation}
giving
$
    H_m = \sum_{\boldsymbol{q}} \omega_{\boldsymbol{q}} A_{\boldsymbol{q}}^\dagger A_{\boldsymbol{q}},
$
with
$
    \omega_{\boldsymbol{q}} = 2\sqrt{[\gamma_1(\boldsymbol{q})]^2-[\gamma_2(\boldsymbol{q})]^2},
$
$
    \gamma_1(\boldsymbol{q}) = 4JS -2\Bar{J}S(\cos q_x + \cos q_y),
$
and
$
    \gamma_2(\boldsymbol{q}) = 2\Delta J S(\cos q_x + \cos q_y).
$

If $\Delta J = 0$ we have $u_{\boldsymbol{q}} = 1$, $v_{\boldsymbol{q}} = 0$, i.e., there is no need for a Bogoliubov transformation. If $\Delta J \neq 0$, $|v_{\boldsymbol{q}}|$ increases. Hence, its magnitude is related to the degree of anisotropy, i.e., the strength of SOC. We find double-magnon EMC on the form
\begin{align}
    H_{\text{em}}^{(2)} &= \frac{J_{\text{sd}}}{N} \sum_{\boldsymbol{k}\boldsymbol{q}\boldsymbol{q}'} (u_{\boldsymbol{q}}u_{\boldsymbol{q}'}A_{\boldsymbol{q}}^\dagger A_{\boldsymbol{q}'} - u_{\boldsymbol{q}}v_{\boldsymbol{q}'}A_{\boldsymbol{q}}^\dagger A_{-\boldsymbol{q}'}^\dagger \nonumber\\
    &-v_{\boldsymbol{q}}u_{\boldsymbol{q}'}A_{-\boldsymbol{q}}A_{\boldsymbol{q}'}+v_{\boldsymbol{q}}v_{\boldsymbol{q}'}A_{-\boldsymbol{q}} A_{-\boldsymbol{q}'}^\dagger)c_{\boldsymbol{k}+\boldsymbol{q}'-\boldsymbol{q},\sigma}^\dagger c_{\boldsymbol{k}\sigma}.
\end{align}
Focusing on spin-polarized Cooper pairing, a Schrieffer-Wolff transformation \cite{Schrieffer1966Wolff} results in
\begin{equation}
    H_{\text{Pair}} = \sum_{\boldsymbol{k}\boldsymbol{k}'} V_{\boldsymbol{k}\boldsymbol{k}'\sigma} c_{\boldsymbol{k}'\sigma}^\dagger c_{-\boldsymbol{k}',\sigma}^\dagger c_{-\boldsymbol{k},\sigma} c_{\boldsymbol{k}\sigma},  
\end{equation}
with
\begin{align}
    V_{\boldsymbol{k}\boldsymbol{k}'\sigma} =& \frac{J_{\text{sd}}^2}{2N^2} \sum_{\boldsymbol{q}} \bigg[ \frac{(u_{\boldsymbol{k}-\boldsymbol{k}'+\boldsymbol{q}}v_{\boldsymbol{q}}v_{\boldsymbol{k}-\boldsymbol{k}'+\boldsymbol{q}}u_{\boldsymbol{q}} + u_{\boldsymbol{k}-\boldsymbol{k}'+\boldsymbol{q}}^2 v_{\boldsymbol{q}}^2)}{\epsilon_{\boldsymbol{k},\sigma}-\epsilon_{\boldsymbol{k}',\sigma}-\omega_{\boldsymbol{k}-\boldsymbol{k}'+\boldsymbol{q}}-\omega_{\boldsymbol{q}}} \nonumber \\
    & -\frac{(u_{\boldsymbol{k}-\boldsymbol{k}'+\boldsymbol{q}}v_{\boldsymbol{q}}v_{\boldsymbol{k}-\boldsymbol{k}'+\boldsymbol{q}}u_{\boldsymbol{q}} + v_{\boldsymbol{k}-\boldsymbol{k}'+\boldsymbol{q}}^2 u_{\boldsymbol{q}}^2)}{\epsilon_{\boldsymbol{k},\sigma}-\epsilon_{\boldsymbol{k}',\sigma}+\omega_{\boldsymbol{k}-\boldsymbol{k}'+\boldsymbol{q}}+\omega_{\boldsymbol{q}}} \bigg], \label{eq:VkksFM2}
\end{align}
Note that we only kept the terms coming from commutators and set all Bose factors to zero, assuming the temperature is lower than the magnon gap. Also note that if $\Delta J = 0$, $v_{\boldsymbol{q}}=0$ and so $V_{\boldsymbol{k}\boldsymbol{k}'\sigma} = 0$.
If $\boldsymbol{k}$ and $\boldsymbol{k}'$ are on the FS for spin $\sigma$ we get
\begin{equation}
    V_{\boldsymbol{k}\boldsymbol{k}'}^{\text{FS}_\sigma} = -\frac{J_{\text{sd}}^2}{2N^2} \sum_{\boldsymbol{q}} \frac{(u_{\boldsymbol{k}-\boldsymbol{k}'+\boldsymbol{q}}v_{\boldsymbol{q}} + u_{\boldsymbol{q}}v_{\boldsymbol{k}-\boldsymbol{k}'+\boldsymbol{q}})^2}{\omega_{\boldsymbol{k}-\boldsymbol{k}'+\boldsymbol{q}}+\omega_{\boldsymbol{q}}}.
\end{equation}
We now symmetrize the interaction as explained in Ref.~\cite{Maeland2023PRL}. We introduce $2\Tilde{V}_{\boldsymbol{k}\boldsymbol{k}'} = V_{\boldsymbol{k}\boldsymbol{k}'}+V_{-\boldsymbol{k},-\boldsymbol{k}'}-V_{-\boldsymbol{k},\boldsymbol{k}'}-V_{\boldsymbol{k},-\boldsymbol{k}'}$. Note that this is a slight change in notation from Ref.~\cite{Brekke2023Aug} where a factor $4$ was used instead. With this change in notation, the spectrum in the superconducting state is $E_{\boldsymbol{k},\sigma} = \pm \sqrt{\epsilon_{\boldsymbol{k},\sigma}^2 + |\Delta_{\boldsymbol{k},\sigma}|^2}$, where $\Delta_{\boldsymbol{k},\sigma} = \Delta_{\boldsymbol{k},\sigma\sigma}$ are the spin-polarized gaps. 
The linearized gap equation is \cite{Maeland2023PRL}
\begin{equation}
    \Delta_{\boldsymbol{k},\sigma} = -\sum_{\boldsymbol{k}'}  \tilde{V}_{\boldsymbol{k}\boldsymbol{k}'\sigma} \frac{\Delta_{\boldsymbol{k}',\sigma}}{2|\epsilon_{\boldsymbol{k}',\sigma}|}\tanh \frac{\beta |\epsilon_{\boldsymbol{k}',\sigma}|}{2}.
\end{equation}
The FS averaged version is derived following Ref.~\cite{Brekke2023Aug} with a slight change in notation, giving
\begin{equation}
    \lambda \Delta_{k_\parallel, \sigma} = -\frac{N S_{\text{FS}_\sigma}}{N_{\text{samp}}A_{\text{BZ}}}\sum_{k'_\parallel} \abs{\frac{\partial \epsilon}{\partial k'_\perp}}^{-1} \tilde{V}_{k_\parallel,k'_\parallel}^{\text{FS}_\sigma} \Delta_{k'_\parallel,\sigma}, \label{eq:linBCS}
\end{equation}
where $A_{\text{BZ}}$ is the area of the 1BZ,
$S_{\text{FS}_\sigma}$ is the length of the FS for spin $\sigma$, $N_{\text{samp}}$ is the number of points we sample on the FS, $k_\parallel$ denotes the momentum dependence parallel to the FS, and $|\partial \epsilon/\partial k'_\perp|$ is the slope of the electron band perpendicular to the FS. The linearized gap equation is an eigenvalue problem and the solution is the eigenvector $\Delta_{k_\parallel, \sigma}$ corresponding to the largest eigenvalue $\lambda$.

The same gap equation is used when finding the BCS prediction $T_c^{\text{BCS}}$ for the AM model in Sec.~\ref{sec:AM}. In that model, we have
\begin{align}
    V_{\boldsymbol{k},\boldsymbol{k}'}^{\text{FS}_\sigma} &= -\frac{J_{sd}^2}{N^2} \sum_{\boldsymbol{q}} \big( |u_{\boldsymbol{k}-\boldsymbol{k}'+\boldsymbol{q}}| |v_{\boldsymbol{q}}|\Omega^A_{\boldsymbol{k}', \boldsymbol{k}, \sigma, \sigma} \nonumber\\
    &-  |u_{\boldsymbol{q}}| |v_{\boldsymbol{k}-\boldsymbol{k}'+\boldsymbol{q}}|\Omega^B_{\boldsymbol{k}', \boldsymbol{k}, \sigma, \sigma}\big)^2/(\omega_{\boldsymbol{k}-\boldsymbol{k}'+\boldsymbol{q}}^\alpha + \omega_{\boldsymbol{q}}^\beta) . \label{eq:VkkAM}
\end{align}

\section{Derivation of self-energies} \label{app:selfenergy}

The renormalized Green's function is $G(\boldsymbol{k},\tau) = -\langle T_\tau \psi_{\boldsymbol{k}}(\tau) \psi_{\boldsymbol{k}}^\dagger (0) S(\beta, 0)\rangle_{\text{conn}}$, where the subscript ``conn'' indicates that it only includes connected diagrams \cite{BruusFlensberg, abrikosov, Schrieffer1964book}, and
\begin{equation}
    S(\tau, 0) = \sum_{n=0}^{\infty} \frac{(-1)^n}{n!} \int_0^{\tau} d\tau_1 \cdots d\tau_n T_\tau [H_{\text{int}}(\tau_1)\cdots H_{\text{int}}(\tau_n)]
\end{equation}
is the $S$ matrix. $H_{\text{int}}(\tau)$ is the interaction Hamiltonian and the interaction picture is used to calculate expectation values \cite{BruusFlensberg}. Here, we present the $S$-matrix expansion to derive the self-energy expressions for double-magnon processes.

The double-magnon EMC terms $H^{(2)}_{\text{em}}$ can in general be written as 
\begin{equation}
    H_{\text{int}} = \sum_{\boldsymbol{k}\boldsymbol{q}\boldsymbol{q}'\sigma\sigma'\gamma\gamma'} f_{\boldsymbol{k},\boldsymbol{q},\boldsymbol{q}'}^{\sigma\sigma'\gamma\gamma'} B_{-\boldsymbol{q}}^{\gamma}B_{\boldsymbol{q}'}^{\gamma'} d_{\boldsymbol{k}+\boldsymbol{q}'-\boldsymbol{q},\sigma}^\dagger d_{\boldsymbol{k}\sigma'},
\end{equation}
where $\boldsymbol{B}_{\boldsymbol{q}}$ is a vector of magnon operators and superscript $\gamma$ is the vector component. $f_{\boldsymbol{k},\boldsymbol{q},\boldsymbol{q}'}^{\sigma\sigma'\gamma\gamma'} $ is only nonzero for $\sigma = \sigma'$ since double-magnon processes do not flip spin.
Symmetrizing, we get
$
    H_{\text{int}} \to \sum_{\boldsymbol{k}\boldsymbol{q}\boldsymbol{q}'\sigma\sigma'\gamma\gamma'} f_{\boldsymbol{k},\boldsymbol{q},\boldsymbol{q}'}^{\sigma\sigma'\gamma\gamma'} B_{-\boldsymbol{q}}^{\gamma}B_{\boldsymbol{q}'}^{\gamma'} (d_{\boldsymbol{k}+\boldsymbol{q}'-\boldsymbol{q},\sigma}^\dagger d_{\boldsymbol{k}\sigma'}-d_{\boldsymbol{k}\sigma'}d_{\boldsymbol{k}+\boldsymbol{q}'-\boldsymbol{q},\sigma}^\dagger).
$
Technically, there should be a factor $1/2$ here. Below, when we calculate expectation values, operators can be contracted either to the left or to the right. If we fix this to only one direction for each operator we get a factor of 2, canceling the factor of $1/2$ from the symmetrization. 
Writing this in terms of the Nambu spinor gives
\begin{equation}
    H_{\text{int}} = \sum_{\boldsymbol{k}\boldsymbol{q}\boldsymbol{q}'\alpha\beta\gamma\gamma'} g_{\boldsymbol{k},\boldsymbol{q},\boldsymbol{q}'}^{\alpha\beta\gamma\gamma'} B_{-\boldsymbol{q}}^{\gamma}B_{\boldsymbol{q}'}^{\gamma'} \psi_{\boldsymbol{k}+\boldsymbol{q}'-\boldsymbol{q},\alpha}^\dagger \psi_{\boldsymbol{k}\beta}.
\end{equation}
Now, the leftmost (rightmost) fermion operator is to be contracted to the left (right).

The Green's function involves an expectation value of a large number of fermion operators.
From Wick's theorem \cite{Wick1950Oct}, the average of a product of fermion operators can be calculated as a product of expectation values of two operators, with all possible contractions of two operators. 
If two ways to contract only differ by $H_{\text{int}}(\tau_i) \leftrightarrow H_{\text{int}}(\tau_j)$ they are equivalent since all $H_{\text{int}}(\tau_i)$ are the same. There are $n!$ ways to order the $n$ interaction Hamiltonians, so we get a factor of $n!$ times one of the orderings,
\begin{widetext}
\begin{equation}
    G(\boldsymbol{k}, \tau) = \sum_{n=0}^{\infty} (-1)^n\int_0^{\beta} d\tau_1 \cdots d\tau_n \langle -T_\tau [\psi_{\boldsymbol{k}}(\tau) H_{\text{int}}(\tau_1)\cdots H_{\text{int}}(\tau_n)\psi_{\boldsymbol{k}}^\dagger(0)]\rangle_{\substack{\text{conn}\\ \text{fixed}}},
\end{equation}
where subscript ``fixed'' indicates that there is a fixed way of contracting the fermion operators. The $n=0$ term is just the bare propagator $G_0$. Letting $n = n_1 + n_2 + 1$, $M = n_1 +1$, and inserting the expression for $H_{\text{int}}(\tau_{M})$ gives
\begin{align}
    G(\boldsymbol{k}, \tau) =& G_0(\boldsymbol{k}, \tau) + \sum_{n_1=0}^{\infty} \sum_{n_2=0}^{\infty} (-1)^{n_1+n_2+1} \int_0^{\beta} d\tau_1 \cdots d\tau_{n_1} d\tau_M \cdots d\tau_{M+n_2} \sum_{\substack{\boldsymbol{k}_2 \boldsymbol{q}_2 \boldsymbol{q}'_2 \\ \alpha_2 \beta_2 \gamma_2 \gamma'_2}} g_{\boldsymbol{k}_2,\boldsymbol{q}_2, \boldsymbol{q}'_2}^{\alpha_2\beta_2\gamma_2\gamma'_2} \nonumber \\
    &\times \langle -T_\tau \{ \psi_{\boldsymbol{k}}(\tau) H_{\text{int}}(\tau_1)\cdots H_{\text{int}}(\tau_{n_1})B_{-\boldsymbol{q}_2}^{\gamma_2}(\tau_M)B_{\boldsymbol{q}'_2}^{\gamma'_2}(\tau_M) \psi_{\boldsymbol{k}_2+\boldsymbol{q}'_2-\boldsymbol{q}_2, \alpha_2}^\dagger(\tau_M)\}\rangle_{\substack{\text{conn}\\ \text{fixed}}}  \nonumber\\
    &\times \langle T_\tau \{ \psi_{\boldsymbol{k}_2 \beta_2}(\tau_M)[H_{\text{int}}(\tau_{M+1})\cdots H_{\text{int}}(\tau_{M+n_2})]\psi_{\boldsymbol{k}}^\dagger(0)\}\rangle_{\substack{\text{conn}\\ \text{fixed}}}.
\end{align}

\subsection{Magnon-loop diagram}
The $n_1 = 0$ term yields the magnon-loop (ML) diagram, named $\Sigma^{\text{ML}}$ and illustrated in Fig.~\ref{fig:Dyson}. We find
\begin{align}
    &(-1)\sum_{\substack{\boldsymbol{k}_2 \boldsymbol{q}_2 \boldsymbol{q}'_2 \\ \alpha_2 \beta_2 \gamma_2 \gamma'_2}} g_{\boldsymbol{k}_2,\boldsymbol{q}_2, \boldsymbol{q}'_2}^{\alpha_2\beta_2\gamma_2\gamma'_2} \int_0^{\beta} d\tau_M \langle -T_\tau \psi_{\boldsymbol{k}\rho}(\tau)\psi_{\boldsymbol{k}_2+\boldsymbol{q}'_2-\boldsymbol{q}_2, \alpha_2}^\dagger(\tau_M)\rangle  \langle -T_\tau B_{-\boldsymbol{q}_2}^{\gamma_2}(\tau_M)B_{\boldsymbol{q}'_2}^{\gamma'_2}(\tau_M) \rangle  \nonumber\\
    &\times \sum_{n_2=0}^{\infty} (-1)^{n_2} \int_0^{\beta} d\tau_{M+1} \cdots d\tau_{M+n_2}  \langle -T_\tau \{ \psi_{\boldsymbol{k}_2 \beta_2}(\tau_M)[H_{\text{int}}(\tau_{M+1})\cdots H_{\text{int}}(\tau_{M+n_2})]\psi_{\boldsymbol{k}\sigma}^\dagger(0)\}\rangle_{\substack{\text{conn}\\ \text{fixed}}}.
\end{align}
We recognize the first term as $G_0$ and the last term as $G$. The term $\langle -T_\tau
B_{-\boldsymbol{q}_2}^{\gamma_2}(\tau_M)B_{\boldsymbol{q}'_2}^{\gamma'_2}(\tau_M) \rangle$ is not a magnon propagator, since the two operators act at equal time. Since only time-differences matter in equilibrium, we have $\langle -T_\tau B_{-\boldsymbol{q}_2}^{\gamma_2}(\tau_M)B_{\boldsymbol{q}'_2}^{\gamma'_2}(\tau_M) \rangle = \langle -T_\tau B_{-\boldsymbol{q}_2}^{\gamma_2}(0)B_{\boldsymbol{q}'_2}^{\gamma'_2}(0) \rangle = \langle - B_{-\boldsymbol{q}_2}^{\gamma_2}B_{\boldsymbol{q}'_2}^{\gamma'_2} \rangle$. 
To get nonzero expectation values, we need the operators to be diagonal in momentum, so $\boldsymbol{k} = \boldsymbol{k}_2+\boldsymbol{q}'_2-\boldsymbol{q}_2$, $\boldsymbol{k} = \boldsymbol{k}_2$, and $\boldsymbol{q}_2 = \boldsymbol{q}'_2 = \boldsymbol{q}$. 
This gives $\sum_{\boldsymbol{q} \alpha \beta \gamma \gamma'} g_{\boldsymbol{k},\boldsymbol{q}, \boldsymbol{q}}^{\alpha\beta\gamma\gamma'} \int_0^{\beta} d\tau_M G_{0,\rho\alpha}(\boldsymbol{k}, \tau-\tau_M) \langle  B_{-\boldsymbol{q}}^{\gamma}B_{\boldsymbol{q}}^{\gamma'} \rangle G_{\beta\sigma}(\boldsymbol{k}, \tau_M).$
From an FT to imaginary frequency, we get 
\begin{align}
     G_{\rho \sigma}(\boldsymbol{k}, i\omega_n) =& G_{0,\rho\sigma} (\boldsymbol{k}, i\omega_n) + \sum_{\boldsymbol{q}  \alpha \beta \gamma \gamma'} g_{\boldsymbol{k},\boldsymbol{q}, \boldsymbol{q}}^{\alpha\beta\gamma\gamma'} G_{0,\rho\alpha}(\boldsymbol{k}, i\omega_n) \langle B_{-\boldsymbol{q}}^{\gamma}B_{\boldsymbol{q}}^{\gamma'} \rangle  G_{\beta\sigma}(\boldsymbol{k}, i\omega_n).
\end{align}
We extract the self-energy from the Dyson equation, giving
$
    \Sigma^{\text{ML}}_{\alpha\beta}(\boldsymbol{k}) = \sum_{\boldsymbol{q}  \gamma \gamma'} g_{\boldsymbol{k},\boldsymbol{q}, \boldsymbol{q}}^{\alpha\beta\gamma\gamma'}\langle B_{-\boldsymbol{q}}^{\gamma}B_{\boldsymbol{q}}^{\gamma'} \rangle.
$
Note that $\Sigma^{\text{ML}}_{\alpha\beta}(\boldsymbol{k})$ is frequency independent since the diagram does not involve any scattering of the electrons. It is however momentum and spin dependent.

\subsection{Electron-loop diagram}
The $n_1 \geq 1$ terms are, after writing out both $H_{\text{int}}(\tau_{1})$ and $H_{\text{int}}(\tau_{M})$, given by
\begin{align}
    &\sum_{n_1=1}^{\infty} \sum_{n_2=0}^{\infty} (-1)^{n_1+n_2+1} \int_0^{\beta} d\tau_1 \cdots d\tau_{n_1} d\tau_M \cdots d\tau_{M+n_2} \sum_{\substack{\boldsymbol{k}_1 \boldsymbol{q}_1 \boldsymbol{q}'_1 \\ \alpha_1 \beta_1 \gamma_1 \gamma'_1}}\sum_{\substack{\boldsymbol{k}_2 \boldsymbol{q}_2 \boldsymbol{q}'_2 \\ \alpha_2 \beta_2 \gamma_2 \gamma'_2}} g_{\boldsymbol{k}_1,\boldsymbol{q}_1, \boldsymbol{q}'_1}^{\alpha_1\beta_1\gamma_1\gamma'_1}  g_{\boldsymbol{k}_2,\boldsymbol{q}_2, \boldsymbol{q}'_2}^{\alpha_2\beta_2\gamma_2\gamma'_2} \nonumber \\
    &\times \langle -T_\tau \{ \psi_{\boldsymbol{k}}(\tau) B_{-\boldsymbol{q}_1}^{\gamma_1}(\tau_1)B_{\boldsymbol{q}'_1}^{\gamma'_1}(\tau_1) \psi_{\boldsymbol{k}_1+\boldsymbol{q}'_1-\boldsymbol{q}_1, \alpha_1}^\dagger(\tau_1)\psi_{\boldsymbol{k}_1 \beta_1}(\tau_1) [H_{\text{int}}(\tau_2)\cdots H_{\text{int}}(\tau_{n_1})] \nonumber \\
    &\times B_{-\boldsymbol{q}_2}^{\gamma_2}(\tau_M)B_{\boldsymbol{q}'_2}^{\gamma'_2}(\tau_M) \psi_{\boldsymbol{k}_2+\boldsymbol{q}'_2-\boldsymbol{q}_2, \alpha_2}^\dagger(\tau_M) \psi_{\boldsymbol{k}_2 \beta_2}(\tau_M)[H_{\text{int}}(\tau_{M+1})\cdots H_{\text{int}}(\tau_{M+n_2})]\psi_{\boldsymbol{k}}^\dagger(0)\}\rangle_{\substack{\text{conn}\\ \text{fixed}}}.
\end{align}
First, we derive the tadpolelike electron-loop (EL) diagram by contracting the two Nambu spinors at $\tau_M$. Such a contraction breaks the rules we set forth after symmetrizing. When contracting at equal time, we should not symmetrize $H_{\text{int}}(\tau_M)$, but leave it in its original form. That is the same as restricting to the first half of the Nambu spinor, i.e., restricting $\alpha_2, \beta_2$ to just 1 and 2.
\begin{align}
    &(-1) \int_0^{\beta} d\tau_1 d\tau_M \sum_{\substack{\boldsymbol{k}_1 \boldsymbol{q}_1 \boldsymbol{q}'_1 \\ \alpha_1 \beta_1 \gamma_1 \gamma'_1}}\sum_{\substack{\boldsymbol{k}_2 \boldsymbol{q}_2 \boldsymbol{q}'_2  \gamma_2 \gamma'_2\\  \alpha_2, \beta_2 \in\{1,2\}}} g_{\boldsymbol{k}_1,\boldsymbol{q}_1, \boldsymbol{q}'_1}^{\alpha_1\beta_1\gamma_1\gamma'_1}  g_{\boldsymbol{k}_2,\boldsymbol{q}_2, \boldsymbol{q}'_2}^{\alpha_2\beta_2\gamma_2\gamma'_2} \langle -T_\tau \psi_{\boldsymbol{k}\rho}(\tau)\psi_{\boldsymbol{k}_1+\boldsymbol{q}'_1-\boldsymbol{q}_1, \alpha_1}^\dagger(\tau_1)\rangle \langle T_\tau \psi_{\boldsymbol{k}_2+\boldsymbol{q}'_2-\boldsymbol{q}_2, \alpha_2}^\dagger(\tau_M) \psi_{\boldsymbol{k}_2 \beta_2}(\tau_M)\rangle \nonumber \\
    &\times \big[ \langle -T_\tau B_{-\boldsymbol{q}_1}^{\gamma_1}(\tau_1) B_{-\boldsymbol{q}_2}^{\gamma_2}(\tau_M) \rangle \langle -T_\tau B_{\boldsymbol{q}'_1}^{\gamma'_1}(\tau_1)B_{\boldsymbol{q}'_2}^{\gamma'_2}(\tau_M)\rangle + \langle -T_\tau B_{-\boldsymbol{q}_1}^{\gamma_1}(\tau_1) B_{\boldsymbol{q}'_2}^{\gamma'_2}(\tau_M) \rangle \langle -T_\tau B_{\boldsymbol{q}'_1}^{\gamma'_1}(\tau_1) B_{-\boldsymbol{q}_2}^{\gamma_2}(\tau_M) \rangle \big] \nonumber \\
    &\times \sum_{n_1=1}^{\infty} \sum_{n_2=0}^{\infty} (-1)^{n_1+n_2-1} \int_0^{\beta}d\tau_2 \cdots d\tau_{n_1} d\tau_{M+1} \cdots d\tau_{M+n_2}\nonumber\\
    &\times \langle -T_\tau \psi_{\boldsymbol{k}_1 \beta_1}(\tau_1) [H_{\text{int}}(\tau_2)\cdots H_{\text{int}}(\tau_{n_1}) H_{\text{int}}(\tau_{M+1})\cdots H_{\text{int}}(\tau_{M+n_2})]\psi_{\boldsymbol{k}\sigma}^\dagger(0)\}\rangle_{\substack{\text{conn}\\ \text{fixed}}}.
\end{align}
To get nonzero expectation values, we need $\boldsymbol{k} = \boldsymbol{k}_1+\boldsymbol{q}'_1-\boldsymbol{q}_1$, $\boldsymbol{k}_2 + \boldsymbol{q}'_2-\boldsymbol{q}_2 = \boldsymbol{k}_2$, $\boldsymbol{k}_1 = \boldsymbol{k}$. So, $\boldsymbol{q}_1 = \boldsymbol{q}'_1$ and $\boldsymbol{q}_2 = \boldsymbol{q}'_2$. For the two choices of magnon contraction we must choose either $\boldsymbol{q}_1 = \boldsymbol{q}_2$ or $\boldsymbol{q}_1 = -\boldsymbol{q}_2$. Then, we can recognize Green's functions, giving
\begin{align}
    &(-1) \int_0^{\beta} d\tau_1 d\tau_M \sum_{\boldsymbol{k}_2 \boldsymbol{q}}\sum_{\substack{\alpha_1 \beta_1 \gamma_1 \gamma'_1 \gamma_2 \gamma'_2\\ \alpha_2, \beta_2 \in \{1,2\} }}  G_{0,\rho\alpha_1}(\boldsymbol{k}, \tau-\tau_1) \langle \psi_{\boldsymbol{k}_2, \alpha_2}^\dagger \psi_{\boldsymbol{k}_2 \beta_2}\rangle  \big[ g_{\boldsymbol{k},\boldsymbol{q}, \boldsymbol{q}}^{\alpha_1\beta_1\gamma_1\gamma'_1}  g_{\boldsymbol{k}_2,-\boldsymbol{q}, -\boldsymbol{q}}^{\alpha_2\beta_2\gamma_2\gamma'_2} D_{\gamma_1 \gamma_2}(-\boldsymbol{q}, \tau_1-\tau_M)\nonumber \\
    &\times  D_{\gamma'_1 \gamma'_2}(\boldsymbol{q}, \tau_1-\tau_M)+ g_{\boldsymbol{k},\boldsymbol{q}, \boldsymbol{q}}^{\alpha_1\beta_1\gamma_1\gamma'_1}  g_{\boldsymbol{k}_2,\boldsymbol{q}, \boldsymbol{q}}^{\alpha_2\beta_2\gamma_2\gamma'_2} D_{\gamma_1 \gamma'_2}(-\boldsymbol{q}, \tau_1-\tau_M) D_{\gamma'_1 \gamma_2}(\boldsymbol{q}, \tau_1-\tau_M) \big]  G_{\beta_1\sigma}(\boldsymbol{k}, \tau_1).
\end{align}

The FT of $F(\tau) = \int_0^\beta d\tau_1 d\tau_M F_1(\tau-\tau_1) F_2(\tau_1-\tau_M) F_3(\tau_1-\tau_M) F_4(\tau_1)$ is $F(i\omega_n) = T\sum_{\nu} F_1(i\omega_n) F_2(-i\omega_\nu)F_3(i\omega_\nu) F_4(i\omega_n)$. This gives
\begin{align}
    G_{\rho \sigma}&(\boldsymbol{k}, i\omega_n) = G_{0,\rho\sigma} (\boldsymbol{k}, i\omega_n) + (-1) \frac{1}{\beta} \sum_{\boldsymbol{k}_2, \boldsymbol{q},i\omega_\nu}\sum_{\substack{\alpha_1 \beta_1 \gamma_1 \gamma'_1 \gamma_2 \gamma'_2\\ \alpha_2, \beta_2 \in \{1,2\} }}  G_{0,\rho\alpha_1}(\boldsymbol{k}, i\omega_n) \langle \psi_{\boldsymbol{k}_2, \alpha_2}^\dagger \psi_{\boldsymbol{k}_2 \beta_2}\rangle  \big[ g_{\boldsymbol{k},\boldsymbol{q}, \boldsymbol{q}}^{\alpha_1\beta_1\gamma_1\gamma'_1}  g_{\boldsymbol{k}_2,-\boldsymbol{q}, -\boldsymbol{q}}^{\alpha_2\beta_2\gamma_2\gamma'_2}  \nonumber \\
    &\times D_{\gamma_1 \gamma_2}(-\boldsymbol{q}, -i\omega_\nu)D_{\gamma'_1 \gamma'_2}(\boldsymbol{q}, i\omega_\nu) + g_{\boldsymbol{k},\boldsymbol{q}, \boldsymbol{q}}^{\alpha_1\beta_1\gamma_1\gamma'_1}  g_{\boldsymbol{k}_2,\boldsymbol{q}, \boldsymbol{q}}^{\alpha_2\beta_2\gamma_2\gamma'_2} D_{\gamma_1 \gamma'_2}(-\boldsymbol{q}, -i\omega_\nu) D_{\gamma'_1 \gamma_2}(\boldsymbol{q}, i\omega_\nu) \big]  G_{\beta_1\sigma}(\boldsymbol{k}, i\omega_n).
\end{align}
We can read off the electron-loop self-energy as
\begin{align}
    \Sigma^{\text{EL}}_{\alpha_1\beta_1}(\boldsymbol{k}) =& - \frac{1}{\beta} \sum_{\boldsymbol{k}_2, \boldsymbol{q},i\omega_\nu} \sum_{\substack{ \gamma_1 \gamma'_1 \gamma_2 \gamma'_2\\ \alpha_2, \beta_2 \in \{1,2\} }} \langle \psi_{\boldsymbol{k}_2, \alpha_2}^\dagger \psi_{\boldsymbol{k}_2 \beta_2}\rangle  \big[ g_{\boldsymbol{k},\boldsymbol{q}, \boldsymbol{q}}^{\alpha_1\beta_1\gamma_1\gamma'_1}  g_{\boldsymbol{k}_2,-\boldsymbol{q}, -\boldsymbol{q}}^{\alpha_2\beta_2\gamma_2\gamma'_2} D_{\gamma_1 \gamma_2}(-\boldsymbol{q}, -i\omega_\nu) D_{\gamma'_1 \gamma'_2}(\boldsymbol{q}, i\omega_\nu) \nonumber \\
    &+ g_{\boldsymbol{k},\boldsymbol{q}, \boldsymbol{q}}^{\alpha_1\beta_1\gamma_1\gamma'_1}  g_{\boldsymbol{k}_2,\boldsymbol{q}, \boldsymbol{q}}^{\alpha_2\beta_2\gamma_2\gamma'_2} D_{\gamma_1 \gamma'_2}(-\boldsymbol{q}, -i\omega_\nu) D_{\gamma'_1 \gamma_2}(\boldsymbol{q}, i\omega_\nu) \big].
\end{align}
As is the case for the magnon-loop self-energy, $\Sigma^{\text{EL}}_{\alpha_1\beta_1}(\boldsymbol{k})$ is frequency independent, since there is no scattering of the external electron. By an alternate derivation, we could get $\langle \psi_{\boldsymbol{k}_2, \alpha_2}^\dagger \psi_{\boldsymbol{k}_2 \beta_2}\rangle \to T\sum_{i\omega_{n_2}} G_{\alpha_2\beta_2}(\boldsymbol{k}_2, i\omega_{n_2})$ which we interpret as a renormalized Fermi-Dirac distribution. For simplicity, we ignore this renormalization.

\subsection{Double-magnon sunset diagram}
Finally, the self-energy in the double-magnon sunset diagram $\Sigma^{\text{S}2}$ is derived by using the symmetrized form of the interaction and avoiding contracting operators at equal time. We find
\begin{align}
    & \int_0^{\beta} d\tau_1 d\tau_M \sum_{\substack{\boldsymbol{k}_1 \boldsymbol{q}_1 \boldsymbol{q}'_1 \\ \alpha_1 \beta_1 \gamma_1 \gamma'_1}}\sum_{\substack{\boldsymbol{k}_2 \boldsymbol{q}_2 \boldsymbol{q}'_2 \\ \alpha_2 \beta_2 \gamma_2 \gamma'_2}} g_{\boldsymbol{k}_1,\boldsymbol{q}_1, \boldsymbol{q}'_1}^{\alpha_1\beta_1\gamma_1\gamma'_1}  g_{\boldsymbol{k}_2,\boldsymbol{q}_2, \boldsymbol{q}'_2}^{\alpha_2\beta_2\gamma_2\gamma'_2} \langle -T_\tau \psi_{\boldsymbol{k}\rho}(\tau)\psi_{\boldsymbol{k}_1+\boldsymbol{q}'_1-\boldsymbol{q}_1, \alpha_1}^\dagger(\tau_1) \rangle \nonumber \\
    &\times \sum_{n_1=1}^{\infty} (-1)^{n_1-1} \int_0^{\beta} d\tau_2 \cdots d\tau_{n_1} \langle -T_\tau \psi_{\boldsymbol{k}_1 \beta_1}(\tau_1) [H_{\text{int}}(\tau_2)\cdots H_{\text{int}}(\tau_{n_1})]\psi_{\boldsymbol{k}_2+\boldsymbol{q}'_2-\boldsymbol{q}_2, \alpha_2}^\dagger(\tau_M) \rangle_{\substack{\text{conn}\\ \text{fixed}}} \nonumber \\
    &\times \big[ \langle -T_\tau B_{-\boldsymbol{q}_1}^{\gamma_1}(\tau_1) B_{-\boldsymbol{q}_2}^{\gamma_2}(\tau_M) \rangle \langle -T_\tau B_{\boldsymbol{q}'_1}^{\gamma'_1}(\tau_1)B_{\boldsymbol{q}'_2}^{\gamma'_2}(\tau_M)\rangle + \langle -T_\tau B_{-\boldsymbol{q}_1}^{\gamma_1}(\tau_1) B_{\boldsymbol{q}'_2}^{\gamma'_2}(\tau_M) \rangle \langle -T_\tau B_{\boldsymbol{q}'_1}^{\gamma'_1}(\tau_1) B_{-\boldsymbol{q}_2}^{\gamma_2}(\tau_M) \rangle \big] \nonumber \\
    &\times  \sum_{n_2=0}^{\infty} (-1)^{n_2} \int_0^{\beta} d\tau_{M+1} \cdots d\tau_{M+n_2} \langle -T_\tau  \psi_{\boldsymbol{k}_2 \beta_2}(\tau_M)[H_{\text{int}}(\tau_{M+1})\cdots H_{\text{int}}(\tau_{M+n_2})]\psi_{\boldsymbol{k}\sigma}^\dagger(0)\rangle_{\substack{\text{conn}\\ \text{fixed}}}.
\end{align}
To get nonzero electron Green's functions, we must choose $\boldsymbol{k} = \boldsymbol{k}_1+\boldsymbol{q}'_1-\boldsymbol{q}_1$, $\boldsymbol{k}_1 = \boldsymbol{k}_2+\boldsymbol{q}'_2-\boldsymbol{q}_2$, $\boldsymbol{k}_2 = \boldsymbol{k}$. So, $\boldsymbol{q}'_1-\boldsymbol{q}_1 = -(\boldsymbol{q}'_2-\boldsymbol{q}_2)$. To get nonzero boson propagators, we must have $\boldsymbol{q}_2 = -\boldsymbol{q}_1$ and $\boldsymbol{q}'_2 = -\boldsymbol{q}'_1$, or, in the second term, $\boldsymbol{q}'_2 = \boldsymbol{q}_1$ and $\boldsymbol{q}_2 = \boldsymbol{q}'_1$.  We choose $\boldsymbol{q}'_2 = \boldsymbol{q}'$ and $\boldsymbol{q}_2 = \boldsymbol{q}$ as free momenta. Then,
\begin{align}
    &\int_0^{\beta} d\tau_1 d\tau_M \sum_{\boldsymbol{q} \boldsymbol{q}'}\sum_{\substack{\alpha_1 \beta_1 \gamma_1 \gamma'_1 \\ \alpha_2 \beta_2 \gamma_2 \gamma'_2}}  G_{0,\rho\alpha_1} (\boldsymbol{k}, \tau-\tau_1) G_{\beta_1\alpha_2}(\boldsymbol{k}+\boldsymbol{q}'-\boldsymbol{q}, \tau_1-\tau_M) \big[ g_{\boldsymbol{k}+\boldsymbol{q}'-\boldsymbol{q},-\boldsymbol{q}, -\boldsymbol{q}'}^{\alpha_1\beta_1\gamma_1\gamma'_1}  g_{\boldsymbol{k},\boldsymbol{q}, \boldsymbol{q}'}^{\alpha_2\beta_2\gamma_2\gamma'_2}  
 D_{\gamma_1\gamma_2}(\boldsymbol{q}, \tau_1-\tau_M)\nonumber \\
    &\times D_{\gamma'_1\gamma'_2}(-\boldsymbol{q}', \tau_1-\tau_M) +g_{\boldsymbol{k}+\boldsymbol{q}'-\boldsymbol{q},\boldsymbol{q}', \boldsymbol{q}}^{\alpha_1\beta_1\gamma_1\gamma'_1}  g_{\boldsymbol{k},\boldsymbol{q}, \boldsymbol{q}'}^{\alpha_2\beta_2\gamma_2\gamma'_2}  D_{\gamma_1\gamma'_2}(-\boldsymbol{q}', \tau_1-\tau_M)D_{\gamma'_1\gamma_2}(\boldsymbol{q}, \tau_1-\tau_M) \big] G_{\beta_2 \sigma}(\boldsymbol{k}, \tau_M).
\end{align}
The FT of $F(\tau) = \int_0^\beta d\tau_1 d\tau_M F_1(\tau-\tau_1) F_2(\tau_1-\tau_M) F_3(\tau_1-\tau_M) F_4(\tau_1-\tau_M) F_5(\tau_M)$ is $F(i\omega_n) = T^2\sum_{\nu_3\nu_4} F_1(i\omega_n) F_2(i\omega_n-i\omega_{\nu_3}-i\omega_{\nu_4})F_3(i\omega_{\nu_3}) F_4(i\omega_{\nu_4}) F_5(i\omega_n)$. We can write this as
\begin{align}
    G_{\rho \sigma}&(\boldsymbol{k}, i\omega_n) = G_{0,\rho\sigma} (\boldsymbol{k}, i\omega_n) +\frac{1}{\beta^2} \sum_{\substack{\boldsymbol{q} \boldsymbol{q}' \\ \nu\nu'}}\sum_{\substack{\alpha_1 \beta_1 \gamma_1 \gamma'_1 \\ \alpha_2 \beta_2 \gamma_2 \gamma'_2}}  G_{0,\rho\alpha_1} (\boldsymbol{k}, i\omega_n) G_{\beta_1\alpha_2}(\boldsymbol{k}+\boldsymbol{q}'-\boldsymbol{q}, i\omega_n+i\omega_{\nu'}-i\omega_{\nu})  \big[ g_{\boldsymbol{k}+\boldsymbol{q}'-\boldsymbol{q},-\boldsymbol{q}, -\boldsymbol{q}'}^{\alpha_1\beta_1\gamma_1\gamma'_1}\nonumber \\
    &\times  g_{\boldsymbol{k},\boldsymbol{q}, \boldsymbol{q}'}^{\alpha_2\beta_2\gamma_2\gamma'_2} D_{\gamma_1\gamma_2}(\boldsymbol{q}, i\omega_\nu) D_{\gamma'_1\gamma'_2}(-\boldsymbol{q}', -i\omega_{\nu'}) +g_{\boldsymbol{k}+\boldsymbol{q}'-\boldsymbol{q},\boldsymbol{q}', \boldsymbol{q}}^{\alpha_1\beta_1\gamma_1\gamma'_1}  g_{\boldsymbol{k},\boldsymbol{q}, \boldsymbol{q}'}^{\alpha_2\beta_2\gamma_2\gamma'_2}  D_{\gamma_1\gamma'_2}(-\boldsymbol{q}', -i\omega_{\nu'})D_{\gamma'_1\gamma_2}(\boldsymbol{q}, i\omega_\nu) \big] G_{\beta_2 \sigma}(\boldsymbol{k}, i\omega_n),
\end{align}
and read off the self-energy
\begin{align}
    \Sigma^{\text{S}2}_{\alpha_1\beta_2}(\boldsymbol{k}, i\omega_n) &= \frac{1}{\beta^2} \sum_{\substack{\boldsymbol{q} \boldsymbol{q}' \\ \nu\nu'}}\sum_{\substack{ \beta_1 \gamma_1 \gamma'_1 \\ \alpha_2 \gamma_2 \gamma'_2}}  G_{\beta_1\alpha_2}(\boldsymbol{k}+\boldsymbol{q}'-\boldsymbol{q}, i\omega_n+i\omega_{\nu'}-i\omega_{\nu})  \big[ g_{\boldsymbol{k}+\boldsymbol{q}'-\boldsymbol{q},-\boldsymbol{q}, -\boldsymbol{q}'}^{\alpha_1\beta_1\gamma_1\gamma'_1}  g_{\boldsymbol{k},\boldsymbol{q}, \boldsymbol{q}'}^{\alpha_2\beta_2\gamma_2\gamma'_2} D_{\gamma_1\gamma_2}(\boldsymbol{q}, i\omega_\nu)\nonumber \\
    &\times D_{\gamma'_1\gamma'_2}(-\boldsymbol{q}', -i\omega_{\nu'}) +g_{\boldsymbol{k}+\boldsymbol{q}'-\boldsymbol{q},\boldsymbol{q}', \boldsymbol{q}}^{\alpha_1\beta_1\gamma_1\gamma'_1}  g_{\boldsymbol{k},\boldsymbol{q}, \boldsymbol{q}'}^{\alpha_2\beta_2\gamma_2\gamma'_2}  D_{\gamma_1\gamma'_2}(-\boldsymbol{q}', -i\omega_{\nu'})D_{\gamma'_1\gamma_2}(\boldsymbol{q}, i\omega_\nu) \big].
\end{align}
This is fully momentum and frequency dependent and involves summing over two sets of independent free momenta and frequencies. It is convenient if the momentum and frequency in $G$ is $k' = \boldsymbol{k}', i\omega_{n'}$ when deriving the Eliashberg equations. So we define $k' = k+q'-q$, $q = k-k'+q'$, and finally rename $q'\to q$ to get Eq.~\eqref{eq:SES2gen}.

\end{widetext}

\section{Self-energies in altermagnets} \label{app:SEAM}
In this appendix, we specialize the self-energy expressions to the AM model in Sec.~\ref{sec:AM}. We also consider the magnon-loop and electron-loop diagrams and show that their combined effect is to renormalize $J_{\text{sd}}S$, and hence the spin splitting in the AM. Therefore we ignore them in the main text, and imagine that our choice of $J_{\text{sd}}S$ corresponds to this renormalized $sd$ exchange interaction.

\subsection{Organizing propagators and sunset diagrams}
For the single-magnon EMC interaction Hamiltonian $H_{\text{em}}^{(1)}$ in Eq.~\eqref{eq:H1em}, the nonzero $g_{\boldsymbol{k}+\boldsymbol{q}, \boldsymbol{k}}^{\alpha\beta\gamma}$ factors are
\begin{align}
    g_{\boldsymbol{k}+\boldsymbol{q}, \boldsymbol{k}}^{1,2,2} =& -J_{\text{sd}} \sqrt{\frac{2S}{N}}(\Omega_{\boldsymbol{k}+\boldsymbol{q}, \boldsymbol{k}, \uparrow, \downarrow}^A u_{\boldsymbol{q}}^* + \Omega_{\boldsymbol{k}+\boldsymbol{q}, \boldsymbol{k}, \uparrow, \downarrow}^B v_{\boldsymbol{q}}),\nonumber \\      g_{\boldsymbol{k}+\boldsymbol{q}, \boldsymbol{k}}^{1,2,3} =& -J_{\text{sd}} \sqrt{\frac{2S}{N}}(\Omega_{\boldsymbol{k}+\boldsymbol{q}, \boldsymbol{k}, \uparrow, \downarrow}^A v_{\boldsymbol{q}}^* + \Omega_{\boldsymbol{k}+\boldsymbol{q}, \boldsymbol{k}, \uparrow, \downarrow}^B u_{\boldsymbol{q}}), \nonumber \\
    g_{\boldsymbol{k}+\boldsymbol{q}, \boldsymbol{k}}^{2,1,1} =& -J_{\text{sd}} \sqrt{\frac{2S}{N}}(\Omega_{\boldsymbol{k}+\boldsymbol{q}, \boldsymbol{k}, \downarrow, \uparrow}^A u_{\boldsymbol{q}} + \Omega_{\boldsymbol{k}+\boldsymbol{q}, \boldsymbol{k}, \downarrow, \uparrow}^B v_{\boldsymbol{q}}^*),\nonumber \\      g_{\boldsymbol{k}+\boldsymbol{q}, \boldsymbol{k}}^{2,1,4} =& -J_{\text{sd}} \sqrt{\frac{2S}{N}}(\Omega_{\boldsymbol{k}+\boldsymbol{q}, \boldsymbol{k}, \downarrow, \uparrow}^A v_{\boldsymbol{q}} + \Omega_{\boldsymbol{k}+\boldsymbol{q}, \boldsymbol{k}, \downarrow, \uparrow}^B u_{\boldsymbol{q}}^*), \nonumber \\
   g_{\boldsymbol{k}+\boldsymbol{q}, \boldsymbol{k}}^{3,4,1} =& J_{\text{sd}} \sqrt{\frac{2S}{N}}(\Omega_{-\boldsymbol{k}, -\boldsymbol{k}-\boldsymbol{q}, \downarrow, \uparrow}^A u_{\boldsymbol{q}} + \Omega_{-\boldsymbol{k}, -\boldsymbol{k}-\boldsymbol{q}, \downarrow, \uparrow}^B v_{\boldsymbol{q}}^*),\nonumber \\      g_{\boldsymbol{k}+\boldsymbol{q}, \boldsymbol{k}}^{3,4,4} =& J_{\text{sd}} \sqrt{\frac{2S}{N}}(\Omega_{-\boldsymbol{k}, -\boldsymbol{k}-\boldsymbol{q}, \downarrow, \uparrow}^A v_{\boldsymbol{q}} + \Omega_{-\boldsymbol{k}, -\boldsymbol{k}-\boldsymbol{q}, \downarrow, \uparrow}^B u_{\boldsymbol{q}}^*) , \nonumber \\
    g_{\boldsymbol{k}+\boldsymbol{q}, \boldsymbol{k}}^{4,3,2} =& J_{\text{sd}} \sqrt{\frac{2S}{N}}(\Omega_{-\boldsymbol{k}, -\boldsymbol{k}-\boldsymbol{q}, \uparrow, \downarrow}^A u_{\boldsymbol{q}}^* + \Omega_{-\boldsymbol{k}, -\boldsymbol{k}-\boldsymbol{q}, \uparrow, \downarrow}^B v_{\boldsymbol{q}}),\nonumber \\      g_{\boldsymbol{k}+\boldsymbol{q}, \boldsymbol{k}}^{4,3,3} =& J_{\text{sd}} \sqrt{\frac{2S}{N}}(\Omega_{-\boldsymbol{k}, -\boldsymbol{k}-\boldsymbol{q}, \uparrow, \downarrow}^A v_{\boldsymbol{q}}^* + \Omega_{-\boldsymbol{k}, -\boldsymbol{k}-\boldsymbol{q}, \uparrow, \downarrow}^B u_{\boldsymbol{q}}).
\end{align}

For the double-magnon EMC interaction Hamiltonian $H_{\text{em}}^{(2)}$ in Eq.~\eqref{eq:H2em}, the nonzero $g_{\boldsymbol{k},\boldsymbol{q},\boldsymbol{q}'}^{\alpha\beta\gamma\gamma'}$ factors are
\begin{align}
    g_{\boldsymbol{k},\boldsymbol{q},\boldsymbol{q}'}^{1,1,1,2} &= -\frac{J_{\text{sd}}}{N} \Omega^B_{\boldsymbol{k}+\boldsymbol{q}'-\boldsymbol{q}, \boldsymbol{k}, \uparrow,\uparrow}v_{\boldsymbol{q}}^* v_{\boldsymbol{q}'}, \nonumber \\
    g_{\boldsymbol{k},\boldsymbol{q},\boldsymbol{q}'}^{1,1,1,3} &= -\frac{J_{\text{sd}}}{N} \Omega^B_{\boldsymbol{k}+\boldsymbol{q}'-\boldsymbol{q}, \boldsymbol{k}, \uparrow,\uparrow}v_{\boldsymbol{q}}^* u_{\boldsymbol{q}'}, \nonumber \\
    g_{\boldsymbol{k},\boldsymbol{q},\boldsymbol{q}'}^{1,1,2,1} &= \frac{J_{\text{sd}}}{N} \Omega^A_{\boldsymbol{k}+\boldsymbol{q}'-\boldsymbol{q}, \boldsymbol{k}, \uparrow,\uparrow}u_{\boldsymbol{q}}^* u_{\boldsymbol{q}'}, \nonumber \\
    g_{\boldsymbol{k},\boldsymbol{q},\boldsymbol{q}'}^{1,1,2,4} &= \frac{J_{\text{sd}}}{N} \Omega^A_{\boldsymbol{k}+\boldsymbol{q}'-\boldsymbol{q}, \boldsymbol{k}, \uparrow,\uparrow}u_{\boldsymbol{q}}^* v_{\boldsymbol{q}'}, \nonumber \\
    g_{\boldsymbol{k},\boldsymbol{q},\boldsymbol{q}'}^{1,1,3,1} &= \frac{J_{\text{sd}}}{N} \Omega^A_{\boldsymbol{k}+\boldsymbol{q}'-\boldsymbol{q}, \boldsymbol{k}, \uparrow,\uparrow}v_{\boldsymbol{q}}^* u_{\boldsymbol{q}'}, \nonumber \\
    g_{\boldsymbol{k},\boldsymbol{q},\boldsymbol{q}'}^{1,1,3,4} &= \frac{J_{\text{sd}}}{N} \Omega^A_{\boldsymbol{k}+\boldsymbol{q}'-\boldsymbol{q}, \boldsymbol{k}, \uparrow,\uparrow}v_{\boldsymbol{q}}^* v_{\boldsymbol{q}'}, \nonumber \\
    g_{\boldsymbol{k},\boldsymbol{q},\boldsymbol{q}'}^{1,1,4,2} &= -\frac{J_{\text{sd}}}{N} \Omega^B_{\boldsymbol{k}+\boldsymbol{q}'-\boldsymbol{q}, \boldsymbol{k}, \uparrow,\uparrow}u_{\boldsymbol{q}}^* v_{\boldsymbol{q}'}, \nonumber \\
    g_{\boldsymbol{k},\boldsymbol{q},\boldsymbol{q}'}^{1,1,4,3} &= -\frac{J_{\text{sd}}}{N} \Omega^B_{\boldsymbol{k}+\boldsymbol{q}'-\boldsymbol{q}, \boldsymbol{k}, \uparrow,\uparrow}u_{\boldsymbol{q}}^* u_{\boldsymbol{q}'},
\end{align}
for $\alpha = \beta = 1$. To get the $\alpha = \beta = 2$ factors: change sign and flip spin $\uparrow \to \downarrow$. To get the $\alpha = \beta = 3$ factors: change sign and let $\boldsymbol{k} \to -\boldsymbol{k}-\boldsymbol{q}'+\boldsymbol{q}$. To get the $\alpha=\beta = 4$ factors: take the $\alpha = \beta = 2$ factors, change sign and let $\boldsymbol{k} \to -\boldsymbol{k}-\boldsymbol{q}'+\boldsymbol{q}$. 

Defining magnon propagators $D_\kappa (q) = D_\kappa (\boldsymbol{q}, i\omega_\nu) = 1/(i\omega_\nu - \omega_{\boldsymbol{q}}^\kappa)$, we have
\begin{equation}
    D(q) = \begin{pmatrix}
        0 & D_\alpha (q) & 0 & 0 \\
        D_\alpha (-q) & 0 & 0 \\
        0 & 0 & 0 & D_\beta (q) \\
        0 & 0 & D_\beta (-q) & 0
    \end{pmatrix}.
\end{equation}
EMC leads to feedback effects which renormalize the magnon propagators. This was considered in Ref.~\cite{ThingstadEliashberg}, and it was argued that the effect is not detrimental as long as the magnon gap is large enough. We consider quite large magnon gaps, as typical for antiferromagnets, and so ignore the magnon renormalizations. Since $J_{\text{sd}}/t$ is expected to be larger in bulk systems than across interfaces, a study of self-energy renormalizations of magnons in AMs seems well worth considering.

Performing the sums over Greek indices in Eq.~\eqref{eq:SES1gen}, within the assumption in Eq.~\eqref{eq:Gassumption}, gives
\begin{align}
    \Sigma_{11}^{\text{S}1}(k) =& -\sum_{k'} \Big(\Big[|g_{\boldsymbol{k}',\boldsymbol{k}}^{(1)}|^2 D_\alpha(k'-k)\nonumber \\
    &+|g_{\boldsymbol{k},\boldsymbol{k}'}^{(3)}|^2 D_\beta(k-k')\Big]G_{22}(k')\Big), \\
    \Sigma_{22}^{\text{S}1}(k) =& -\sum_{k'} \Big(\Big[|g_{\boldsymbol{k},\boldsymbol{k}'}^{(1)}|^2 D_\alpha(k-k')\nonumber \\
    &+|g_{\boldsymbol{k}',\boldsymbol{k}}^{(3)}|^2 D_\beta(k'-k)\Big]G_{11}(k')\Big), \\
    \Sigma_{33}^{\text{S}1}(k) =& -\sum_{k'} \Big(\Big[|g_{-\boldsymbol{k}',-\boldsymbol{k}}^{(1)}|^2 D_\alpha(k-k')\nonumber \\
    &+|g_{-\boldsymbol{k},-\boldsymbol{k}'}^{(3)}|^2 D_\beta(k'-k)\Big]G_{44}(k')\Big), \\
    \Sigma_{44}^{\text{S}1}(k) =& -\sum_{k'} \Big(\Big[|g_{-\boldsymbol{k},-\boldsymbol{k}'}^{(1)}|^2 D_\alpha(k'-k)\nonumber \\
    &+|g_{-\boldsymbol{k}',-\boldsymbol{k}}^{(3)}|^2 D_\beta(k-k')\Big]G_{33}(k')\Big).
\end{align}
We defined $g_{\boldsymbol{k},\boldsymbol{k}'}^{(1)} = g_{\boldsymbol{k},\boldsymbol{k}'}^{2,1,1}$ and $g_{\boldsymbol{k},\boldsymbol{k}'}^{(3)} = g_{\boldsymbol{k},\boldsymbol{k}'}^{1,2,3}$. Comparing to Eqs.~\eqref{eq:S1lamstart}-\eqref{eq:S1Lamend} the definition of $\Lambda_{\alpha\beta}^{\text{S}1}(k,k')$ is clear. Performing the sums over Greek indices in Eq.~\eqref{eq:SES2gen} gives
\begin{widetext}
\begin{align}
    \Sigma^{\text{S}2}_{\alpha\beta}(k) =& \sum_{k',q}  G_{\alpha\beta}(k')\Big[ \pqty{g_{\boldsymbol{k}',-\boldsymbol{k}+\boldsymbol{k}'-\boldsymbol{q}, -\boldsymbol{q}}^{\alpha,\alpha,1,2}+g_{\boldsymbol{k}',\boldsymbol{q}, \boldsymbol{k}-\boldsymbol{k}'+\boldsymbol{q}}^{\alpha,\alpha,2,1}}g_{\boldsymbol{k},\boldsymbol{k}-\boldsymbol{k}'+\boldsymbol{q}, \boldsymbol{q}}^{\beta,\beta,2,1} D_{12}(k-k'+q) D_{21}(-q)  \nonumber \\
    &+\pqty{g_{\boldsymbol{k}',-\boldsymbol{k}+\boldsymbol{k}'-\boldsymbol{q}, -\boldsymbol{q}}^{\alpha,\alpha,2,1}+g_{\boldsymbol{k}',\boldsymbol{q}, \boldsymbol{k}-\boldsymbol{k}'+\boldsymbol{q}}^{\alpha,\alpha,1,2}}g_{\boldsymbol{k},\boldsymbol{k}-\boldsymbol{k}'+\boldsymbol{q}, \boldsymbol{q}}^{\beta,\beta,1,2} D_{21}(k-k'+q) D_{12}(-q)  \nonumber \\
    &+\pqty{g_{\boldsymbol{k}',-\boldsymbol{k}+\boldsymbol{k}'-\boldsymbol{q}, -\boldsymbol{q}}^{\alpha,\alpha,1,3}+g_{\boldsymbol{k}',\boldsymbol{q}, \boldsymbol{k}-\boldsymbol{k}'+\boldsymbol{q}}^{\alpha,\alpha,3,1}}g_{\boldsymbol{k},\boldsymbol{k}-\boldsymbol{k}'+\boldsymbol{q}, \boldsymbol{q}}^{\beta,\beta,2,4} D_{12}(k-k'+q) D_{34}(-q)  \nonumber \\
    &+\pqty{g_{\boldsymbol{k}',-\boldsymbol{k}+\boldsymbol{k}'-\boldsymbol{q}, -\boldsymbol{q}}^{\alpha,\alpha,3,1}+g_{\boldsymbol{k}',\boldsymbol{q}, \boldsymbol{k}-\boldsymbol{k}'+\boldsymbol{q}}^{\alpha,\alpha,1,3}}g_{\boldsymbol{k},\boldsymbol{k}-\boldsymbol{k}'+\boldsymbol{q}, \boldsymbol{q}}^{\beta,\beta,4,2} D_{34}(k-k'+q) D_{12}(-q)  \nonumber \\
    &+\pqty{g_{\boldsymbol{k}',-\boldsymbol{k}+\boldsymbol{k}'-\boldsymbol{q}, -\boldsymbol{q}}^{\alpha,\alpha,2,4}+g_{\boldsymbol{k}',\boldsymbol{q}, \boldsymbol{k}-\boldsymbol{k}'+\boldsymbol{q}}^{\alpha,\alpha,4,2}}g_{\boldsymbol{k},\boldsymbol{k}-\boldsymbol{k}'+\boldsymbol{q}, \boldsymbol{q}}^{\beta,\beta,1,3} D_{21}(k-k'+q) D_{43}(-q)  \nonumber \\
    &+\pqty{g_{\boldsymbol{k}',-\boldsymbol{k}+\boldsymbol{k}'-\boldsymbol{q}, -\boldsymbol{q}}^{\alpha,\alpha,4,2}+g_{\boldsymbol{k}',\boldsymbol{q}, \boldsymbol{k}-\boldsymbol{k}'+\boldsymbol{q}}^{\alpha,\alpha,2,4}}g_{\boldsymbol{k},\boldsymbol{k}-\boldsymbol{k}'+\boldsymbol{q}, \boldsymbol{q}}^{\beta,\beta,3,1} D_{43}(k-k'+q) D_{21}(-q)  \nonumber \\
    &+\pqty{g_{\boldsymbol{k}',-\boldsymbol{k}+\boldsymbol{k}'-\boldsymbol{q}, -\boldsymbol{q}}^{\alpha,\alpha,3,4}+g_{\boldsymbol{k}',\boldsymbol{q}, \boldsymbol{k}-\boldsymbol{k}'+\boldsymbol{q}}^{\alpha,\alpha,4,3}}g_{\boldsymbol{k},\boldsymbol{k}-\boldsymbol{k}'+\boldsymbol{q}, \boldsymbol{q}}^{\beta,\beta,4,3} D_{34}(k-k'+q) D_{43}(-q)  \nonumber \\
    &+\pqty{g_{\boldsymbol{k}',-\boldsymbol{k}+\boldsymbol{k}'-\boldsymbol{q}, -\boldsymbol{q}}^{\alpha,\alpha,4,3}+g_{\boldsymbol{k}',\boldsymbol{q}, \boldsymbol{k}-\boldsymbol{k}'+\boldsymbol{q}}^{\alpha,\alpha,3,4}}g_{\boldsymbol{k},\boldsymbol{k}-\boldsymbol{k}'+\boldsymbol{q}, \boldsymbol{q}}^{\beta,\beta,3,4} D_{43}(k-k'+q) D_{34}(-q)    \Big].
\end{align}
\end{widetext}
We define $\Lambda_{\alpha\beta}^{\text{S}2}(k,k',q)$ as the expression inside the square brackets, giving Eq.~\eqref{eq:S2lam}.

\subsection{Definition of logarithmic average} \label{app:omlogdef}
$\omega_{\text{log}}$ is defined as the average of the logarithm of $\omega$ weighted by $\lambda_2(0)$. $\lambda_2(i\omega_\nu)$ is defined in Eq.~\eqref{eq:lam2} and contains $\Lambda_{13}^{\text{S}2}(k,k')$. The frequency sum in $\Lambda_{13}^{\text{S}2}(k,k')$ can be computed analytically. At low temperatures $T \ll \omega_0$ such that $n_B(\omega_{\boldsymbol{q}}^{\alpha/\beta}) \approx 0$, we get
\begin{align}
    \Lambda_{13}^{\text{S}2}(k,k') =& \sum_{\boldsymbol{q}} \bigg[ \pqty{g_{\boldsymbol{k}',-\boldsymbol{k}+\boldsymbol{k}'-\boldsymbol{q}, -\boldsymbol{q}}^{1,1,1,3}+g_{\boldsymbol{k}',\boldsymbol{q}, \boldsymbol{k}-\boldsymbol{k}'+\boldsymbol{q}}^{1,1,3,1}} \nonumber \\
    &\times g_{\boldsymbol{k},\boldsymbol{k}-\boldsymbol{k}'+\boldsymbol{q}, \boldsymbol{q}}^{3,3,2,4} \frac{-1}{i\omega_\nu-\omega_{\boldsymbol{k}-\boldsymbol{k}'+\boldsymbol{q}}^\alpha-\omega_{\boldsymbol{q}}^\beta}  \nonumber \\
    &+\pqty{g_{\boldsymbol{k}',-\boldsymbol{k}+\boldsymbol{k}'-\boldsymbol{q}, -\boldsymbol{q}}^{1,1,3,1}+g_{\boldsymbol{k}',\boldsymbol{q}, \boldsymbol{k}-\boldsymbol{k}'+\boldsymbol{q}}^{1,1,1,3}} \nonumber \\
    &\times g_{\boldsymbol{k},\boldsymbol{k}-\boldsymbol{k}'+\boldsymbol{q}, \boldsymbol{q}}^{3,3,4,2} \frac{-1}{i\omega_\nu-\omega_{\boldsymbol{k}-\boldsymbol{k}'+\boldsymbol{q}}^\beta-\omega_{\boldsymbol{q}}^\alpha}  \nonumber \\
    &+\pqty{g_{\boldsymbol{k}',-\boldsymbol{k}+\boldsymbol{k}'-\boldsymbol{q}, -\boldsymbol{q}}^{1,1,2,4}+g_{\boldsymbol{k}',\boldsymbol{q}, \boldsymbol{k}-\boldsymbol{k}'+\boldsymbol{q}}^{1,1,4,2}} \nonumber \\
    &\times g_{\boldsymbol{k},\boldsymbol{k}-\boldsymbol{k}'+\boldsymbol{q}, \boldsymbol{q}}^{3,3,1,3} \frac{1}{i\omega_\nu+\omega_{\boldsymbol{k}-\boldsymbol{k}'+\boldsymbol{q}}^\alpha + \omega_{\boldsymbol{q}}^\beta}  \nonumber \\
    &+\pqty{g_{\boldsymbol{k}',-\boldsymbol{k}+\boldsymbol{k}'-\boldsymbol{q}, -\boldsymbol{q}}^{1,1,4,2}+g_{\boldsymbol{k}',\boldsymbol{q}, \boldsymbol{k}-\boldsymbol{k}'+\boldsymbol{q}}^{1,1,2,4}} \nonumber \\
    &\times g_{\boldsymbol{k},\boldsymbol{k}-\boldsymbol{k}'+\boldsymbol{q}, \boldsymbol{q}}^{3,3,3,1} \frac{1}{i\omega_\nu + \omega_{\boldsymbol{k}-\boldsymbol{k}'+\boldsymbol{q}}^\beta + \omega_{\boldsymbol{q}}^\alpha} \bigg].
\end{align}
From \cite{Marsiglio1988AnalyticCont}
\begin{equation}
    \lambda_2(i\omega_\nu) = -\int_{-\infty}^{\infty} d\omega \alpha^2 F(\omega) \frac{1}{i\omega_\nu-\omega}
\end{equation}
and Eq.~\eqref{eq:lam2}, we define
\begin{align}
    \alpha^2F(\omega) =& \frac{-1}{N_{F,\uparrow}\langle \psi_\uparrow^2 (\boldsymbol{k})\rangle_{\text{FS}_\uparrow}} \sum_{\boldsymbol{k}\boldsymbol{k}'} \delta(\epsilon_{\boldsymbol{k}\uparrow})\delta(\epsilon_{\boldsymbol{k}'\uparrow})\nonumber\\
    &\times\psi_\uparrow(\boldsymbol{k}) A_{13}^{\text{S}2}(k,k')\psi_\uparrow(\boldsymbol{k}'),
\end{align}
with
\begin{align}
    A_{13}^{\text{S}2}(k,k') &= \sum_{\boldsymbol{q}} \bigg[ \pqty{g_{\boldsymbol{k}',-\boldsymbol{k}+\boldsymbol{k}'-\boldsymbol{q}, -\boldsymbol{q}}^{1,1,1,3}+g_{\boldsymbol{k}',\boldsymbol{q}, \boldsymbol{k}-\boldsymbol{k}'+\boldsymbol{q}}^{1,1,3,1}} \nonumber \\
    &\times g_{\boldsymbol{k},\boldsymbol{k}-\boldsymbol{k}'+\boldsymbol{q}, \boldsymbol{q}}^{3,3,2,4} \delta(\omega-\omega_{\boldsymbol{k}-\boldsymbol{k}'+\boldsymbol{q}}^\alpha-\omega_{\boldsymbol{q}}^\beta)  \nonumber \\
    &+\pqty{g_{\boldsymbol{k}',-\boldsymbol{k}+\boldsymbol{k}'-\boldsymbol{q}, -\boldsymbol{q}}^{1,1,3,1}+g_{\boldsymbol{k}',\boldsymbol{q}, \boldsymbol{k}-\boldsymbol{k}'+\boldsymbol{q}}^{1,1,1,3}} \nonumber \\
    &\times g_{\boldsymbol{k},\boldsymbol{k}-\boldsymbol{k}'+\boldsymbol{q}, \boldsymbol{q}}^{3,3,4,2} \delta(\omega-\omega_{\boldsymbol{k}-\boldsymbol{k}'+\boldsymbol{q}}^\beta-\omega_{\boldsymbol{q}}^\alpha)  \nonumber \\
    &-\pqty{g_{\boldsymbol{k}',-\boldsymbol{k}+\boldsymbol{k}'-\boldsymbol{q}, -\boldsymbol{q}}^{1,1,2,4}+g_{\boldsymbol{k}',\boldsymbol{q}, \boldsymbol{k}-\boldsymbol{k}'+\boldsymbol{q}}^{1,1,4,2}} \nonumber \\
    &\times g_{\boldsymbol{k},\boldsymbol{k}-\boldsymbol{k}'+\boldsymbol{q}, \boldsymbol{q}}^{3,3,1,3} \delta(\omega+\omega_{\boldsymbol{k}-\boldsymbol{k}'+\boldsymbol{q}}^\alpha + \omega_{\boldsymbol{q}}^\beta)  \nonumber \\
    &-\pqty{g_{\boldsymbol{k}',-\boldsymbol{k}+\boldsymbol{k}'-\boldsymbol{q}, -\boldsymbol{q}}^{1,1,4,2}+g_{\boldsymbol{k}',\boldsymbol{q}, \boldsymbol{k}-\boldsymbol{k}'+\boldsymbol{q}}^{1,1,2,4}} \nonumber \\
    &\times g_{\boldsymbol{k},\boldsymbol{k}-\boldsymbol{k}'+\boldsymbol{q}, \boldsymbol{q}}^{3,3,3,1} \delta(\omega+ \omega_{\boldsymbol{k}-\boldsymbol{k}'+\boldsymbol{q}}^\beta + \omega_{\boldsymbol{q}}^\alpha) \bigg].
\end{align}
Note that
$
    \lambda_2(0) = \int_{-\infty}^{\infty} d\omega  \alpha^2 F(\omega)/\omega,
$
so we define
\begin{equation}
    \omega_{\text{log}} = \omega_a \exp[ \frac{1}{\lambda_2(0)} \int_{-\infty}^{\infty} d\omega \ln(\frac{|\omega|}{\omega_a}) \frac{\alpha^2F(\omega)}{\omega} ].
\end{equation}
Here, $\omega_a$ is an arbitrary energy scale.

\subsection{Magnon-loop and electron-loop diagrams}
Generalizing to three bands, and focusing on the normal state, the magnon-loop diagram becomes a $6\times 6$ matrix in spin and band indices,
\begin{align}
    \Sigma^{\text{ML}}_{n'n\sigma}(\boldsymbol{k}) =& \sum_{\boldsymbol{q}  \gamma \gamma'} g_{\boldsymbol{k},\boldsymbol{q}, \boldsymbol{q}}^{n',n,\sigma,\gamma,\gamma'}\langle B_{-\boldsymbol{q}}^{\gamma}B_{\boldsymbol{q}}^{\gamma'} \rangle \nonumber\\
    =& -\frac{J_{\text{sd}}\sigma}{N}\sum_{\boldsymbol{q}}\Big[ \Omega_{n',n,\boldsymbol{k}, \boldsymbol{k},\sigma,\sigma}^B |v_{\boldsymbol{q}}|^2 [1+n_B(\omega_{\boldsymbol{q}}^\alpha)]\nonumber\\
    &-\Omega_{n',n,\boldsymbol{k}, \boldsymbol{k},\sigma,\sigma}^A |u_{\boldsymbol{q}}|^2 n_B(\omega_{\boldsymbol{q}}^\alpha)  \nonumber \\
    &-\Omega_{n',n,\boldsymbol{k}, \boldsymbol{k},\sigma,\sigma}^A |v_{\boldsymbol{q}}|^2 [1+n_B(\omega_{\boldsymbol{q}}^\beta)] \nonumber\\
    &+ \Omega_{n',n,\boldsymbol{k}, \boldsymbol{k},\sigma,\sigma}^B |u_{\boldsymbol{q}}|^2 n_B(\omega_{\boldsymbol{q}}^\beta) \Big].
\end{align}
We define $g^{n',n,\sigma,\gamma,\gamma'}_{\boldsymbol{k}, \boldsymbol{q}, \boldsymbol{q}'}$ from the one-band $g^{\alpha, \beta,\gamma,\gamma'}_{\boldsymbol{k}, \boldsymbol{q}, \boldsymbol{q}'}$ by $\sigma = \uparrow \leftrightarrow \alpha = \beta = 1$, $\sigma = \downarrow \leftrightarrow \alpha = \beta = 2$, and by adding band indices $n',n$ in $\Omega^\ell_{n',n,\boldsymbol{k}',\boldsymbol{k},\sigma', \sigma} \equiv q^*_{n',\ell, \boldsymbol{k}',\sigma'} q_{n,\ell, \boldsymbol{k},\sigma}$.

Similarly, the electron-loop diagram becomes
\begin{align}
    &\Sigma^{\text{EL}}_{n'n\sigma}(\boldsymbol{k}) = -\sum_{q, \boldsymbol{k}_2} \sum_{\substack{ \gamma_1 \gamma'_1 \gamma_2 \gamma'_2\\ n_2, \sigma_2 }} \langle d_{n_2,\boldsymbol{k}_2, \sigma_2}^\dagger d_{n_2, \boldsymbol{k}_2, \sigma_2}\rangle  \nonumber\\
    &\times\big[ g_{\boldsymbol{k},\boldsymbol{q}, \boldsymbol{q}}^{n',n,\sigma,\gamma_1,\gamma'_1}  g_{\boldsymbol{k}_2,-\boldsymbol{q}, -\boldsymbol{q}}^{n_2,n_2,\sigma_2, \gamma_2,\gamma'_2} D_{\gamma_1 \gamma_2}(-q) D_{\gamma'_1 \gamma'_2}(q) \nonumber \\
    &+ g_{\boldsymbol{k},\boldsymbol{q}, \boldsymbol{q}}^{n',n,\sigma,\gamma_1,\gamma'_1}  g_{\boldsymbol{k}_2,\boldsymbol{q}, \boldsymbol{q}}^{n_2,n_2,\sigma_2,\gamma_2,\gamma'_2} D_{\gamma_1 \gamma'_2}(-q) D_{\gamma'_1 \gamma_2}(q) \big].
\end{align}
The sum over bosonic frequencies can be computed analytically, and $\langle d_{n_2,\boldsymbol{k}_2, \sigma_2}^\dagger d_{n_2, \boldsymbol{k}_2, \sigma_2}\rangle =  n_F(\epsilon_{n_2, \boldsymbol{k}_2, \sigma_2})$, where $n_F(z)$ is the Fermi-Dirac distribution.

Since the self-energies $\Sigma^{\text{ML}}$ and $\Sigma^{\text{EL}}$ are frequency independent, we can analytically continue to the real frequency axis $\omega$. From the Dyson equation, focusing solely on their contributions, we get
\begin{equation}
    G(\boldsymbol{k}, \omega) = \big[ G_0^{-1}(\boldsymbol{k}, \omega) - \Sigma^{\text{ML}}(\boldsymbol{k}) -\Sigma^{\text{EL}}(\boldsymbol{k}) \big]^{-1}.
\end{equation}
The renormalized bands are the zeros of the real part of the denominators in $G$. This is the zeros of $\det[ G_0^{-1}(\boldsymbol{k}, \omega) - \Sigma^{\text{ML}}(\boldsymbol{k}) -\Sigma^{\text{EL}}(\boldsymbol{k})]$ which are the eigenvalues of $\operatorname{diag}(\epsilon_{1,\boldsymbol{k},\uparrow}, \epsilon_{1,\boldsymbol{k},\downarrow}, \epsilon_{2,\boldsymbol{k},\uparrow}, \epsilon_{2,\boldsymbol{k},\downarrow}, \epsilon_{3,\boldsymbol{k},\uparrow}, \epsilon_{3,\boldsymbol{k},\downarrow}) + \Sigma^{\text{ML}}(\boldsymbol{k}) +\Sigma^{\text{EL}}(\boldsymbol{k})$. This is in the basis of diagonalized electrons, $d_{n,\boldsymbol{k},\sigma}$. The spin sectors decouple, so we can consider two $3\times 3$ matrices separately, and transform back to the original electron operators $c_{\ell, \boldsymbol{k}, \sigma}$. 
We have $U_{\boldsymbol{k},\sigma}D_{\boldsymbol{k},\sigma}U_{\boldsymbol{k},\sigma}^\dagger = \mathcal{H}_{\boldsymbol{k},\sigma}$, where $D_{\boldsymbol{k},\sigma}$ is diagonal with $[D_{\boldsymbol{k},\sigma}]_{nn} = \epsilon_{n,\boldsymbol{k},\sigma}$, and $\mathcal{H}_{\boldsymbol{k},\sigma}$ the electron Hamiltonian on matrix form. 
Numerically, we find
\begin{equation}
    U_{\boldsymbol{k},\sigma} \Sigma^{\text{ML}}_\sigma(\boldsymbol{k}) U_{\boldsymbol{k},\sigma}^\dagger = \begin{pmatrix}
        \sigma C^{\text{ML}} & 0 & 0 \\
        0 & 0 &0\\
        0 & 0 & -\sigma C^{\text{ML}}
    \end{pmatrix},
\end{equation}
with $C^{\text{ML}} > 0$ and
\begin{equation}
    U_{\boldsymbol{k},\sigma} \Sigma^{\text{EL}}_\sigma(\boldsymbol{k}) U_{\boldsymbol{k},\sigma}^\dagger = \begin{pmatrix}
        -\sigma C^{\text{EL}} & 0 & 0 \\
        0 & 0 &0\\
        0 & 0 & \sigma C^{\text{EL}}
    \end{pmatrix},
\end{equation}
with $C^{\text{EL}} > 0$, i.e., the contributions have opposite sign and are momentum independent. Comparing to
\begin{align}
    \mathcal{H}_{\boldsymbol{k}, \sigma} =& \begin{pmatrix} 
    -\sigma J_{\text{sd}} S & -2t\cos{k_x } & -4 t_2\cos k_x \cos k_y \\ 
    -2t\cos{k_x } & 0  & -2t\cos{k_y } \\ 
    -4 t_2\cos k_x \cos k_y & -2t\cos{k_y } & \sigma J_{\text{sd}} S 
    \end{pmatrix} \nonumber\\
    &-[\mu+2t_3(\cos2k_x + \cos2k_y)]I, \label{eq:Hk}
\end{align}
where $I$ is a unit matrix,
we see that the effects of the diagrams $\Sigma^{\text{ML}}$ and $\Sigma^{\text{EL}}$ are simply to renormalize $J_{\text{sd}}S$, i.e., the spin splitting in the AM. Instead of explicitly including them in our calculations, we instead assume that our choice of $J_{\text{sd}}S$ includes this renormalization.

The magnon-loop diagram in fact has the exact same effect as including corrections to the magnetization, i.e., letting \cite{Maeland2022QSk}
\begin{align}
    S \to& \frac{1}{N}\sum_{i\in A (B)} \langle S_{i,z}\rangle = S- \frac{1}{N}\sum_{\boldsymbol{q}}\Big(|u_{\boldsymbol{q}}|^2 n_B(\omega_{\boldsymbol{q}}^{\alpha (\beta)}) \nonumber \\
    & +|v_{\boldsymbol{q}}|^2 [1+n_B(\omega_{\boldsymbol{q}}^{\beta (\alpha)})]\Big)
\end{align}
in the upper left (lower right) corner of $\mathcal{H}_{\boldsymbol{k}, \sigma}$.
$\Sigma^{\text{ML}}$ is the diagrammatic version of this correction.

\bibliography{main.bbl}

\begin{thebibliography}{113}%
\makeatletter
\providecommand \@ifxundefined [1]{%
 \@ifx{#1\undefined}
}%
\providecommand \@ifnum [1]{%
 \ifnum #1\expandafter \@firstoftwo
 \else \expandafter \@secondoftwo
 \fi
}%
\providecommand \@ifx [1]{%
 \ifx #1\expandafter \@firstoftwo
 \else \expandafter \@secondoftwo
 \fi
}%
\providecommand \natexlab [1]{#1}%
\providecommand \enquote  [1]{``#1''}%
\providecommand \bibnamefont  [1]{#1}%
\providecommand \bibfnamefont [1]{#1}%
\providecommand \citenamefont [1]{#1}%
\providecommand \href@noop [0]{\@secondoftwo}%
\providecommand \href [0]{\begingroup \@sanitize@url \@href}%
\providecommand \@href[1]{\@@startlink{#1}\@@href}%
\providecommand \@@href[1]{\endgroup#1\@@endlink}%
\providecommand \@sanitize@url [0]{\catcode `\\12\catcode `\$12\catcode `\&12\catcode `\#12\catcode `\^12\catcode `\_12\catcode `\%12\relax}%
\providecommand \@@startlink[1]{}%
\providecommand \@@endlink[0]{}%
\providecommand \url  [0]{\begingroup\@sanitize@url \@url }%
\providecommand \@url [1]{\endgroup\@href {#1}{\urlprefix }}%
\providecommand \urlprefix  [0]{URL }%
\providecommand \Eprint [0]{\href }%
\providecommand \doibase [0]{https://doi.org/}%
\providecommand \selectlanguage [0]{\@gobble}%
\providecommand \bibinfo  [0]{\@secondoftwo}%
\providecommand \bibfield  [0]{\@secondoftwo}%
\providecommand \translation [1]{[#1]}%
\providecommand \BibitemOpen [0]{}%
\providecommand \bibitemStop [0]{}%
\providecommand \bibitemNoStop [0]{.\EOS\space}%
\providecommand \EOS [0]{\spacefactor3000\relax}%
\providecommand \BibitemShut  [1]{\csname bibitem#1\endcsname}%
\let\auto@bib@innerbib\@empty
\bibitem [{\citenamefont {Brataas}\ \emph {et~al.}(2020)\citenamefont {Brataas}, \citenamefont {van Wees}, \citenamefont {Klein}, \citenamefont {de~Loubens},\ and\ \citenamefont {Viret}}]{Brataas2020MagnonCurrent}%
  \BibitemOpen
  \bibfield  {author} {\bibinfo {author} {\bibfnamefont {A.}~\bibnamefont {Brataas}}, \bibinfo {author} {\bibfnamefont {B.}~\bibnamefont {van Wees}}, \bibinfo {author} {\bibfnamefont {O.}~\bibnamefont {Klein}}, \bibinfo {author} {\bibfnamefont {G.}~\bibnamefont {de~Loubens}},\ and\ \bibinfo {author} {\bibfnamefont {M.}~\bibnamefont {Viret}},\ }\bibfield  {title} {\bibinfo {title} {{Spin insulatronics}},\ }\href {https://doi.org/10.1016/j.physrep.2020.08.006} {\bibfield  {journal} {\bibinfo  {journal} {Phys. Rep.}\ }\textbf {\bibinfo {volume} {885}},\ \bibinfo {pages} {1} (\bibinfo {year} {2020})}\BibitemShut {NoStop}%
\bibitem [{\citenamefont {Chumak}\ \emph {et~al.}(2015)\citenamefont {Chumak}, \citenamefont {Vasyuchka}, \citenamefont {Serga},\ and\ \citenamefont {Hillebrands}}]{Chumak2015MagnonCurrent}%
  \BibitemOpen
  \bibfield  {author} {\bibinfo {author} {\bibfnamefont {A.~V.}\ \bibnamefont {Chumak}}, \bibinfo {author} {\bibfnamefont {V.~I.}\ \bibnamefont {Vasyuchka}}, \bibinfo {author} {\bibfnamefont {A.~A.}\ \bibnamefont {Serga}},\ and\ \bibinfo {author} {\bibfnamefont {B.}~\bibnamefont {Hillebrands}},\ }\bibfield  {title} {\bibinfo {title} {{Magnon spintronics}},\ }\href {https://doi.org/10.1038/nphys3347} {\bibfield  {journal} {\bibinfo  {journal} {Nat. Phys.}\ }\textbf {\bibinfo {volume} {11}},\ \bibinfo {pages} {453} (\bibinfo {year} {2015})}\BibitemShut {NoStop}%
\bibitem [{\citenamefont {Tserkovnyak}\ \emph {et~al.}(2002)\citenamefont {Tserkovnyak}, \citenamefont {Brataas},\ and\ \citenamefont {Bauer}}]{TserkovnyakPRL2002}%
  \BibitemOpen
  \bibfield  {author} {\bibinfo {author} {\bibfnamefont {Y.}~\bibnamefont {Tserkovnyak}}, \bibinfo {author} {\bibfnamefont {A.}~\bibnamefont {Brataas}},\ and\ \bibinfo {author} {\bibfnamefont {G.~E.~W.}\ \bibnamefont {Bauer}},\ }\bibfield  {title} {\bibinfo {title} {{Enhanced Gilbert Damping in Thin Ferromagnetic Films}},\ }\href {https://doi.org/10.1103/PhysRevLett.88.117601} {\bibfield  {journal} {\bibinfo  {journal} {Phys. Rev. Lett.}\ }\textbf {\bibinfo {volume} {88}},\ \bibinfo {pages} {117601} (\bibinfo {year} {2002})}\BibitemShut {NoStop}%
\bibitem [{\citenamefont {Ralph}\ and\ \citenamefont {Stiles}(2008)}]{RalphJMM2008}%
  \BibitemOpen
  \bibfield  {author} {\bibinfo {author} {\bibfnamefont {D.~C.}\ \bibnamefont {Ralph}}\ and\ \bibinfo {author} {\bibfnamefont {M.~D.}\ \bibnamefont {Stiles}},\ }\bibfield  {title} {\bibinfo {title} {{Spin transfer torques}},\ }\href {https://doi.org/10.1016/j.jmmm.2007.12.019} {\bibfield  {journal} {\bibinfo  {journal} {J. Magn. Magn. Mater.}\ }\textbf {\bibinfo {volume} {320}},\ \bibinfo {pages} {1190} (\bibinfo {year} {2008})}\BibitemShut {NoStop}%
\bibitem [{\citenamefont {Han}\ \emph {et~al.}(2020)\citenamefont {Han}, \citenamefont {Maekawa},\ and\ \citenamefont {Xie}}]{Han2020MagnonCurrent}%
  \BibitemOpen
  \bibfield  {author} {\bibinfo {author} {\bibfnamefont {W.}~\bibnamefont {Han}}, \bibinfo {author} {\bibfnamefont {S.}~\bibnamefont {Maekawa}},\ and\ \bibinfo {author} {\bibfnamefont {X.-C.}\ \bibnamefont {Xie}},\ }\bibfield  {title} {\bibinfo {title} {{Spin current as a probe of quantum materials}},\ }\href {https://doi.org/10.1038/s41563-019-0456-7} {\bibfield  {journal} {\bibinfo  {journal} {Nat. Mater.}\ }\textbf {\bibinfo {volume} {19}},\ \bibinfo {pages} {139} (\bibinfo {year} {2020})}\BibitemShut {NoStop}%
\bibitem [{\citenamefont {Baltz}\ \emph {et~al.}(2018)\citenamefont {Baltz}, \citenamefont {Manchon}, \citenamefont {Tsoi}, \citenamefont {Moriyama}, \citenamefont {Ono},\ and\ \citenamefont {Tserkovnyak}}]{Baltz2018YaroslavAFMrev}%
  \BibitemOpen
  \bibfield  {author} {\bibinfo {author} {\bibfnamefont {V.}~\bibnamefont {Baltz}}, \bibinfo {author} {\bibfnamefont {A.}~\bibnamefont {Manchon}}, \bibinfo {author} {\bibfnamefont {M.}~\bibnamefont {Tsoi}}, \bibinfo {author} {\bibfnamefont {T.}~\bibnamefont {Moriyama}}, \bibinfo {author} {\bibfnamefont {T.}~\bibnamefont {Ono}},\ and\ \bibinfo {author} {\bibfnamefont {Y.}~\bibnamefont {Tserkovnyak}},\ }\bibfield  {title} {\bibinfo {title} {{Antiferromagnetic spintronics}},\ }\href {https://doi.org/10.1103/RevModPhys.90.015005} {\bibfield  {journal} {\bibinfo  {journal} {Rev. Mod. Phys.}\ }\textbf {\bibinfo {volume} {90}},\ \bibinfo {pages} {015005} (\bibinfo {year} {2018})}\BibitemShut {NoStop}%
\bibitem [{\citenamefont {Fiebig}\ \emph {et~al.}(2008)\citenamefont {Fiebig}, \citenamefont {Duong}, \citenamefont {Satoh}, \citenamefont {Van~Aken}, \citenamefont {Miyano}, \citenamefont {Tomioka},\ and\ \citenamefont {Tokura}}]{Fiebig2008AFMexpUltrafast}%
  \BibitemOpen
  \bibfield  {author} {\bibinfo {author} {\bibfnamefont {M.}~\bibnamefont {Fiebig}}, \bibinfo {author} {\bibfnamefont {N.~P.}\ \bibnamefont {Duong}}, \bibinfo {author} {\bibfnamefont {T.}~\bibnamefont {Satoh}}, \bibinfo {author} {\bibfnamefont {B.~B.}\ \bibnamefont {Van~Aken}}, \bibinfo {author} {\bibfnamefont {K.}~\bibnamefont {Miyano}}, \bibinfo {author} {\bibfnamefont {Y.}~\bibnamefont {Tomioka}},\ and\ \bibinfo {author} {\bibfnamefont {Y.}~\bibnamefont {Tokura}},\ }\bibfield  {title} {\bibinfo {title} {{Ultrafast magnetization dynamics of antiferromagnetic compounds}},\ }\href {https://doi.org/10.1088/0022-3727/41/16/164005} {\bibfield  {journal} {\bibinfo  {journal} {J. Phys. D: Appl. Phys.}\ }\textbf {\bibinfo {volume} {41}},\ \bibinfo {pages} {164005} (\bibinfo {year} {2008})}\BibitemShut {NoStop}%
\bibitem [{\citenamefont {Mazin}(2022{\natexlab{a}})}]{Mazin2022PRX}%
  \BibitemOpen
  \bibfield  {author} {\bibinfo {author} {\bibfnamefont {I.}~\bibnamefont {Mazin}} (\bibinfo {collaboration} {The PRX Editors}),\ }\bibfield  {title} {\bibinfo {title} {Editorial: Altermagnetism---a new punch line of fundamental magnetism},\ }\href {https://doi.org/10.1103/PhysRevX.12.040002} {\bibfield  {journal} {\bibinfo  {journal} {Phys. Rev. X}\ }\textbf {\bibinfo {volume} {12}},\ \bibinfo {pages} {040002} (\bibinfo {year} {2022}{\natexlab{a}})}\BibitemShut {NoStop}%
\bibitem [{\citenamefont {{\ifmmode\check{S}\else\v{S}\fi}mejkal}\ \emph {et~al.}(2022{\natexlab{a}})\citenamefont {{\ifmmode\check{S}\else\v{S}\fi}mejkal}, \citenamefont {Sinova},\ and\ \citenamefont {Jungwirth}}]{Smejkal2022Sep}%
  \BibitemOpen
  \bibfield  {author} {\bibinfo {author} {\bibfnamefont {L.}~\bibnamefont {{\ifmmode\check{S}\else\v{S}\fi}mejkal}}, \bibinfo {author} {\bibfnamefont {J.}~\bibnamefont {Sinova}},\ and\ \bibinfo {author} {\bibfnamefont {T.}~\bibnamefont {Jungwirth}},\ }\bibfield  {title} {\bibinfo {title} {{Beyond Conventional Ferromagnetism and Antiferromagnetism: A Phase with Nonrelativistic Spin and Crystal Rotation Symmetry}},\ }\href {https://doi.org/10.1103/PhysRevX.12.031042} {\bibfield  {journal} {\bibinfo  {journal} {Phys. Rev. X}\ }\textbf {\bibinfo {volume} {12}},\ \bibinfo {pages} {031042} (\bibinfo {year} {2022}{\natexlab{a}})}\BibitemShut {NoStop}%
\bibitem [{\citenamefont {{\ifmmode\check{S}\else\v{S}\fi}mejkal}\ \emph {et~al.}(2022{\natexlab{b}})\citenamefont {{\ifmmode\check{S}\else\v{S}\fi}mejkal}, \citenamefont {Sinova},\ and\ \citenamefont {Jungwirth}}]{Smejkal2022Dec}%
  \BibitemOpen
  \bibfield  {author} {\bibinfo {author} {\bibfnamefont {L.}~\bibnamefont {{\ifmmode\check{S}\else\v{S}\fi}mejkal}}, \bibinfo {author} {\bibfnamefont {J.}~\bibnamefont {Sinova}},\ and\ \bibinfo {author} {\bibfnamefont {T.}~\bibnamefont {Jungwirth}},\ }\bibfield  {title} {\bibinfo {title} {{Emerging Research Landscape of Altermagnetism}},\ }\href {https://doi.org/10.1103/PhysRevX.12.040501} {\bibfield  {journal} {\bibinfo  {journal} {Phys. Rev. X}\ }\textbf {\bibinfo {volume} {12}},\ \bibinfo {pages} {040501} (\bibinfo {year} {2022}{\natexlab{b}})}\BibitemShut {NoStop}%
\bibitem [{\citenamefont {Mazin}(2022{\natexlab{b}})}]{mazin2022notes}%
  \BibitemOpen
  \bibfield  {author} {\bibinfo {author} {\bibfnamefont {I.~I.}\ \bibnamefont {Mazin}},\ }\bibfield  {title} {\bibinfo {title} {Notes on altermagnetism and superconductivity},\ }\href {https://arxiv.org/abs/2203.05000} {\bibfield  {journal} {\bibinfo  {journal} {arXiv:2203.05000}\ } (\bibinfo {year} {2022}{\natexlab{b}})}\BibitemShut {NoStop}%
\bibitem [{\citenamefont {Ouassou}\ \emph {et~al.}(2023)\citenamefont {Ouassou}, \citenamefont {Brataas},\ and\ \citenamefont {Linder}}]{OuassouPRL2023}%
  \BibitemOpen
  \bibfield  {author} {\bibinfo {author} {\bibfnamefont {J.~A.}\ \bibnamefont {Ouassou}}, \bibinfo {author} {\bibfnamefont {A.}~\bibnamefont {Brataas}},\ and\ \bibinfo {author} {\bibfnamefont {J.}~\bibnamefont {Linder}},\ }\bibfield  {title} {\bibinfo {title} {{dc Josephson Effect in Altermagnets}},\ }\href {https://doi.org/10.1103/PhysRevLett.131.076003} {\bibfield  {journal} {\bibinfo  {journal} {Phys. Rev. Lett.}\ }\textbf {\bibinfo {volume} {131}},\ \bibinfo {pages} {076003} (\bibinfo {year} {2023})}\BibitemShut {NoStop}%
\bibitem [{\citenamefont {Beenakker}\ and\ \citenamefont {Vakhtel}(2023)}]{BeenakkerPRB2023}%
  \BibitemOpen
  \bibfield  {author} {\bibinfo {author} {\bibfnamefont {C.~W.~J.}\ \bibnamefont {Beenakker}}\ and\ \bibinfo {author} {\bibfnamefont {T.}~\bibnamefont {Vakhtel}},\ }\bibfield  {title} {\bibinfo {title} {{Phase-shifted Andreev levels in an altermagnet Josephson junction}},\ }\href {https://doi.org/10.1103/PhysRevB.108.075425} {\bibfield  {journal} {\bibinfo  {journal} {Phys. Rev. B}\ }\textbf {\bibinfo {volume} {108}},\ \bibinfo {pages} {075425} (\bibinfo {year} {2023})}\BibitemShut {NoStop}%
\bibitem [{\citenamefont {Sun}\ \emph {et~al.}(2023{\natexlab{a}})\citenamefont {Sun}, \citenamefont {Brataas},\ and\ \citenamefont {Linder}}]{Sun2023AndreevAM}%
  \BibitemOpen
  \bibfield  {author} {\bibinfo {author} {\bibfnamefont {C.}~\bibnamefont {Sun}}, \bibinfo {author} {\bibfnamefont {A.}~\bibnamefont {Brataas}},\ and\ \bibinfo {author} {\bibfnamefont {J.}~\bibnamefont {Linder}},\ }\bibfield  {title} {\bibinfo {title} {{Andreev reflection in altermagnets}},\ }\href {https://doi.org/10.1103/PhysRevB.108.054511} {\bibfield  {journal} {\bibinfo  {journal} {Phys. Rev. B}\ }\textbf {\bibinfo {volume} {108}},\ \bibinfo {pages} {054511} (\bibinfo {year} {2023}{\natexlab{a}})}\BibitemShut {NoStop}%
\bibitem [{\citenamefont {Papaj}(2023)}]{PapajPRB2023}%
  \BibitemOpen
  \bibfield  {author} {\bibinfo {author} {\bibfnamefont {M.}~\bibnamefont {Papaj}},\ }\bibfield  {title} {\bibinfo {title} {Andreev reflection at the altermagnet-superconductor interface},\ }\href {https://doi.org/10.1103/PhysRevB.108.L060508} {\bibfield  {journal} {\bibinfo  {journal} {Phys. Rev. B}\ }\textbf {\bibinfo {volume} {108}},\ \bibinfo {pages} {L060508} (\bibinfo {year} {2023})}\BibitemShut {NoStop}%
\bibitem [{\citenamefont {Giil}\ and\ \citenamefont {Linder}(2023)}]{giil2023superconductor}%
  \BibitemOpen
  \bibfield  {author} {\bibinfo {author} {\bibfnamefont {H.~G.}\ \bibnamefont {Giil}}\ and\ \bibinfo {author} {\bibfnamefont {J.}~\bibnamefont {Linder}},\ }\bibfield  {title} {\bibinfo {title} {{Superconductor-altermagnet memory functionality without stray fields}},\ }\href {https://doi.org/10.48550/arXiv.2308.10939} {\bibfield  {journal} {\bibinfo  {journal} {arXiv:2308.10939}\ } (\bibinfo {year} {2023})}\BibitemShut {NoStop}%
\bibitem [{\citenamefont {Ghorashi}\ \emph {et~al.}(2023)\citenamefont {Ghorashi}, \citenamefont {Hughes},\ and\ \citenamefont {Cano}}]{Ghorashi2023AMTSC}%
  \BibitemOpen
  \bibfield  {author} {\bibinfo {author} {\bibfnamefont {S.~A.~A.}\ \bibnamefont {Ghorashi}}, \bibinfo {author} {\bibfnamefont {T.~L.}\ \bibnamefont {Hughes}},\ and\ \bibinfo {author} {\bibfnamefont {J.}~\bibnamefont {Cano}},\ }\bibfield  {title} {\bibinfo {title} {{Altermagnetic Routes to Majorana Modes in Zero Net Magnetization}},\ }\href {https://doi.org/10.48550/arXiv.2306.09413} {\bibfield  {journal} {\bibinfo  {journal} {arXiv:2306.09413}\ } (\bibinfo {year} {2023})}\BibitemShut {NoStop}%
\bibitem [{\citenamefont {Zhang}\ \emph {et~al.}(2024)\citenamefont {Zhang}, \citenamefont {Hu},\ and\ \citenamefont {Neupert}}]{Zhang2024FFLOAM}%
  \BibitemOpen
  \bibfield  {author} {\bibinfo {author} {\bibfnamefont {S.-B.}\ \bibnamefont {Zhang}}, \bibinfo {author} {\bibfnamefont {L.-H.}\ \bibnamefont {Hu}},\ and\ \bibinfo {author} {\bibfnamefont {T.}~\bibnamefont {Neupert}},\ }\bibfield  {title} {\bibinfo {title} {{Finite-momentum Cooper pairing in proximitized altermagnets}},\ }\href {https://doi.org/10.1038/s41467-024-45951-3} {\bibfield  {journal} {\bibinfo  {journal} {Nat. Commun.}\ }\textbf {\bibinfo {volume} {15}},\ \bibinfo {pages} {1801} (\bibinfo {year} {2024})}\BibitemShut {NoStop}%
\bibitem [{\citenamefont {Zhu}\ \emph {et~al.}(2023)\citenamefont {Zhu}, \citenamefont {Zhuang}, \citenamefont {Wu},\ and\ \citenamefont {Yan}}]{ZhuPRB2023}%
  \BibitemOpen
  \bibfield  {author} {\bibinfo {author} {\bibfnamefont {D.}~\bibnamefont {Zhu}}, \bibinfo {author} {\bibfnamefont {Z.-Y.}\ \bibnamefont {Zhuang}}, \bibinfo {author} {\bibfnamefont {Z.}~\bibnamefont {Wu}},\ and\ \bibinfo {author} {\bibfnamefont {Z.}~\bibnamefont {Yan}},\ }\bibfield  {title} {\bibinfo {title} {Topological superconductivity in two-dimensional altermagnetic metals},\ }\href {https://doi.org/10.1103/PhysRevB.108.184505} {\bibfield  {journal} {\bibinfo  {journal} {Phys. Rev. B}\ }\textbf {\bibinfo {volume} {108}},\ \bibinfo {pages} {184505} (\bibinfo {year} {2023})}\BibitemShut {NoStop}%
\bibitem [{\citenamefont {Chakraborty}\ and\ \citenamefont {Black-Schaffer}(2023)}]{Chakraborty2023AMSCd}%
  \BibitemOpen
  \bibfield  {author} {\bibinfo {author} {\bibfnamefont {D.}~\bibnamefont {Chakraborty}}\ and\ \bibinfo {author} {\bibfnamefont {A.~M.}\ \bibnamefont {Black-Schaffer}},\ }\bibfield  {title} {\bibinfo {title} {{Zero-field finite-momentum and field-induced superconductivity in altermagnets}},\ }\href {https://doi.org/10.48550/arXiv.2309.14427} {\bibfield  {journal} {\bibinfo  {journal} {arXiv:2309.14427}\ } (\bibinfo {year} {2023})}\BibitemShut {NoStop}%
\bibitem [{\citenamefont {Bose}\ \emph {et~al.}(2024)\citenamefont {Bose}, \citenamefont {Vadnais},\ and\ \citenamefont {Paramekanti}}]{Bose2024AMSC}%
  \BibitemOpen
  \bibfield  {author} {\bibinfo {author} {\bibfnamefont {A.}~\bibnamefont {Bose}}, \bibinfo {author} {\bibfnamefont {S.}~\bibnamefont {Vadnais}},\ and\ \bibinfo {author} {\bibfnamefont {A.}~\bibnamefont {Paramekanti}},\ }\bibfield  {title} {\bibinfo {title} {{Altermagnetism and superconductivity in a multiorbital t-J model}},\ }\href {https://doi.org/10.48550/arXiv.2403.17050} {\bibfield  {journal} {\bibinfo  {journal} {arXiv:2403.17050}\ } (\bibinfo {year} {2024})}\BibitemShut {NoStop}%
\bibitem [{\citenamefont {Fedchenko}\ \emph {et~al.}(2024)\citenamefont {Fedchenko}, \citenamefont {Min{\ifmmode\acute{a}\else\'{a}\fi}r}, \citenamefont {Akashdeep}, \citenamefont {D{'}Souza}, \citenamefont {Vasilyev}, \citenamefont {Tkach}, \citenamefont {Odenbreit}, \citenamefont {Nguyen}, \citenamefont {Kutnyakhov}, \citenamefont {Wind}, \citenamefont {Wenthaus}, \citenamefont {Scholz}, \citenamefont {Rossnagel}, \citenamefont {Hoesch}, \citenamefont {Aeschlimann}, \citenamefont {Stadtm{\ifmmode\ddot{u}\else\"{u}\fi}ller}, \citenamefont {Kl{\ifmmode\ddot{a}\else\"{a}\fi}ui}, \citenamefont {Sch{\ifmmode\ddot{o}\else\"{o}\fi}nhense}, \citenamefont {Jungwirth}, \citenamefont {Hellenes}, \citenamefont {Jakob}, \citenamefont {{\ifmmode\check{S}\else\v{S}\fi}mejkal}, \citenamefont {Sinova},\ and\ \citenamefont {Elmers}}]{fedchenko2024observation}%
  \BibitemOpen
  \bibfield  {author} {\bibinfo {author} {\bibfnamefont {O.}~\bibnamefont {Fedchenko}}, \bibinfo {author} {\bibfnamefont {J.}~\bibnamefont {Min{\ifmmode\acute{a}\else\'{a}\fi}r}}, \bibinfo {author} {\bibfnamefont {A.}~\bibnamefont {Akashdeep}}, \bibinfo {author} {\bibfnamefont {S.~W.}\ \bibnamefont {D{'}Souza}}, \bibinfo {author} {\bibfnamefont {D.}~\bibnamefont {Vasilyev}}, \bibinfo {author} {\bibfnamefont {O.}~\bibnamefont {Tkach}}, \bibinfo {author} {\bibfnamefont {L.}~\bibnamefont {Odenbreit}}, \bibinfo {author} {\bibfnamefont {Q.}~\bibnamefont {Nguyen}}, \bibinfo {author} {\bibfnamefont {D.}~\bibnamefont {Kutnyakhov}}, \bibinfo {author} {\bibfnamefont {N.}~\bibnamefont {Wind}}, \bibinfo {author} {\bibfnamefont {L.}~\bibnamefont {Wenthaus}}, \bibinfo {author} {\bibfnamefont {M.}~\bibnamefont {Scholz}}, \bibinfo {author} {\bibfnamefont {K.}~\bibnamefont {Rossnagel}}, \bibinfo {author} {\bibfnamefont {M.}~\bibnamefont {Hoesch}}, \bibinfo {author} {\bibfnamefont {M.}~\bibnamefont {Aeschlimann}}, \bibinfo
  {author} {\bibfnamefont {B.}~\bibnamefont {Stadtm{\ifmmode\ddot{u}\else\"{u}\fi}ller}}, \bibinfo {author} {\bibfnamefont {M.}~\bibnamefont {Kl{\ifmmode\ddot{a}\else\"{a}\fi}ui}}, \bibinfo {author} {\bibfnamefont {G.}~\bibnamefont {Sch{\ifmmode\ddot{o}\else\"{o}\fi}nhense}}, \bibinfo {author} {\bibfnamefont {T.}~\bibnamefont {Jungwirth}}, \bibinfo {author} {\bibfnamefont {A.~B.}\ \bibnamefont {Hellenes}}, \bibinfo {author} {\bibfnamefont {G.}~\bibnamefont {Jakob}}, \bibinfo {author} {\bibfnamefont {L.}~\bibnamefont {{\ifmmode\check{S}\else\v{S}\fi}mejkal}}, \bibinfo {author} {\bibfnamefont {J.}~\bibnamefont {Sinova}},\ and\ \bibinfo {author} {\bibfnamefont {H.-J.}\ \bibnamefont {Elmers}},\ }\bibfield  {title} {\bibinfo {title} {{Observation of time-reversal symmetry breaking in the band structure of altermagnetic RuO$_2$}},\ }\href {https://doi.org/10.1126/sciadv.adj4883} {\bibfield  {journal} {\bibinfo  {journal} {Sci. Adv.}\ }\textbf {\bibinfo {volume} {10}},\ \bibinfo {pages} {eadj4883} (\bibinfo {year}
  {2024})}\BibitemShut {NoStop}%
\bibitem [{\citenamefont {Lin}\ \emph {et~al.}(2024)\citenamefont {Lin}, \citenamefont {Chen}, \citenamefont {Lu}, \citenamefont {Liang}, \citenamefont {Feng}, \citenamefont {Yamagami}, \citenamefont {Osiecki}, \citenamefont {Leandersson}, \citenamefont {Thiagarajan}, \citenamefont {Liu}, \citenamefont {Felser},\ and\ \citenamefont {Ma}}]{Lin2024RuO2}%
  \BibitemOpen
  \bibfield  {author} {\bibinfo {author} {\bibfnamefont {Z.}~\bibnamefont {Lin}}, \bibinfo {author} {\bibfnamefont {D.}~\bibnamefont {Chen}}, \bibinfo {author} {\bibfnamefont {W.}~\bibnamefont {Lu}}, \bibinfo {author} {\bibfnamefont {X.}~\bibnamefont {Liang}}, \bibinfo {author} {\bibfnamefont {S.}~\bibnamefont {Feng}}, \bibinfo {author} {\bibfnamefont {K.}~\bibnamefont {Yamagami}}, \bibinfo {author} {\bibfnamefont {J.}~\bibnamefont {Osiecki}}, \bibinfo {author} {\bibfnamefont {M.}~\bibnamefont {Leandersson}}, \bibinfo {author} {\bibfnamefont {B.}~\bibnamefont {Thiagarajan}}, \bibinfo {author} {\bibfnamefont {J.}~\bibnamefont {Liu}}, \bibinfo {author} {\bibfnamefont {C.}~\bibnamefont {Felser}},\ and\ \bibinfo {author} {\bibfnamefont {J.}~\bibnamefont {Ma}},\ }\bibfield  {title} {\bibinfo {title} {{Observation of Giant Spin Splitting and d-wave Spin Texture in Room Temperature Altermagnet RuO2}},\ }\href {https://doi.org/10.48550/arXiv.2402.04995} {\bibfield  {journal} {\bibinfo  {journal} {arXiv:2402.04995}\ }
  (\bibinfo {year} {2024})}\BibitemShut {NoStop}%
\bibitem [{\citenamefont {Lee}\ \emph {et~al.}(2024)\citenamefont {Lee}, \citenamefont {Lee}, \citenamefont {Jung}, \citenamefont {Jung}, \citenamefont {Kim}, \citenamefont {Lee}, \citenamefont {Seok}, \citenamefont {Kim}, \citenamefont {Park}, \citenamefont {{\ifmmode\check{S}\else\v{S}\fi}mejkal}, \citenamefont {Kang},\ and\ \citenamefont {Kim}}]{LeePRL2024}%
  \BibitemOpen
  \bibfield  {author} {\bibinfo {author} {\bibfnamefont {S.}~\bibnamefont {Lee}}, \bibinfo {author} {\bibfnamefont {S.}~\bibnamefont {Lee}}, \bibinfo {author} {\bibfnamefont {S.}~\bibnamefont {Jung}}, \bibinfo {author} {\bibfnamefont {J.}~\bibnamefont {Jung}}, \bibinfo {author} {\bibfnamefont {D.}~\bibnamefont {Kim}}, \bibinfo {author} {\bibfnamefont {Y.}~\bibnamefont {Lee}}, \bibinfo {author} {\bibfnamefont {B.}~\bibnamefont {Seok}}, \bibinfo {author} {\bibfnamefont {J.}~\bibnamefont {Kim}}, \bibinfo {author} {\bibfnamefont {B.~G.}\ \bibnamefont {Park}}, \bibinfo {author} {\bibfnamefont {L.}~\bibnamefont {{\ifmmode\check{S}\else\v{S}\fi}mejkal}}, \bibinfo {author} {\bibfnamefont {C.-J.}\ \bibnamefont {Kang}},\ and\ \bibinfo {author} {\bibfnamefont {C.}~\bibnamefont {Kim}},\ }\bibfield  {title} {\bibinfo {title} {{Broken Kramers Degeneracy in Altermagnetic MnTe}},\ }\href {https://doi.org/10.1103/PhysRevLett.132.036702} {\bibfield  {journal} {\bibinfo  {journal} {Phys. Rev. Lett.}\ }\textbf {\bibinfo {volume}
  {132}},\ \bibinfo {pages} {036702} (\bibinfo {year} {2024})}\BibitemShut {NoStop}%
\bibitem [{\citenamefont {Osumi}\ \emph {et~al.}(2024)\citenamefont {Osumi}, \citenamefont {Souma}, \citenamefont {Aoyama}, \citenamefont {Yamauchi}, \citenamefont {Honma}, \citenamefont {Nakayama}, \citenamefont {Takahashi}, \citenamefont {Ohgushi},\ and\ \citenamefont {Sato}}]{osumi2023observation}%
  \BibitemOpen
  \bibfield  {author} {\bibinfo {author} {\bibfnamefont {T.}~\bibnamefont {Osumi}}, \bibinfo {author} {\bibfnamefont {S.}~\bibnamefont {Souma}}, \bibinfo {author} {\bibfnamefont {T.}~\bibnamefont {Aoyama}}, \bibinfo {author} {\bibfnamefont {K.}~\bibnamefont {Yamauchi}}, \bibinfo {author} {\bibfnamefont {A.}~\bibnamefont {Honma}}, \bibinfo {author} {\bibfnamefont {K.}~\bibnamefont {Nakayama}}, \bibinfo {author} {\bibfnamefont {T.}~\bibnamefont {Takahashi}}, \bibinfo {author} {\bibfnamefont {K.}~\bibnamefont {Ohgushi}},\ and\ \bibinfo {author} {\bibfnamefont {T.}~\bibnamefont {Sato}},\ }\bibfield  {title} {\bibinfo {title} {{Observation of a giant band splitting in altermagnetic MnTe}},\ }\href {https://doi.org/10.1103/PhysRevB.109.115102} {\bibfield  {journal} {\bibinfo  {journal} {Phys. Rev. B}\ }\textbf {\bibinfo {volume} {109}},\ \bibinfo {pages} {115102} (\bibinfo {year} {2024})}\BibitemShut {NoStop}%
\bibitem [{\citenamefont {Krempask{\ifmmode\acute{y}\else\'{y}\fi}}\ \emph {et~al.}(2024)\citenamefont {Krempask{\ifmmode\acute{y}\else\'{y}\fi}}, \citenamefont {{\ifmmode\check{S}\else\v{S}\fi}mejkal}, \citenamefont {D{'}Souza}, \citenamefont {Hajlaoui}, \citenamefont {Springholz}, \citenamefont {Uhl{\ifmmode\acute{\imath}\else\'{\i}\fi}{\ifmmode\check{r}\else\v{r}\fi}ov{\ifmmode\acute{a}\else\'{a}\fi}}, \citenamefont {Alarab}, \citenamefont {Constantinou}, \citenamefont {Strocov}, \citenamefont {Usanov}, \citenamefont {Pudelko}, \citenamefont {Gonz{\ifmmode\acute{a}\else\'{a}\fi}lez-Hern{\ifmmode\acute{a}\else\'{a}\fi}ndez}, \citenamefont {Birk~Hellenes}, \citenamefont {Jansa}, \citenamefont {Reichlov{\ifmmode\acute{a}\else\'{a}\fi}}, \citenamefont {{\ifmmode\check{S}\else\v{S}\fi}ob{\ifmmode\acute{a}\else\'{a}\fi}{\ifmmode\check{n}\else\v{n}\fi}}, \citenamefont {Gonzalez~Betancourt}, \citenamefont {Wadley}, \citenamefont {Sinova}, \citenamefont {Kriegner}, \citenamefont
  {Min{\ifmmode\acute{a}\else\'{a}\fi}r}, \citenamefont {Dil},\ and\ \citenamefont {Jungwirth}}]{krempaskyNature2024}%
  \BibitemOpen
  \bibfield  {author} {\bibinfo {author} {\bibfnamefont {J.}~\bibnamefont {Krempask{\ifmmode\acute{y}\else\'{y}\fi}}}, \bibinfo {author} {\bibfnamefont {L.}~\bibnamefont {{\ifmmode\check{S}\else\v{S}\fi}mejkal}}, \bibinfo {author} {\bibfnamefont {S.~W.}\ \bibnamefont {D{'}Souza}}, \bibinfo {author} {\bibfnamefont {M.}~\bibnamefont {Hajlaoui}}, \bibinfo {author} {\bibfnamefont {G.}~\bibnamefont {Springholz}}, \bibinfo {author} {\bibfnamefont {K.}~\bibnamefont {Uhl{\ifmmode\acute{\imath}\else\'{\i}\fi}{\ifmmode\check{r}\else\v{r}\fi}ov{\ifmmode\acute{a}\else\'{a}\fi}}}, \bibinfo {author} {\bibfnamefont {F.}~\bibnamefont {Alarab}}, \bibinfo {author} {\bibfnamefont {P.~C.}\ \bibnamefont {Constantinou}}, \bibinfo {author} {\bibfnamefont {V.}~\bibnamefont {Strocov}}, \bibinfo {author} {\bibfnamefont {D.}~\bibnamefont {Usanov}}, \bibinfo {author} {\bibfnamefont {W.~R.}\ \bibnamefont {Pudelko}}, \bibinfo {author} {\bibfnamefont {R.}~\bibnamefont
  {Gonz{\ifmmode\acute{a}\else\'{a}\fi}lez-Hern{\ifmmode\acute{a}\else\'{a}\fi}ndez}}, \bibinfo {author} {\bibfnamefont {A.}~\bibnamefont {Birk~Hellenes}}, \bibinfo {author} {\bibfnamefont {Z.}~\bibnamefont {Jansa}}, \bibinfo {author} {\bibfnamefont {H.}~\bibnamefont {Reichlov{\ifmmode\acute{a}\else\'{a}\fi}}}, \bibinfo {author} {\bibfnamefont {Z.}~\bibnamefont {{\ifmmode\check{S}\else\v{S}\fi}ob{\ifmmode\acute{a}\else\'{a}\fi}{\ifmmode\check{n}\else\v{n}\fi}}}, \bibinfo {author} {\bibfnamefont {R.~D.}\ \bibnamefont {Gonzalez~Betancourt}}, \bibinfo {author} {\bibfnamefont {P.}~\bibnamefont {Wadley}}, \bibinfo {author} {\bibfnamefont {J.}~\bibnamefont {Sinova}}, \bibinfo {author} {\bibfnamefont {D.}~\bibnamefont {Kriegner}}, \bibinfo {author} {\bibfnamefont {J.}~\bibnamefont {Min{\ifmmode\acute{a}\else\'{a}\fi}r}}, \bibinfo {author} {\bibfnamefont {J.~H.}\ \bibnamefont {Dil}},\ and\ \bibinfo {author} {\bibfnamefont {T.}~\bibnamefont {Jungwirth}},\ }\bibfield  {title} {\bibinfo {title} {{Altermagnetic lifting
  of Kramers spin degeneracy}},\ }\href {https://doi.org/10.1038/s41586-023-06907-7} {\bibfield  {journal} {\bibinfo  {journal} {Nature}\ }\textbf {\bibinfo {volume} {626}},\ \bibinfo {pages} {517} (\bibinfo {year} {2024})}\BibitemShut {NoStop}%
\bibitem [{\citenamefont {Reimers}\ \emph {et~al.}(2024)\citenamefont {Reimers}, \citenamefont {Odenbreit}, \citenamefont {{\ifmmode\check{S}\else\v{S}\fi}mejkal}, \citenamefont {Strocov}, \citenamefont {Constantinou}, \citenamefont {Hellenes}, \citenamefont {Jaeschke~Ubiergo}, \citenamefont {Campos}, \citenamefont {Bharadwaj}, \citenamefont {Chakraborty}, \citenamefont {Denneulin}, \citenamefont {Shi}, \citenamefont {Dunin-Borkowski}, \citenamefont {Das}, \citenamefont {Kl{\ifmmode\ddot{a}\else\"{a}\fi}ui}, \citenamefont {Sinova},\ and\ \citenamefont {Jourdan}}]{reimers2023direct}%
  \BibitemOpen
  \bibfield  {author} {\bibinfo {author} {\bibfnamefont {S.}~\bibnamefont {Reimers}}, \bibinfo {author} {\bibfnamefont {L.}~\bibnamefont {Odenbreit}}, \bibinfo {author} {\bibfnamefont {L.}~\bibnamefont {{\ifmmode\check{S}\else\v{S}\fi}mejkal}}, \bibinfo {author} {\bibfnamefont {V.~N.}\ \bibnamefont {Strocov}}, \bibinfo {author} {\bibfnamefont {P.}~\bibnamefont {Constantinou}}, \bibinfo {author} {\bibfnamefont {A.~B.}\ \bibnamefont {Hellenes}}, \bibinfo {author} {\bibfnamefont {R.}~\bibnamefont {Jaeschke~Ubiergo}}, \bibinfo {author} {\bibfnamefont {W.~H.}\ \bibnamefont {Campos}}, \bibinfo {author} {\bibfnamefont {V.~K.}\ \bibnamefont {Bharadwaj}}, \bibinfo {author} {\bibfnamefont {A.}~\bibnamefont {Chakraborty}}, \bibinfo {author} {\bibfnamefont {T.}~\bibnamefont {Denneulin}}, \bibinfo {author} {\bibfnamefont {W.}~\bibnamefont {Shi}}, \bibinfo {author} {\bibfnamefont {R.~E.}\ \bibnamefont {Dunin-Borkowski}}, \bibinfo {author} {\bibfnamefont {S.}~\bibnamefont {Das}}, \bibinfo {author} {\bibfnamefont
  {M.}~\bibnamefont {Kl{\ifmmode\ddot{a}\else\"{a}\fi}ui}}, \bibinfo {author} {\bibfnamefont {J.}~\bibnamefont {Sinova}},\ and\ \bibinfo {author} {\bibfnamefont {M.}~\bibnamefont {Jourdan}},\ }\bibfield  {title} {\bibinfo {title} {{Direct observation of altermagnetic band splitting in CrSb thin films}},\ }\href {https://doi.org/10.1038/s41467-024-46476-5} {\bibfield  {journal} {\bibinfo  {journal} {Nat. Commun.}\ }\textbf {\bibinfo {volume} {15}},\ \bibinfo {pages} {2116} (\bibinfo {year} {2024})}\BibitemShut {NoStop}%
\bibitem [{\citenamefont {Brekke}\ \emph {et~al.}(2023)\citenamefont {Brekke}, \citenamefont {Brataas},\ and\ \citenamefont {Sudb{\o}}}]{Brekke2023Aug}%
  \BibitemOpen
  \bibfield  {author} {\bibinfo {author} {\bibfnamefont {B.}~\bibnamefont {Brekke}}, \bibinfo {author} {\bibfnamefont {A.}~\bibnamefont {Brataas}},\ and\ \bibinfo {author} {\bibfnamefont {A.}~\bibnamefont {Sudb{\o}}},\ }\bibfield  {title} {\bibinfo {title} {{Two-dimensional altermagnets: Superconductivity in a minimal microscopic model}},\ }\href {https://doi.org/10.1103/PhysRevB.108.224421} {\bibfield  {journal} {\bibinfo  {journal} {Phys. Rev. B}\ }\textbf {\bibinfo {volume} {108}},\ \bibinfo {pages} {224421} (\bibinfo {year} {2023})}\BibitemShut {NoStop}%
\bibitem [{\citenamefont {Maier}\ and\ \citenamefont {Okamoto}(2023)}]{Maier2023Sep}%
  \BibitemOpen
  \bibfield  {author} {\bibinfo {author} {\bibfnamefont {T.~A.}\ \bibnamefont {Maier}}\ and\ \bibinfo {author} {\bibfnamefont {S.}~\bibnamefont {Okamoto}},\ }\bibfield  {title} {\bibinfo {title} {{Weak-coupling theory of neutron scattering as a probe of altermagnetism}},\ }\href {https://doi.org/10.1103/PhysRevB.108.L100402} {\bibfield  {journal} {\bibinfo  {journal} {Phys. Rev. B}\ }\textbf {\bibinfo {volume} {108}},\ \bibinfo {pages} {L100402} (\bibinfo {year} {2023})}\BibitemShut {NoStop}%
\bibitem [{\citenamefont {Roig}\ \emph {et~al.}(2024)\citenamefont {Roig}, \citenamefont {Kreisel}, \citenamefont {Yu}, \citenamefont {Andersen},\ and\ \citenamefont {Agterberg}}]{Roig2024Feb}%
  \BibitemOpen
  \bibfield  {author} {\bibinfo {author} {\bibfnamefont {M.}~\bibnamefont {Roig}}, \bibinfo {author} {\bibfnamefont {A.}~\bibnamefont {Kreisel}}, \bibinfo {author} {\bibfnamefont {Y.}~\bibnamefont {Yu}}, \bibinfo {author} {\bibfnamefont {B.~M.}\ \bibnamefont {Andersen}},\ and\ \bibinfo {author} {\bibfnamefont {D.~F.}\ \bibnamefont {Agterberg}},\ }\bibfield  {title} {\bibinfo {title} {{Minimal Models for Altermagnetism}},\ }\href {https://doi.org/10.48550/arXiv.2402.15616} {\bibfield  {journal} {\bibinfo  {journal} {arXiv:2402.15616}\ } (\bibinfo {year} {2024})}\BibitemShut {NoStop}%
\bibitem [{\citenamefont {Lieb}(1989)}]{Lieb1989Mar}%
  \BibitemOpen
  \bibfield  {author} {\bibinfo {author} {\bibfnamefont {E.~H.}\ \bibnamefont {Lieb}},\ }\bibfield  {title} {\bibinfo {title} {{Two theorems on the Hubbard model}},\ }\href {https://doi.org/10.1103/PhysRevLett.62.1201} {\bibfield  {journal} {\bibinfo  {journal} {Phys. Rev. Lett.}\ }\textbf {\bibinfo {volume} {62}},\ \bibinfo {pages} {1201} (\bibinfo {year} {1989})}\BibitemShut {NoStop}%
\bibitem [{\citenamefont {Bardeen}\ \emph {et~al.}(1957)\citenamefont {Bardeen}, \citenamefont {Cooper},\ and\ \citenamefont {Schrieffer}}]{BCS}%
  \BibitemOpen
  \bibfield  {author} {\bibinfo {author} {\bibfnamefont {J.}~\bibnamefont {Bardeen}}, \bibinfo {author} {\bibfnamefont {L.~N.}\ \bibnamefont {Cooper}},\ and\ \bibinfo {author} {\bibfnamefont {J.~R.}\ \bibnamefont {Schrieffer}},\ }\bibfield  {title} {\bibinfo {title} {Theory of {Superconductivity}},\ }\href {https://doi.org/10.1103/PhysRev.108.1175} {\bibfield  {journal} {\bibinfo  {journal} {Phys. Rev.}\ }\textbf {\bibinfo {volume} {108}},\ \bibinfo {pages} {1175} (\bibinfo {year} {1957})}\BibitemShut {NoStop}%
\bibitem [{\citenamefont {Monthoux}\ and\ \citenamefont {Pines}(1993)}]{Pines1993}%
  \BibitemOpen
  \bibfield  {author} {\bibinfo {author} {\bibfnamefont {P.}~\bibnamefont {Monthoux}}\ and\ \bibinfo {author} {\bibfnamefont {D.}~\bibnamefont {Pines}},\ }\bibfield  {title} {\bibinfo {title} {{${\mathrm{YBa}}_{2}$${\mathrm{Cu}}_{3}$${\mathrm{O}}_{7}$: A nearly antiferromagnetic Fermi liquid}},\ }\href {https://doi.org/10.1103/PhysRevB.47.6069} {\bibfield  {journal} {\bibinfo  {journal} {Phys. Rev. B}\ }\textbf {\bibinfo {volume} {47}},\ \bibinfo {pages} {6069} (\bibinfo {year} {1993})}\BibitemShut {NoStop}%
\bibitem [{\citenamefont {Scalapino}(1999)}]{SCspinHistory_Scalapino1999}%
  \BibitemOpen
  \bibfield  {author} {\bibinfo {author} {\bibfnamefont {D.~J.}\ \bibnamefont {Scalapino}},\ }\bibfield  {title} {\bibinfo {title} {{Superconductivity and Spin Fluctuations}},\ }\href {https://doi.org/10.1023/A:1022559920049} {\bibfield  {journal} {\bibinfo  {journal} {J. Low Temp. Phys.}\ }\textbf {\bibinfo {volume} {117}},\ \bibinfo {pages} {179} (\bibinfo {year} {1999})}\BibitemShut {NoStop}%
\bibitem [{\citenamefont {Moriya}\ and\ \citenamefont {Ueda}(2003)}]{SCAFMspinMoriya2003Jul}%
  \BibitemOpen
  \bibfield  {author} {\bibinfo {author} {\bibfnamefont {T.}~\bibnamefont {Moriya}}\ and\ \bibinfo {author} {\bibfnamefont {K.}~\bibnamefont {Ueda}},\ }\bibfield  {title} {\bibinfo {title} {{Antiferromagnetic spin fluctuation and superconductivity}},\ }\href {https://doi.org/10.1088/0034-4885/66/8/202} {\bibfield  {journal} {\bibinfo  {journal} {Rep. Prog. Phys.}\ }\textbf {\bibinfo {volume} {66}},\ \bibinfo {pages} {1299} (\bibinfo {year} {2003})}\BibitemShut {NoStop}%
\bibitem [{\citenamefont {Hirschfeld}\ \emph {et~al.}(2011)\citenamefont {Hirschfeld}, \citenamefont {Korshunov},\ and\ \citenamefont {Mazin}}]{SCspinGapSym_Hirschfeld2011}%
  \BibitemOpen
  \bibfield  {author} {\bibinfo {author} {\bibfnamefont {P.~J.}\ \bibnamefont {Hirschfeld}}, \bibinfo {author} {\bibfnamefont {M.~M.}\ \bibnamefont {Korshunov}},\ and\ \bibinfo {author} {\bibfnamefont {I.~I.}\ \bibnamefont {Mazin}},\ }\bibfield  {title} {\bibinfo {title} {{Gap symmetry and structure of Fe-based superconductors}},\ }\href {https://doi.org/10.1088/0034-4885/74/12/124508} {\bibfield  {journal} {\bibinfo  {journal} {Rep. Prog. Phys.}\ }\textbf {\bibinfo {volume} {74}},\ \bibinfo {pages} {124508} (\bibinfo {year} {2011})}\BibitemShut {NoStop}%
\bibitem [{\citenamefont {Kargarian}\ \emph {et~al.}(2016)\citenamefont {Kargarian}, \citenamefont {Efimkin},\ and\ \citenamefont {Galitski}}]{KargarianFMTI}%
  \BibitemOpen
  \bibfield  {author} {\bibinfo {author} {\bibfnamefont {M.}~\bibnamefont {Kargarian}}, \bibinfo {author} {\bibfnamefont {D.~K.}\ \bibnamefont {Efimkin}},\ and\ \bibinfo {author} {\bibfnamefont {V.}~\bibnamefont {Galitski}},\ }\bibfield  {title} {\bibinfo {title} {{Amperean Pairing at the Surface of Topological Insulators}},\ }\href {https://doi.org/10.1103/PhysRevLett.117.076806} {\bibfield  {journal} {\bibinfo  {journal} {Phys. Rev. Lett.}\ }\textbf {\bibinfo {volume} {117}},\ \bibinfo {pages} {076806} (\bibinfo {year} {2016})}\BibitemShut {NoStop}%
\bibitem [{\citenamefont {Rohling}\ \emph {et~al.}(2018)\citenamefont {Rohling}, \citenamefont {Fj{\ae}rbu},\ and\ \citenamefont {Brataas}}]{ArneFMNM}%
  \BibitemOpen
  \bibfield  {author} {\bibinfo {author} {\bibfnamefont {N.}~\bibnamefont {Rohling}}, \bibinfo {author} {\bibfnamefont {E.~L.}\ \bibnamefont {Fj{\ae}rbu}},\ and\ \bibinfo {author} {\bibfnamefont {A.}~\bibnamefont {Brataas}},\ }\bibfield  {title} {\bibinfo {title} {{Superconductivity induced by interfacial coupling to magnons}},\ }\href {https://doi.org/10.1103/PhysRevB.97.115401} {\bibfield  {journal} {\bibinfo  {journal} {Phys. Rev. B}\ }\textbf {\bibinfo {volume} {97}},\ \bibinfo {pages} {115401} (\bibinfo {year} {2018})}\BibitemShut {NoStop}%
\bibitem [{\citenamefont {Brekke}\ \emph {et~al.}(2024)\citenamefont {Brekke}, \citenamefont {Sudb{\o}},\ and\ \citenamefont {Brataas}}]{brekke2023interfacial}%
  \BibitemOpen
  \bibfield  {author} {\bibinfo {author} {\bibfnamefont {B.}~\bibnamefont {Brekke}}, \bibinfo {author} {\bibfnamefont {A.}~\bibnamefont {Sudb{\o}}},\ and\ \bibinfo {author} {\bibfnamefont {A.}~\bibnamefont {Brataas}},\ }\bibfield  {title} {\bibinfo {title} {{Interfacial magnon-mediated superconductivity in twisted bilayer graphene}},\ }\href {https://doi.org/10.1088/1367-2630/ad2ffd} {\bibfield  {journal} {\bibinfo  {journal} {New J. Phys.}\ }\textbf {\bibinfo {volume} {26}},\ \bibinfo {pages} {033014} (\bibinfo {year} {2024})}\BibitemShut {NoStop}%
\bibitem [{\citenamefont {Hugdal}\ and\ \citenamefont {Sudb{\o}}(2020)}]{Hugdal2020FMTI}%
  \BibitemOpen
  \bibfield  {author} {\bibinfo {author} {\bibfnamefont {H.~G.}\ \bibnamefont {Hugdal}}\ and\ \bibinfo {author} {\bibfnamefont {A.}~\bibnamefont {Sudb{\o}}},\ }\bibfield  {title} {\bibinfo {title} {{Possible odd-frequency Amperean magnon-mediated superconductivity in topological insulator--ferromagnetic insulator bilayer}},\ }\href {https://doi.org/10.1103/PhysRevB.102.125429} {\bibfield  {journal} {\bibinfo  {journal} {Phys. Rev. B}\ }\textbf {\bibinfo {volume} {102}},\ \bibinfo {pages} {125429} (\bibinfo {year} {2020})}\BibitemShut {NoStop}%
\bibitem [{\citenamefont {Hugdal}\ \emph {et~al.}(2018)\citenamefont {Hugdal}, \citenamefont {Rex}, \citenamefont {Nogueira},\ and\ \citenamefont {Sudb{\o}}}]{Hugdal2018May}%
  \BibitemOpen
  \bibfield  {author} {\bibinfo {author} {\bibfnamefont {H.~G.}\ \bibnamefont {Hugdal}}, \bibinfo {author} {\bibfnamefont {S.}~\bibnamefont {Rex}}, \bibinfo {author} {\bibfnamefont {F.~S.}\ \bibnamefont {Nogueira}},\ and\ \bibinfo {author} {\bibfnamefont {A.}~\bibnamefont {Sudb{\o}}},\ }\bibfield  {title} {\bibinfo {title} {{Magnon-induced superconductivity in a topological insulator coupled to ferromagnetic and antiferromagnetic insulators}},\ }\href {https://doi.org/10.1103/PhysRevB.97.195438} {\bibfield  {journal} {\bibinfo  {journal} {Phys. Rev. B}\ }\textbf {\bibinfo {volume} {97}},\ \bibinfo {pages} {195438} (\bibinfo {year} {2018})}\BibitemShut {NoStop}%
\bibitem [{\citenamefont {Erlandsen}\ \emph {et~al.}(2020)\citenamefont {Erlandsen}, \citenamefont {Brataas},\ and\ \citenamefont {Sudb{\o}}}]{Erlandsen2020TIAFM}%
  \BibitemOpen
  \bibfield  {author} {\bibinfo {author} {\bibfnamefont {E.}~\bibnamefont {Erlandsen}}, \bibinfo {author} {\bibfnamefont {A.}~\bibnamefont {Brataas}},\ and\ \bibinfo {author} {\bibfnamefont {A.}~\bibnamefont {Sudb{\o}}},\ }\bibfield  {title} {\bibinfo {title} {{Magnon-mediated superconductivity on the surface of a topological insulator}},\ }\href {https://doi.org/10.1103/PhysRevB.101.094503} {\bibfield  {journal} {\bibinfo  {journal} {Phys. Rev. B}\ }\textbf {\bibinfo {volume} {101}},\ \bibinfo {pages} {094503} (\bibinfo {year} {2020})}\BibitemShut {NoStop}%
\bibitem [{\citenamefont {Erlandsen}\ \emph {et~al.}(2019)\citenamefont {Erlandsen}, \citenamefont {Kamra}, \citenamefont {Brataas},\ and\ \citenamefont {Sudb{\o}}}]{EirikAFMNM}%
  \BibitemOpen
  \bibfield  {author} {\bibinfo {author} {\bibfnamefont {E.}~\bibnamefont {Erlandsen}}, \bibinfo {author} {\bibfnamefont {A.}~\bibnamefont {Kamra}}, \bibinfo {author} {\bibfnamefont {A.}~\bibnamefont {Brataas}},\ and\ \bibinfo {author} {\bibfnamefont {A.}~\bibnamefont {Sudb{\o}}},\ }\bibfield  {title} {\bibinfo {title} {{Enhancement of superconductivity mediated by antiferromagnetic squeezed magnons}},\ }\href {https://doi.org/10.1103/PhysRevB.100.100503} {\bibfield  {journal} {\bibinfo  {journal} {Phys. Rev. B}\ }\textbf {\bibinfo {volume} {100}},\ \bibinfo {pages} {100503} (\bibinfo {year} {2019})}\BibitemShut {NoStop}%
\bibitem [{\citenamefont {Fj{\ae}rbu}\ \emph {et~al.}(2019)\citenamefont {Fj{\ae}rbu}, \citenamefont {Rohling},\ and\ \citenamefont {Brataas}}]{Fjaerbu2019arneAFMNM}%
  \BibitemOpen
  \bibfield  {author} {\bibinfo {author} {\bibfnamefont {E.~L.}\ \bibnamefont {Fj{\ae}rbu}}, \bibinfo {author} {\bibfnamefont {N.}~\bibnamefont {Rohling}},\ and\ \bibinfo {author} {\bibfnamefont {A.}~\bibnamefont {Brataas}},\ }\bibfield  {title} {\bibinfo {title} {{Superconductivity at metal-antiferromagnetic insulator interfaces}},\ }\href {https://doi.org/10.1103/PhysRevB.100.125432} {\bibfield  {journal} {\bibinfo  {journal} {Phys. Rev. B}\ }\textbf {\bibinfo {volume} {100}},\ \bibinfo {pages} {125432} (\bibinfo {year} {2019})}\BibitemShut {NoStop}%
\bibitem [{\citenamefont {Thingstad}\ \emph {et~al.}(2021)\citenamefont {Thingstad}, \citenamefont {Erlandsen},\ and\ \citenamefont {Sudb{\o}}}]{ThingstadEliashberg}%
  \BibitemOpen
  \bibfield  {author} {\bibinfo {author} {\bibfnamefont {E.}~\bibnamefont {Thingstad}}, \bibinfo {author} {\bibfnamefont {E.}~\bibnamefont {Erlandsen}},\ and\ \bibinfo {author} {\bibfnamefont {A.}~\bibnamefont {Sudb{\o}}},\ }\bibfield  {title} {\bibinfo {title} {{Eliashberg study of superconductivity induced by interfacial coupling to antiferromagnets}},\ }\href {https://doi.org/10.1103/PhysRevB.104.014508} {\bibfield  {journal} {\bibinfo  {journal} {Phys. Rev. B}\ }\textbf {\bibinfo {volume} {104}},\ \bibinfo {pages} {014508} (\bibinfo {year} {2021})}\BibitemShut {NoStop}%
\bibitem [{\citenamefont {Sun}\ \emph {et~al.}(2023{\natexlab{b}})\citenamefont {Sun}, \citenamefont {M{\ae}land},\ and\ \citenamefont {Sudb{\o}}}]{Sun2023Aug}%
  \BibitemOpen
  \bibfield  {author} {\bibinfo {author} {\bibfnamefont {C.}~\bibnamefont {Sun}}, \bibinfo {author} {\bibfnamefont {K.}~\bibnamefont {M{\ae}land}},\ and\ \bibinfo {author} {\bibfnamefont {A.}~\bibnamefont {Sudb{\o}}},\ }\bibfield  {title} {\bibinfo {title} {{Stability of superconducting gap symmetries arising from antiferromagnetic magnons}},\ }\href {https://doi.org/10.1103/PhysRevB.108.054520} {\bibfield  {journal} {\bibinfo  {journal} {Phys. Rev. B}\ }\textbf {\bibinfo {volume} {108}},\ \bibinfo {pages} {054520} (\bibinfo {year} {2023}{\natexlab{b}})}\BibitemShut {NoStop}%
\bibitem [{\citenamefont {M{\ae}land}\ and\ \citenamefont {Sudb{\o}}(2023{\natexlab{a}})}]{Maeland2023PRL}%
  \BibitemOpen
  \bibfield  {author} {\bibinfo {author} {\bibfnamefont {K.}~\bibnamefont {M{\ae}land}}\ and\ \bibinfo {author} {\bibfnamefont {A.}~\bibnamefont {Sudb{\o}}},\ }\bibfield  {title} {\bibinfo {title} {{Topological Superconductivity Mediated by Skyrmionic Magnons}},\ }\href {https://doi.org/10.1103/PhysRevLett.130.156002} {\bibfield  {journal} {\bibinfo  {journal} {Phys. Rev. Lett.}\ }\textbf {\bibinfo {volume} {130}},\ \bibinfo {pages} {156002} (\bibinfo {year} {2023}{\natexlab{a}})}\BibitemShut {NoStop}%
\bibitem [{\citenamefont {M{\ae}land}\ \emph {et~al.}(2023)\citenamefont {M{\ae}land}, \citenamefont {Abnar}, \citenamefont {Benestad},\ and\ \citenamefont {Sudb{\o}}}]{Maeland2023Dec}%
  \BibitemOpen
  \bibfield  {author} {\bibinfo {author} {\bibfnamefont {K.}~\bibnamefont {M{\ae}land}}, \bibinfo {author} {\bibfnamefont {S.}~\bibnamefont {Abnar}}, \bibinfo {author} {\bibfnamefont {J.}~\bibnamefont {Benestad}},\ and\ \bibinfo {author} {\bibfnamefont {A.}~\bibnamefont {Sudb{\o}}},\ }\bibfield  {title} {\bibinfo {title} {{Topological superconductivity mediated by magnons of helical magnetic states}},\ }\href {https://doi.org/10.1103/PhysRevB.108.224515} {\bibfield  {journal} {\bibinfo  {journal} {Phys. Rev. B}\ }\textbf {\bibinfo {volume} {108}},\ \bibinfo {pages} {224515} (\bibinfo {year} {2023})}\BibitemShut {NoStop}%
\bibitem [{\citenamefont {Bostr{\ifmmode\ddot{o}\else\"{o}\fi}m}\ and\ \citenamefont {Bostr{\ifmmode\ddot{o}\else\"{o}\fi}m}(2023)}]{Bostrom2023Dec}%
  \BibitemOpen
  \bibfield  {author} {\bibinfo {author} {\bibfnamefont {F.~V.}\ \bibnamefont {Bostr{\ifmmode\ddot{o}\else\"{o}\fi}m}}\ and\ \bibinfo {author} {\bibfnamefont {E.~V.}\ \bibnamefont {Bostr{\ifmmode\ddot{o}\else\"{o}\fi}m}},\ }\bibfield  {title} {\bibinfo {title} {{Topological superconductivity in quantum wires mediated by helical magnons}},\ }\href {https://arxiv.org/abs/2312.02655v1} {\bibfield  {journal} {\bibinfo  {journal} {arXiv:2312.02655}\ } (\bibinfo {year} {2023})}\BibitemShut {NoStop}%
\bibitem [{\citenamefont {Fossheim}\ and\ \citenamefont {Sudb{\o}}(2004)}]{SFsuperconductivity}%
  \BibitemOpen
  \bibfield  {author} {\bibinfo {author} {\bibfnamefont {K.}~\bibnamefont {Fossheim}}\ and\ \bibinfo {author} {\bibfnamefont {A.}~\bibnamefont {Sudb{\o}}},\ }\href@noop {} {\emph {\bibinfo {title} {Superconductivity: Physics and Applications}}}\ (\bibinfo  {publisher} {Wiley, Chichester, UK},\ \bibinfo {year} {2004})\BibitemShut {NoStop}%
\bibitem [{\citenamefont {Eliashberg}(1960{\natexlab{a}})}]{Eliashberg1960Sep}%
  \BibitemOpen
  \bibfield  {author} {\bibinfo {author} {\bibfnamefont {G.~M.}\ \bibnamefont {Eliashberg}},\ }\bibfield  {title} {\bibinfo {title} {{Interactions between electrons and lattice vibrations in a superconductor}},\ }\href@noop {} {\bibfield  {journal} {\bibinfo  {journal} {Zh. Eksp. Teor. Fiz.}\ }\textbf {\bibinfo {volume} {38}},\ \bibinfo {pages} {966} (\bibinfo {year} {1960}{\natexlab{a}})},\ \bibinfo {note} {[Sov. Phys. JETP \textbf{11}, 696 (1960)]}\BibitemShut {NoStop}%
\bibitem [{\citenamefont {Eliashberg}(1960{\natexlab{b}})}]{Eliashberg1961}%
  \BibitemOpen
  \bibfield  {author} {\bibinfo {author} {\bibfnamefont {G.~M.}\ \bibnamefont {Eliashberg}},\ }\bibfield  {title} {\bibinfo {title} {Temperature {Green’s} function for electrons in a superconductor},\ }\href@noop {} {\bibfield  {journal} {\bibinfo  {journal} {Zh. Eksp. Teor. Fiz.}\ }\textbf {\bibinfo {volume} {39}},\ \bibinfo {pages} {1437} (\bibinfo {year} {1960}{\natexlab{b}})},\ \bibinfo {note} {[Sov. Phys. JETP \textbf{12}, 1000 (1961)]}\BibitemShut {NoStop}%
\bibitem [{\citenamefont {Schrieffer}(1964)}]{Schrieffer1964book}%
  \BibitemOpen
  \bibfield  {author} {\bibinfo {author} {\bibfnamefont {J.~R.}\ \bibnamefont {Schrieffer}},\ }\href@noop {} {\emph {\bibinfo {title} {{Theory of Superconductivity}}}}\ (\bibinfo  {publisher} {Benjamin},\ \bibinfo {address} {Reading, MA},\ \bibinfo {year} {1964})\BibitemShut {NoStop}%
\bibitem [{\citenamefont {Carbotte}(1990)}]{Carbotte1990FreeEnergy}%
  \BibitemOpen
  \bibfield  {author} {\bibinfo {author} {\bibfnamefont {J.~P.}\ \bibnamefont {Carbotte}},\ }\bibfield  {title} {\bibinfo {title} {{Properties of boson-exchange superconductors}},\ }\href {https://doi.org/10.1103/RevModPhys.62.1027} {\bibfield  {journal} {\bibinfo  {journal} {Rev. Mod. Phys.}\ }\textbf {\bibinfo {volume} {62}},\ \bibinfo {pages} {1027} (\bibinfo {year} {1990})}\BibitemShut {NoStop}%
\bibitem [{\citenamefont {Marsiglio}(2020)}]{EliashbergRevMarsiglio2020}%
  \BibitemOpen
  \bibfield  {author} {\bibinfo {author} {\bibfnamefont {F.}~\bibnamefont {Marsiglio}},\ }\bibfield  {title} {\bibinfo {title} {{Eliashberg theory: A short review}},\ }\href {https://doi.org/10.1016/j.aop.2020.168102} {\bibfield  {journal} {\bibinfo  {journal} {Ann. Phys.}\ }\textbf {\bibinfo {volume} {417}},\ \bibinfo {pages} {168102} (\bibinfo {year} {2020})}\BibitemShut {NoStop}%
\bibitem [{\citenamefont {Chubukov}\ \emph {et~al.}(2020)\citenamefont {Chubukov}, \citenamefont {Abanov}, \citenamefont {Esterlis},\ and\ \citenamefont {Kivelson}}]{Chubukov2020FEspecialized}%
  \BibitemOpen
  \bibfield  {author} {\bibinfo {author} {\bibfnamefont {A.~V.}\ \bibnamefont {Chubukov}}, \bibinfo {author} {\bibfnamefont {A.}~\bibnamefont {Abanov}}, \bibinfo {author} {\bibfnamefont {I.}~\bibnamefont {Esterlis}},\ and\ \bibinfo {author} {\bibfnamefont {S.~A.}\ \bibnamefont {Kivelson}},\ }\bibfield  {title} {\bibinfo {title} {{Eliashberg theory of phonon-mediated superconductivity {\ifmmode---\else\textemdash\fi} When it is valid and how it breaks down}},\ }\href {https://doi.org/10.1016/j.aop.2020.168190} {\bibfield  {journal} {\bibinfo  {journal} {Ann. Phys.}\ }\textbf {\bibinfo {volume} {417}},\ \bibinfo {pages} {168190} (\bibinfo {year} {2020})}\BibitemShut {NoStop}%
\bibitem [{\citenamefont {Ginzburg}(1956)}]{ginzburg1957FMSC}%
  \BibitemOpen
  \bibfield  {author} {\bibinfo {author} {\bibfnamefont {V.~L.}\ \bibnamefont {Ginzburg}},\ }\bibfield  {title} {\bibinfo {title} {Ferromagnetic superconductors},\ }\href@noop {} {\bibfield  {journal} {\bibinfo  {journal} {Zh. Eksp. Teor. Fiz.}\ }\textbf {\bibinfo {volume} {31}},\ \bibinfo {pages} {202} (\bibinfo {year} {1956})},\ \bibinfo {note} {[Sov. Phys. JETP, \textbf{4}, 153 (1957)]}\BibitemShut {NoStop}%
\bibitem [{\citenamefont {Fay}\ and\ \citenamefont {Appel}(1980)}]{Fay1980FMSC}%
  \BibitemOpen
  \bibfield  {author} {\bibinfo {author} {\bibfnamefont {D.}~\bibnamefont {Fay}}\ and\ \bibinfo {author} {\bibfnamefont {J.}~\bibnamefont {Appel}},\ }\bibfield  {title} {\bibinfo {title} {{Coexistence of $p$-state superconductivity and itinerant ferromagnetism}},\ }\href {https://doi.org/10.1103/PhysRevB.22.3173} {\bibfield  {journal} {\bibinfo  {journal} {Phys. Rev. B}\ }\textbf {\bibinfo {volume} {22}},\ \bibinfo {pages} {3173} (\bibinfo {year} {1980})}\BibitemShut {NoStop}%
\bibitem [{\citenamefont {Linder}\ and\ \citenamefont {Sudb{\o}}(2007)}]{Linder2007FMSC}%
  \BibitemOpen
  \bibfield  {author} {\bibinfo {author} {\bibfnamefont {J.}~\bibnamefont {Linder}}\ and\ \bibinfo {author} {\bibfnamefont {A.}~\bibnamefont {Sudb{\o}}},\ }\bibfield  {title} {\bibinfo {title} {{Quantum transport in noncentrosymmetric superconductors and thermodynamics of ferromagnetic superconductors}},\ }\href {https://doi.org/10.1103/PhysRevB.76.054511} {\bibfield  {journal} {\bibinfo  {journal} {Phys. Rev. B}\ }\textbf {\bibinfo {volume} {76}},\ \bibinfo {pages} {054511} (\bibinfo {year} {2007})}\BibitemShut {NoStop}%
\bibitem [{\citenamefont {Jian}\ \emph {et~al.}(2009)\citenamefont {Jian}, \citenamefont {Zhang}, \citenamefont {Gu},\ and\ \citenamefont {Klemm}}]{Jian2009FMSC}%
  \BibitemOpen
  \bibfield  {author} {\bibinfo {author} {\bibfnamefont {X.}~\bibnamefont {Jian}}, \bibinfo {author} {\bibfnamefont {J.}~\bibnamefont {Zhang}}, \bibinfo {author} {\bibfnamefont {Q.}~\bibnamefont {Gu}},\ and\ \bibinfo {author} {\bibfnamefont {R.~A.}\ \bibnamefont {Klemm}},\ }\bibfield  {title} {\bibinfo {title} {{Enhancement of ferromagnetism by $p$-wave Cooper pairing in superconducting ferromagnets}},\ }\href {https://doi.org/10.1103/PhysRevB.80.224514} {\bibfield  {journal} {\bibinfo  {journal} {Phys. Rev. B}\ }\textbf {\bibinfo {volume} {80}},\ \bibinfo {pages} {224514} (\bibinfo {year} {2009})}\BibitemShut {NoStop}%
\bibitem [{\citenamefont {Saxena}\ \emph {et~al.}(2000)\citenamefont {Saxena}, \citenamefont {Agarwal}, \citenamefont {Ahilan}, \citenamefont {Grosche}, \citenamefont {Haselwimmer}, \citenamefont {Steiner}, \citenamefont {Pugh}, \citenamefont {Walker}, \citenamefont {Julian}, \citenamefont {Monthoux}, \citenamefont {Lonzarich}, \citenamefont {Huxley}, \citenamefont {Sheikin}, \citenamefont {Braithwaite},\ and\ \citenamefont {Flouquet}}]{Saxena2000FMSC}%
  \BibitemOpen
  \bibfield  {author} {\bibinfo {author} {\bibfnamefont {S.~S.}\ \bibnamefont {Saxena}}, \bibinfo {author} {\bibfnamefont {P.}~\bibnamefont {Agarwal}}, \bibinfo {author} {\bibfnamefont {K.}~\bibnamefont {Ahilan}}, \bibinfo {author} {\bibfnamefont {F.~M.}\ \bibnamefont {Grosche}}, \bibinfo {author} {\bibfnamefont {R.~K.~W.}\ \bibnamefont {Haselwimmer}}, \bibinfo {author} {\bibfnamefont {M.~J.}\ \bibnamefont {Steiner}}, \bibinfo {author} {\bibfnamefont {E.}~\bibnamefont {Pugh}}, \bibinfo {author} {\bibfnamefont {I.~R.}\ \bibnamefont {Walker}}, \bibinfo {author} {\bibfnamefont {S.~R.}\ \bibnamefont {Julian}}, \bibinfo {author} {\bibfnamefont {P.}~\bibnamefont {Monthoux}}, \bibinfo {author} {\bibfnamefont {G.~G.}\ \bibnamefont {Lonzarich}}, \bibinfo {author} {\bibfnamefont {A.}~\bibnamefont {Huxley}}, \bibinfo {author} {\bibfnamefont {I.}~\bibnamefont {Sheikin}}, \bibinfo {author} {\bibfnamefont {D.}~\bibnamefont {Braithwaite}},\ and\ \bibinfo {author} {\bibfnamefont {J.}~\bibnamefont {Flouquet}},\ }\bibfield
  {title} {\bibinfo {title} {{Superconductivity on the border of itinerant-electron ferromagnetism in UGe$_2$}},\ }\href {https://doi.org/10.1038/35020500} {\bibfield  {journal} {\bibinfo  {journal} {Nature}\ }\textbf {\bibinfo {volume} {406}},\ \bibinfo {pages} {587} (\bibinfo {year} {2000})}\BibitemShut {NoStop}%
\bibitem [{\citenamefont {Huy}\ \emph {et~al.}(2007)\citenamefont {Huy}, \citenamefont {Gasparini}, \citenamefont {de~Nijs}, \citenamefont {Huang}, \citenamefont {Klaasse}, \citenamefont {Gortenmulder}, \citenamefont {de~Visser}, \citenamefont {Hamann}, \citenamefont {G{\ifmmode\ddot{o}\else\"{o}\fi}rlach},\ and\ \citenamefont {L{\ifmmode\ddot{o}\else\"{o}\fi}hneysen}}]{Huy2007FMSCexp}%
  \BibitemOpen
  \bibfield  {author} {\bibinfo {author} {\bibfnamefont {N.~T.}\ \bibnamefont {Huy}}, \bibinfo {author} {\bibfnamefont {A.}~\bibnamefont {Gasparini}}, \bibinfo {author} {\bibfnamefont {D.~E.}\ \bibnamefont {de~Nijs}}, \bibinfo {author} {\bibfnamefont {Y.}~\bibnamefont {Huang}}, \bibinfo {author} {\bibfnamefont {J.~C.~P.}\ \bibnamefont {Klaasse}}, \bibinfo {author} {\bibfnamefont {T.}~\bibnamefont {Gortenmulder}}, \bibinfo {author} {\bibfnamefont {A.}~\bibnamefont {de~Visser}}, \bibinfo {author} {\bibfnamefont {A.}~\bibnamefont {Hamann}}, \bibinfo {author} {\bibfnamefont {T.}~\bibnamefont {G{\ifmmode\ddot{o}\else\"{o}\fi}rlach}},\ and\ \bibinfo {author} {\bibfnamefont {H.~v.}\ \bibnamefont {L{\ifmmode\ddot{o}\else\"{o}\fi}hneysen}},\ }\bibfield  {title} {\bibinfo {title} {{Superconductivity on the Border of Weak Itinerant Ferromagnetism in UCoGe}},\ }\href {https://doi.org/10.1103/PhysRevLett.99.067006} {\bibfield  {journal} {\bibinfo  {journal} {Phys. Rev. Lett.}\ }\textbf {\bibinfo {volume} {99}},\ \bibinfo
  {pages} {067006} (\bibinfo {year} {2007})}\BibitemShut {NoStop}%
\bibitem [{\citenamefont {Schrieffer}\ \emph {et~al.}(1989)\citenamefont {Schrieffer}, \citenamefont {Wen},\ and\ \citenamefont {Zhang}}]{Schrieffer1989AFMSC}%
  \BibitemOpen
  \bibfield  {author} {\bibinfo {author} {\bibfnamefont {J.~R.}\ \bibnamefont {Schrieffer}}, \bibinfo {author} {\bibfnamefont {X.~G.}\ \bibnamefont {Wen}},\ and\ \bibinfo {author} {\bibfnamefont {S.~C.}\ \bibnamefont {Zhang}},\ }\bibfield  {title} {\bibinfo {title} {{Dynamic spin fluctuations and the bag mechanism of high-${T}_{c}$ superconductivity}},\ }\href {https://doi.org/10.1103/PhysRevB.39.11663} {\bibfield  {journal} {\bibinfo  {journal} {Phys. Rev. B}\ }\textbf {\bibinfo {volume} {39}},\ \bibinfo {pages} {11663} (\bibinfo {year} {1989})}\BibitemShut {NoStop}%
\bibitem [{\citenamefont {Frenkel}\ and\ \citenamefont {Hanke}(1990)}]{Frenkel1990AFMSC}%
  \BibitemOpen
  \bibfield  {author} {\bibinfo {author} {\bibfnamefont {D.~M.}\ \bibnamefont {Frenkel}}\ and\ \bibinfo {author} {\bibfnamefont {W.}~\bibnamefont {Hanke}},\ }\bibfield  {title} {\bibinfo {title} {{Spirals and spin bags: A link between the weak- and the strong-coupling limits of the Hubbard model}},\ }\href {https://doi.org/10.1103/PhysRevB.42.6711} {\bibfield  {journal} {\bibinfo  {journal} {Phys. Rev. B}\ }\textbf {\bibinfo {volume} {42}},\ \bibinfo {pages} {6711} (\bibinfo {year} {1990})}\BibitemShut {NoStop}%
\bibitem [{\citenamefont {Capone}\ and\ \citenamefont {Kotliar}(2006)}]{Capone2006AFMSC}%
  \BibitemOpen
  \bibfield  {author} {\bibinfo {author} {\bibfnamefont {M.}~\bibnamefont {Capone}}\ and\ \bibinfo {author} {\bibfnamefont {G.}~\bibnamefont {Kotliar}},\ }\bibfield  {title} {\bibinfo {title} {{Competition between $d$-wave superconductivity and antiferromagnetism in the two-dimensional Hubbard model}},\ }\href {https://doi.org/10.1103/PhysRevB.74.054513} {\bibfield  {journal} {\bibinfo  {journal} {Phys. Rev. B}\ }\textbf {\bibinfo {volume} {74}},\ \bibinfo {pages} {054513} (\bibinfo {year} {2006})}\BibitemShut {NoStop}%
\bibitem [{\citenamefont {Ismer}\ \emph {et~al.}(2010)\citenamefont {Ismer}, \citenamefont {Eremin}, \citenamefont {Rossi}, \citenamefont {Morr},\ and\ \citenamefont {Blumberg}}]{Ismer2010AFMSC}%
  \BibitemOpen
  \bibfield  {author} {\bibinfo {author} {\bibfnamefont {J.-P.}\ \bibnamefont {Ismer}}, \bibinfo {author} {\bibfnamefont {I.}~\bibnamefont {Eremin}}, \bibinfo {author} {\bibfnamefont {E.}~\bibnamefont {Rossi}}, \bibinfo {author} {\bibfnamefont {D.~K.}\ \bibnamefont {Morr}},\ and\ \bibinfo {author} {\bibfnamefont {G.}~\bibnamefont {Blumberg}},\ }\bibfield  {title} {\bibinfo {title} {{Theory of Multiband Superconductivity in Spin-Density-Wave Metals}},\ }\href {https://doi.org/10.1103/PhysRevLett.105.037003} {\bibfield  {journal} {\bibinfo  {journal} {Phys. Rev. Lett.}\ }\textbf {\bibinfo {volume} {105}},\ \bibinfo {pages} {037003} (\bibinfo {year} {2010})}\BibitemShut {NoStop}%
\bibitem [{\citenamefont {R{\o}mer}\ \emph {et~al.}(2016)\citenamefont {R{\o}mer}, \citenamefont {Eremin}, \citenamefont {Hirschfeld},\ and\ \citenamefont {Andersen}}]{Romer2016AFMSC}%
  \BibitemOpen
  \bibfield  {author} {\bibinfo {author} {\bibfnamefont {A.~T.}\ \bibnamefont {R{\o}mer}}, \bibinfo {author} {\bibfnamefont {I.}~\bibnamefont {Eremin}}, \bibinfo {author} {\bibfnamefont {P.~J.}\ \bibnamefont {Hirschfeld}},\ and\ \bibinfo {author} {\bibfnamefont {B.~M.}\ \bibnamefont {Andersen}},\ }\bibfield  {title} {\bibinfo {title} {{Superconducting phase diagram of itinerant antiferromagnets}},\ }\href {https://doi.org/10.1103/PhysRevB.93.174519} {\bibfield  {journal} {\bibinfo  {journal} {Phys. Rev. B}\ }\textbf {\bibinfo {volume} {93}},\ \bibinfo {pages} {174519} (\bibinfo {year} {2016})}\BibitemShut {NoStop}%
\bibitem [{\citenamefont {Berger}(1996)}]{Berger1996Magnondrag}%
  \BibitemOpen
  \bibfield  {author} {\bibinfo {author} {\bibfnamefont {L.}~\bibnamefont {Berger}},\ }\bibfield  {title} {\bibinfo {title} {{Emission of spin waves by a magnetic multilayer traversed by a current}},\ }\href {https://doi.org/10.1103/PhysRevB.54.9353} {\bibfield  {journal} {\bibinfo  {journal} {Phys. Rev. B}\ }\textbf {\bibinfo {volume} {54}},\ \bibinfo {pages} {9353} (\bibinfo {year} {1996})}\BibitemShut {NoStop}%
\bibitem [{\citenamefont {Takahashi}\ \emph {et~al.}(2010)\citenamefont {Takahashi}, \citenamefont {Saitoh},\ and\ \citenamefont {Maekawa}}]{Takahashi2010MagnonCurrent}%
  \BibitemOpen
  \bibfield  {author} {\bibinfo {author} {\bibfnamefont {S.}~\bibnamefont {Takahashi}}, \bibinfo {author} {\bibfnamefont {E.}~\bibnamefont {Saitoh}},\ and\ \bibinfo {author} {\bibfnamefont {S.}~\bibnamefont {Maekawa}},\ }\bibfield  {title} {\bibinfo {title} {{Spin current through a normal-metal/insulating-ferromagnet junction}},\ }\href {https://doi.org/10.1088/1742-6596/200/6/062030} {\bibfield  {journal} {\bibinfo  {journal} {J. Phys. Conf. Ser.}\ }\textbf {\bibinfo {volume} {200}},\ \bibinfo {pages} {062030} (\bibinfo {year} {2010})}\BibitemShut {NoStop}%
\bibitem [{\citenamefont {Zhang}\ and\ \citenamefont {Zhang}(2012)}]{Zhang2012MagnonCurrent}%
  \BibitemOpen
  \bibfield  {author} {\bibinfo {author} {\bibfnamefont {S.~S.-L.}\ \bibnamefont {Zhang}}\ and\ \bibinfo {author} {\bibfnamefont {S.}~\bibnamefont {Zhang}},\ }\bibfield  {title} {\bibinfo {title} {{Spin convertance at magnetic interfaces}},\ }\href {https://doi.org/10.1103/PhysRevB.86.214424} {\bibfield  {journal} {\bibinfo  {journal} {Phys. Rev. B}\ }\textbf {\bibinfo {volume} {86}},\ \bibinfo {pages} {214424} (\bibinfo {year} {2012})}\BibitemShut {NoStop}%
\bibitem [{\citenamefont {M{\ae}land}\ \emph {et~al.}(2021)\citenamefont {M{\ae}land}, \citenamefont {R{\o}st}, \citenamefont {Wells},\ and\ \citenamefont {Sudb{\o}}}]{Maeland2021Sep}%
  \BibitemOpen
  \bibfield  {author} {\bibinfo {author} {\bibfnamefont {K.}~\bibnamefont {M{\ae}land}}, \bibinfo {author} {\bibfnamefont {H.~I.}\ \bibnamefont {R{\o}st}}, \bibinfo {author} {\bibfnamefont {J.~W.}\ \bibnamefont {Wells}},\ and\ \bibinfo {author} {\bibfnamefont {A.}~\bibnamefont {Sudb{\o}}},\ }\bibfield  {title} {\bibinfo {title} {{Electron-magnon coupling and quasiparticle lifetimes on the surface of a topological insulator}},\ }\href {https://doi.org/10.1103/PhysRevB.104.125125} {\bibfield  {journal} {\bibinfo  {journal} {Phys. Rev. B}\ }\textbf {\bibinfo {volume} {104}},\ \bibinfo {pages} {125125} (\bibinfo {year} {2021})}\BibitemShut {NoStop}%
\bibitem [{\citenamefont {Nakosai}\ \emph {et~al.}(2013)\citenamefont {Nakosai}, \citenamefont {Tanaka},\ and\ \citenamefont {Nagaosa}}]{SkTopoSCNagaosa}%
  \BibitemOpen
  \bibfield  {author} {\bibinfo {author} {\bibfnamefont {S.}~\bibnamefont {Nakosai}}, \bibinfo {author} {\bibfnamefont {Y.}~\bibnamefont {Tanaka}},\ and\ \bibinfo {author} {\bibfnamefont {N.}~\bibnamefont {Nagaosa}},\ }\bibfield  {title} {\bibinfo {title} {Two-dimensional $p$-wave superconducting states with magnetic moments on a conventional $s$-wave superconductor},\ }\href {https://doi.org/10.1103/PhysRevB.88.180503} {\bibfield  {journal} {\bibinfo  {journal} {Phys. Rev. B}\ }\textbf {\bibinfo {volume} {88}},\ \bibinfo {pages} {180503(R)} (\bibinfo {year} {2013})}\BibitemShut {NoStop}%
\bibitem [{\citenamefont {Petrovi\ifmmode~\acute{c}\else \'{c}\fi{}}\ \emph {et~al.}(2021)\citenamefont {Petrovi\ifmmode~\acute{c}\else \'{c}\fi{}}, \citenamefont {Raju}, \citenamefont {Tee}, \citenamefont {Louat}, \citenamefont {Maggio-Aprile}, \citenamefont {Menezes}, \citenamefont {Wyszy\ifmmode~\acute{n}\else \'{n}\fi{}ski}, \citenamefont {Duong}, \citenamefont {Reznikov}, \citenamefont {Renner}, \citenamefont {Milo\ifmmode \check{s}\else \v{s}\fi{}evi\ifmmode~\acute{c}\else \'{c}\fi{}},\ and\ \citenamefont {Panagopoulos}}]{ExpSkHeterostructure}%
  \BibitemOpen
  \bibfield  {author} {\bibinfo {author} {\bibfnamefont {A.~P.}\ \bibnamefont {Petrovi\ifmmode~\acute{c}\else \'{c}\fi{}}}, \bibinfo {author} {\bibfnamefont {M.}~\bibnamefont {Raju}}, \bibinfo {author} {\bibfnamefont {X.~Y.}\ \bibnamefont {Tee}}, \bibinfo {author} {\bibfnamefont {A.}~\bibnamefont {Louat}}, \bibinfo {author} {\bibfnamefont {I.}~\bibnamefont {Maggio-Aprile}}, \bibinfo {author} {\bibfnamefont {R.~M.}\ \bibnamefont {Menezes}}, \bibinfo {author} {\bibfnamefont {M.~J.}\ \bibnamefont {Wyszy\ifmmode~\acute{n}\else \'{n}\fi{}ski}}, \bibinfo {author} {\bibfnamefont {N.~K.}\ \bibnamefont {Duong}}, \bibinfo {author} {\bibfnamefont {M.}~\bibnamefont {Reznikov}}, \bibinfo {author} {\bibfnamefont {C.}~\bibnamefont {Renner}}, \bibinfo {author} {\bibfnamefont {M.~V.}\ \bibnamefont {Milo\ifmmode \check{s}\else \v{s}\fi{}evi\ifmmode~\acute{c}\else \'{c}\fi{}}},\ and\ \bibinfo {author} {\bibfnamefont {C.}~\bibnamefont {Panagopoulos}},\ }\bibfield  {title} {\bibinfo {title} {{Skyrmion-(Anti)Vortex Coupling in a
  Chiral Magnet-Superconductor Heterostructure}},\ }\href {https://doi.org/10.1103/PhysRevLett.126.117205} {\bibfield  {journal} {\bibinfo  {journal} {Phys. Rev. Lett.}\ }\textbf {\bibinfo {volume} {126}},\ \bibinfo {pages} {117205} (\bibinfo {year} {2021})}\BibitemShut {NoStop}%
\bibitem [{\citenamefont {Kajiwara}\ \emph {et~al.}(2010)\citenamefont {Kajiwara}, \citenamefont {Harii}, \citenamefont {Takahashi}, \citenamefont {Ohe}, \citenamefont {Uchida}, \citenamefont {Mizuguchi}, \citenamefont {Umezawa}, \citenamefont {Kawai}, \citenamefont {Ando}, \citenamefont {Takanashi}, \citenamefont {Maekawa},\ and\ \citenamefont {Saitoh}}]{ExpInterfaceExchange}%
  \BibitemOpen
  \bibfield  {author} {\bibinfo {author} {\bibfnamefont {Y.}~\bibnamefont {Kajiwara}}, \bibinfo {author} {\bibfnamefont {K.}~\bibnamefont {Harii}}, \bibinfo {author} {\bibfnamefont {S.}~\bibnamefont {Takahashi}}, \bibinfo {author} {\bibfnamefont {J.}~\bibnamefont {Ohe}}, \bibinfo {author} {\bibfnamefont {K.}~\bibnamefont {Uchida}}, \bibinfo {author} {\bibfnamefont {M.}~\bibnamefont {Mizuguchi}}, \bibinfo {author} {\bibfnamefont {H.}~\bibnamefont {Umezawa}}, \bibinfo {author} {\bibfnamefont {H.}~\bibnamefont {Kawai}}, \bibinfo {author} {\bibfnamefont {K.}~\bibnamefont {Ando}}, \bibinfo {author} {\bibfnamefont {K.}~\bibnamefont {Takanashi}}, \bibinfo {author} {\bibfnamefont {S.}~\bibnamefont {Maekawa}},\ and\ \bibinfo {author} {\bibfnamefont {E.}~\bibnamefont {Saitoh}},\ }\bibfield  {title} {\bibinfo {title} {Transmission of electrical signals by spin-wave interconversion in a magnetic insulator},\ }\href {https://doi.org/10.1038/nature08876} {\bibfield  {journal} {\bibinfo  {journal} {Nature}\ }\textbf
  {\bibinfo {volume} {464}},\ \bibinfo {pages} {262} (\bibinfo {year} {2010})}\BibitemShut {NoStop}%
\bibitem [{\citenamefont {Li}\ \emph {et~al.}(2016)\citenamefont {Li}, \citenamefont {Xu}, \citenamefont {Aldosary}, \citenamefont {Tang}, \citenamefont {Lin}, \citenamefont {Zhang}, \citenamefont {Lake},\ and\ \citenamefont {Shi}}]{Li2016MagnonDragExp}%
  \BibitemOpen
  \bibfield  {author} {\bibinfo {author} {\bibfnamefont {J.}~\bibnamefont {Li}}, \bibinfo {author} {\bibfnamefont {Y.}~\bibnamefont {Xu}}, \bibinfo {author} {\bibfnamefont {M.}~\bibnamefont {Aldosary}}, \bibinfo {author} {\bibfnamefont {C.}~\bibnamefont {Tang}}, \bibinfo {author} {\bibfnamefont {Z.}~\bibnamefont {Lin}}, \bibinfo {author} {\bibfnamefont {S.}~\bibnamefont {Zhang}}, \bibinfo {author} {\bibfnamefont {R.}~\bibnamefont {Lake}},\ and\ \bibinfo {author} {\bibfnamefont {J.}~\bibnamefont {Shi}},\ }\bibfield  {title} {\bibinfo {title} {{Observation of magnon-mediated current drag in Pt/yttrium iron garnet/Pt(Ta) trilayers}},\ }\href {https://doi.org/10.1038/ncomms10858} {\bibfield  {journal} {\bibinfo  {journal} {Nat. Commun.}\ }\textbf {\bibinfo {volume} {7}},\ \bibinfo {pages} {10858} (\bibinfo {year} {2016})}\BibitemShut {NoStop}%
\bibitem [{\citenamefont {Wu}\ \emph {et~al.}(2016)\citenamefont {Wu}, \citenamefont {Wan}, \citenamefont {Zhang}, \citenamefont {Yuan}, \citenamefont {Zhang}, \citenamefont {Qin}, \citenamefont {Wei}, \citenamefont {Han},\ and\ \citenamefont {Zhang}}]{Wu2016MagnonDragExp}%
  \BibitemOpen
  \bibfield  {author} {\bibinfo {author} {\bibfnamefont {H.}~\bibnamefont {Wu}}, \bibinfo {author} {\bibfnamefont {C.~H.}\ \bibnamefont {Wan}}, \bibinfo {author} {\bibfnamefont {X.}~\bibnamefont {Zhang}}, \bibinfo {author} {\bibfnamefont {Z.~H.}\ \bibnamefont {Yuan}}, \bibinfo {author} {\bibfnamefont {Q.~T.}\ \bibnamefont {Zhang}}, \bibinfo {author} {\bibfnamefont {J.~Y.}\ \bibnamefont {Qin}}, \bibinfo {author} {\bibfnamefont {H.~X.}\ \bibnamefont {Wei}}, \bibinfo {author} {\bibfnamefont {X.~F.}\ \bibnamefont {Han}},\ and\ \bibinfo {author} {\bibfnamefont {S.}~\bibnamefont {Zhang}},\ }\bibfield  {title} {\bibinfo {title} {{Observation of magnon-mediated electric current drag at room temperature}},\ }\href {https://doi.org/10.1103/PhysRevB.93.060403} {\bibfield  {journal} {\bibinfo  {journal} {Phys. Rev. B}\ }\textbf {\bibinfo {volume} {93}},\ \bibinfo {pages} {060403(R)} (\bibinfo {year} {2016})}\BibitemShut {NoStop}%
\bibitem [{\citenamefont {Garate}\ and\ \citenamefont {Franz}(2010)}]{Garate2010Jsdbulk}%
  \BibitemOpen
  \bibfield  {author} {\bibinfo {author} {\bibfnamefont {I.}~\bibnamefont {Garate}}\ and\ \bibinfo {author} {\bibfnamefont {M.}~\bibnamefont {Franz}},\ }\bibfield  {title} {\bibinfo {title} {Inverse spin-galvanic effect in the interface between a topological insulator and a ferromagnet},\ }\href {https://doi.org/10.1103/PhysRevLett.104.146802} {\bibfield  {journal} {\bibinfo  {journal} {Phys. Rev. Lett.}\ }\textbf {\bibinfo {volume} {104}},\ \bibinfo {pages} {146802} (\bibinfo {year} {2010})}\BibitemShut {NoStop}%
\bibitem [{\citenamefont {Dyck}\ \emph {et~al.}(2002)\citenamefont {Dyck}, \citenamefont {H{\ifmmode\acute{a}\else\'{a}\fi}jek}, \citenamefont {Lo{\ifmmode\check{s}\else\v{s}\fi}t{'}{\ifmmode\acute{a}\else\'{a}\fi}k},\ and\ \citenamefont {Uher}}]{ExpStrongJbar}%
  \BibitemOpen
  \bibfield  {author} {\bibinfo {author} {\bibfnamefont {J.~S.}\ \bibnamefont {Dyck}}, \bibinfo {author} {\bibfnamefont {P.}~\bibnamefont {H{\ifmmode\acute{a}\else\'{a}\fi}jek}}, \bibinfo {author} {\bibfnamefont {P.}~\bibnamefont {Lo{\ifmmode\check{s}\else\v{s}\fi}t{'}{\ifmmode\acute{a}\else\'{a}\fi}k}},\ and\ \bibinfo {author} {\bibfnamefont {C.}~\bibnamefont {Uher}},\ }\bibfield  {title} {\bibinfo {title} {{Diluted magnetic semiconductors based on ${\mathrm{Sb}}_{2\ensuremath{-}x}{\mathrm{V}}_{x}{\mathrm{Te}}_{3}$ $(0.01<~x<~0.03)$}},\ }\href {https://doi.org/10.1103/PhysRevB.65.115212} {\bibfield  {journal} {\bibinfo  {journal} {Phys. Rev. B}\ }\textbf {\bibinfo {volume} {65}},\ \bibinfo {pages} {115212} (\bibinfo {year} {2002})}\BibitemShut {NoStop}%
\bibitem [{\citenamefont {Liu}\ \emph {et~al.}(2009)\citenamefont {Liu}, \citenamefont {Liu}, \citenamefont {Xu}, \citenamefont {Qi},\ and\ \citenamefont {Zhang}}]{Liu2009strongJbarPRL}%
  \BibitemOpen
  \bibfield  {author} {\bibinfo {author} {\bibfnamefont {Q.}~\bibnamefont {Liu}}, \bibinfo {author} {\bibfnamefont {C.-X.}\ \bibnamefont {Liu}}, \bibinfo {author} {\bibfnamefont {C.}~\bibnamefont {Xu}}, \bibinfo {author} {\bibfnamefont {X.-L.}\ \bibnamefont {Qi}},\ and\ \bibinfo {author} {\bibfnamefont {S.-C.}\ \bibnamefont {Zhang}},\ }\bibfield  {title} {\bibinfo {title} {Magnetic impurities on the surface of a topological insulator},\ }\href {https://doi.org/10.1103/PhysRevLett.102.156603} {\bibfield  {journal} {\bibinfo  {journal} {Phys. Rev. Lett.}\ }\textbf {\bibinfo {volume} {102}},\ \bibinfo {pages} {156603} (\bibinfo {year} {2009})}\BibitemShut {NoStop}%
\bibitem [{\citenamefont {Holstein}\ and\ \citenamefont {Primakoff}(1940)}]{Holstein1940Dec}%
  \BibitemOpen
  \bibfield  {author} {\bibinfo {author} {\bibfnamefont {T.}~\bibnamefont {Holstein}}\ and\ \bibinfo {author} {\bibfnamefont {H.}~\bibnamefont {Primakoff}},\ }\bibfield  {title} {\bibinfo {title} {{Field Dependence of the Intrinsic Domain Magnetization of a Ferromagnet}},\ }\href {https://doi.org/10.1103/PhysRev.58.1098} {\bibfield  {journal} {\bibinfo  {journal} {Phys. Rev.}\ }\textbf {\bibinfo {volume} {58}},\ \bibinfo {pages} {1098} (\bibinfo {year} {1940})}\BibitemShut {NoStop}%
\bibitem [{\citenamefont {M{\ae}land}\ and\ \citenamefont {Sudb{\o}}(2022)}]{Maeland2022QSk}%
  \BibitemOpen
  \bibfield  {author} {\bibinfo {author} {\bibfnamefont {K.}~\bibnamefont {M{\ae}land}}\ and\ \bibinfo {author} {\bibfnamefont {A.}~\bibnamefont {Sudb{\o}}},\ }\bibfield  {title} {\bibinfo {title} {{Quantum fluctuations in the order parameter of quantum skyrmion crystals}},\ }\href {https://doi.org/10.1103/PhysRevB.105.224416} {\bibfield  {journal} {\bibinfo  {journal} {Phys. Rev. B}\ }\textbf {\bibinfo {volume} {105}},\ \bibinfo {pages} {224416} (\bibinfo {year} {2022})}\BibitemShut {NoStop}%
\bibitem [{\citenamefont {Hashimoto}(2000)}]{Hashimoto2000DoubleMagnon}%
  \BibitemOpen
  \bibfield  {author} {\bibinfo {author} {\bibfnamefont {K.}~\bibnamefont {Hashimoto}},\ }\bibfield  {title} {\bibinfo {title} {{Effects of Double Magnon Scattering on the Particle-Particle Interaction in the Two-Dimensional Hubbard Model with Next-Nearest-Neighbor Hopping}},\ }\href {https://doi.org/10.1143/JPSJ.69.2229} {\bibfield  {journal} {\bibinfo  {journal} {J. Phys. Soc. Jpn.}\ }\textbf {\bibinfo {volume} {69}},\ \bibinfo {pages} {2229} (\bibinfo {year} {2000})}\BibitemShut {NoStop}%
\bibitem [{\citenamefont {Erlandsen}\ and\ \citenamefont {Sudb{\o}}(2022)}]{Erlandsen2022MagnonDrag}%
  \BibitemOpen
  \bibfield  {author} {\bibinfo {author} {\bibfnamefont {E.}~\bibnamefont {Erlandsen}}\ and\ \bibinfo {author} {\bibfnamefont {A.}~\bibnamefont {Sudb{\o}}},\ }\bibfield  {title} {\bibinfo {title} {{Magnon drag in a metal--insulating antiferromagnet bilayer}},\ }\href {https://doi.org/10.1103/PhysRevB.105.184434} {\bibfield  {journal} {\bibinfo  {journal} {Phys. Rev. B}\ }\textbf {\bibinfo {volume} {105}},\ \bibinfo {pages} {184434} (\bibinfo {year} {2022})}\BibitemShut {NoStop}%
\bibitem [{\citenamefont {Lucassen}\ \emph {et~al.}(2011)\citenamefont {Lucassen}, \citenamefont {Wong}, \citenamefont {Duine},\ and\ \citenamefont {Tserkovnyak}}]{Lucassen2011MagnonDrag}%
  \BibitemOpen
  \bibfield  {author} {\bibinfo {author} {\bibfnamefont {M.~E.}\ \bibnamefont {Lucassen}}, \bibinfo {author} {\bibfnamefont {C.~H.}\ \bibnamefont {Wong}}, \bibinfo {author} {\bibfnamefont {R.~A.}\ \bibnamefont {Duine}},\ and\ \bibinfo {author} {\bibfnamefont {Y.}~\bibnamefont {Tserkovnyak}},\ }\bibfield  {title} {\bibinfo {title} {{Spin-transfer mechanism for magnon-drag thermopower}},\ }\href {https://doi.org/10.1063/1.3672207} {\bibfield  {journal} {\bibinfo  {journal} {Appl. Phys. Lett.}\ }\textbf {\bibinfo {volume} {99}},\ \bibinfo {pages} {262506} (\bibinfo {year} {2011})}\BibitemShut {NoStop}%
\bibitem [{\citenamefont {Yamaguchi}\ \emph {et~al.}(2019)\citenamefont {Yamaguchi}, \citenamefont {Kohno},\ and\ \citenamefont {Duine}}]{Yamaguchi2019MagnonDrag}%
  \BibitemOpen
  \bibfield  {author} {\bibinfo {author} {\bibfnamefont {T.}~\bibnamefont {Yamaguchi}}, \bibinfo {author} {\bibfnamefont {H.}~\bibnamefont {Kohno}},\ and\ \bibinfo {author} {\bibfnamefont {R.~A.}\ \bibnamefont {Duine}},\ }\bibfield  {title} {\bibinfo {title} {{Microscopic theory of magnon-drag electron flow in ferromagnetic metals}},\ }\href {https://doi.org/10.1103/PhysRevB.99.094425} {\bibfield  {journal} {\bibinfo  {journal} {Phys. Rev. B}\ }\textbf {\bibinfo {volume} {99}},\ \bibinfo {pages} {094425} (\bibinfo {year} {2019})}\BibitemShut {NoStop}%
\bibitem [{\citenamefont {R{\o}st}\ \emph {et~al.}(2024)\citenamefont {R{\o}st}, \citenamefont {Mazzola}, \citenamefont {Bakkelund}, \citenamefont {{\AA}sland}, \citenamefont {Hu}, \citenamefont {Cooil}, \citenamefont {Polley},\ and\ \citenamefont {Wells}}]{Rost2024Jan}%
  \BibitemOpen
  \bibfield  {author} {\bibinfo {author} {\bibfnamefont {H.~I.}\ \bibnamefont {R{\o}st}}, \bibinfo {author} {\bibfnamefont {F.}~\bibnamefont {Mazzola}}, \bibinfo {author} {\bibfnamefont {J.}~\bibnamefont {Bakkelund}}, \bibinfo {author} {\bibfnamefont {A.~C.}\ \bibnamefont {{\AA}sland}}, \bibinfo {author} {\bibfnamefont {J.}~\bibnamefont {Hu}}, \bibinfo {author} {\bibfnamefont {S.~P.}\ \bibnamefont {Cooil}}, \bibinfo {author} {\bibfnamefont {C.~M.}\ \bibnamefont {Polley}},\ and\ \bibinfo {author} {\bibfnamefont {J.~W.}\ \bibnamefont {Wells}},\ }\bibfield  {title} {\bibinfo {title} {{Disentangling electron-boson interactions on the surface of a familiar ferromagnet}},\ }\href {https://doi.org/10.1103/PhysRevB.109.035137} {\bibfield  {journal} {\bibinfo  {journal} {Phys. Rev. B}\ }\textbf {\bibinfo {volume} {109}},\ \bibinfo {pages} {035137} (\bibinfo {year} {2024})}\BibitemShut {NoStop}%
\bibitem [{\citenamefont {Aldaihan}\ \emph {et~al.}(2017)\citenamefont {Aldaihan}, \citenamefont {Krause}, \citenamefont {Long},\ and\ \citenamefont {Snow}}]{Aldaihan2017DoubleBoson}%
  \BibitemOpen
  \bibfield  {author} {\bibinfo {author} {\bibfnamefont {S.}~\bibnamefont {Aldaihan}}, \bibinfo {author} {\bibfnamefont {D.~E.}\ \bibnamefont {Krause}}, \bibinfo {author} {\bibfnamefont {J.~C.}\ \bibnamefont {Long}},\ and\ \bibinfo {author} {\bibfnamefont {W.~M.}\ \bibnamefont {Snow}},\ }\bibfield  {title} {\bibinfo {title} {{Calculations of the dominant long-range, spin-independent contributions to the interaction energy between two nonrelativistic Dirac fermions from double-boson exchange of spin-0 and spin-1 bosons with spin-dependent couplings}},\ }\href {https://doi.org/10.1103/PhysRevD.95.096005} {\bibfield  {journal} {\bibinfo  {journal} {Phys. Rev. D}\ }\textbf {\bibinfo {volume} {95}},\ \bibinfo {pages} {096005} (\bibinfo {year} {2017})}\BibitemShut {NoStop}%
\bibitem [{\citenamefont {Drell}\ and\ \citenamefont {Huang}(1953)}]{Drell1953DoubleBoson}%
  \BibitemOpen
  \bibfield  {author} {\bibinfo {author} {\bibfnamefont {S.~D.}\ \bibnamefont {Drell}}\ and\ \bibinfo {author} {\bibfnamefont {K.}~\bibnamefont {Huang}},\ }\bibfield  {title} {\bibinfo {title} {{Many-Body Forces and Nuclear Saturation}},\ }\href {https://doi.org/10.1103/PhysRev.91.1527} {\bibfield  {journal} {\bibinfo  {journal} {Phys. Rev.}\ }\textbf {\bibinfo {volume} {91}},\ \bibinfo {pages} {1527} (\bibinfo {year} {1953})}\BibitemShut {NoStop}%
\bibitem [{\citenamefont {Kummer}\ and\ \citenamefont {Lane}(1973)}]{Kummer1973DoubleBoson}%
  \BibitemOpen
  \bibfield  {author} {\bibinfo {author} {\bibfnamefont {W.}~\bibnamefont {Kummer}}\ and\ \bibinfo {author} {\bibfnamefont {K.}~\bibnamefont {Lane}},\ }\bibfield  {title} {\bibinfo {title} {{Divergence Cancellations in Spontaneously Broken Gauge Theories}},\ }\href {https://doi.org/10.1103/PhysRevD.7.1910} {\bibfield  {journal} {\bibinfo  {journal} {Phys. Rev. D}\ }\textbf {\bibinfo {volume} {7}},\ \bibinfo {pages} {1910} (\bibinfo {year} {1973})}\BibitemShut {NoStop}%
\bibitem [{\citenamefont {Ferrer}\ and\ \citenamefont {Nowakowski}(1999)}]{Ferrer1999DoubleBoson}%
  \BibitemOpen
  \bibfield  {author} {\bibinfo {author} {\bibfnamefont {F.}~\bibnamefont {Ferrer}}\ and\ \bibinfo {author} {\bibfnamefont {M.}~\bibnamefont {Nowakowski}},\ }\bibfield  {title} {\bibinfo {title} {{Higgs- and Goldstone-boson-mediated long range forces}},\ }\href {https://doi.org/10.1103/PhysRevD.59.075009} {\bibfield  {journal} {\bibinfo  {journal} {Phys. Rev. D}\ }\textbf {\bibinfo {volume} {59}},\ \bibinfo {pages} {075009} (\bibinfo {year} {1999})}\BibitemShut {NoStop}%
\bibitem [{\citenamefont {Villegas}\ \emph {et~al.}(2019)\citenamefont {Villegas}, \citenamefont {Sun}, \citenamefont {Kovalev},\ and\ \citenamefont {Savenko}}]{Villegas2019DoubleBogolon}%
  \BibitemOpen
  \bibfield  {author} {\bibinfo {author} {\bibfnamefont {K.~H.~A.}\ \bibnamefont {Villegas}}, \bibinfo {author} {\bibfnamefont {M.}~\bibnamefont {Sun}}, \bibinfo {author} {\bibfnamefont {V.~M.}\ \bibnamefont {Kovalev}},\ and\ \bibinfo {author} {\bibfnamefont {I.~G.}\ \bibnamefont {Savenko}},\ }\bibfield  {title} {\bibinfo {title} {{Unconventional Bloch-Gr{\"u}neisen Scattering in Hybrid Bose-Fermi Systems}},\ }\href {https://doi.org/10.1103/PhysRevLett.123.095301} {\bibfield  {journal} {\bibinfo  {journal} {Phys. Rev. Lett.}\ }\textbf {\bibinfo {volume} {123}},\ \bibinfo {pages} {095301} (\bibinfo {year} {2019})}\BibitemShut {NoStop}%
\bibitem [{\citenamefont {Sun}\ \emph {et~al.}(2021{\natexlab{a}})\citenamefont {Sun}, \citenamefont {Parafilo}, \citenamefont {Villegas}, \citenamefont {Kovalev},\ and\ \citenamefont {Savenko}}]{Sun2021DoubleBogolon}%
  \BibitemOpen
  \bibfield  {author} {\bibinfo {author} {\bibfnamefont {M.}~\bibnamefont {Sun}}, \bibinfo {author} {\bibfnamefont {A.~V.}\ \bibnamefont {Parafilo}}, \bibinfo {author} {\bibfnamefont {K.~H.~A.}\ \bibnamefont {Villegas}}, \bibinfo {author} {\bibfnamefont {V.~M.}\ \bibnamefont {Kovalev}},\ and\ \bibinfo {author} {\bibfnamefont {I.~G.}\ \bibnamefont {Savenko}},\ }\bibfield  {title} {\bibinfo {title} {{Bose{\textendash}Einstein condensate-mediated superconductivity in graphene}},\ }\href {https://doi.org/10.1088/2053-1583/ac0b49} {\bibfield  {journal} {\bibinfo  {journal} {2D Mater.}\ }\textbf {\bibinfo {volume} {8}},\ \bibinfo {pages} {031004} (\bibinfo {year} {2021}{\natexlab{a}})}\BibitemShut {NoStop}%
\bibitem [{\citenamefont {Sun}\ \emph {et~al.}(2021{\natexlab{b}})\citenamefont {Sun}, \citenamefont {Parafilo}, \citenamefont {Villegas}, \citenamefont {Kovalev},\ and\ \citenamefont {Savenko}}]{Sun2021DoubleBogolonBCS}%
  \BibitemOpen
  \bibfield  {author} {\bibinfo {author} {\bibfnamefont {M.}~\bibnamefont {Sun}}, \bibinfo {author} {\bibfnamefont {A.~V.}\ \bibnamefont {Parafilo}}, \bibinfo {author} {\bibfnamefont {K.~H.~A.}\ \bibnamefont {Villegas}}, \bibinfo {author} {\bibfnamefont {V.~M.}\ \bibnamefont {Kovalev}},\ and\ \bibinfo {author} {\bibfnamefont {I.~G.}\ \bibnamefont {Savenko}},\ }\bibfield  {title} {\bibinfo {title} {{Theory of BCS-like bogolon-mediated superconductivity in transition metal dichalcogenides}},\ }\href {https://doi.org/10.1088/1367-2630/abe285} {\bibfield  {journal} {\bibinfo  {journal} {New J. Phys.}\ }\textbf {\bibinfo {volume} {23}},\ \bibinfo {pages} {023023} (\bibinfo {year} {2021}{\natexlab{b}})}\BibitemShut {NoStop}%
\bibitem [{\citenamefont {Sun}\ \emph {et~al.}(2021{\natexlab{c}})\citenamefont {Sun}, \citenamefont {Parafilo}, \citenamefont {Kovalev},\ and\ \citenamefont {Savenko}}]{Sun2021DoubleBogolonEliashberg}%
  \BibitemOpen
  \bibfield  {author} {\bibinfo {author} {\bibfnamefont {M.}~\bibnamefont {Sun}}, \bibinfo {author} {\bibfnamefont {A.~V.}\ \bibnamefont {Parafilo}}, \bibinfo {author} {\bibfnamefont {V.~M.}\ \bibnamefont {Kovalev}},\ and\ \bibinfo {author} {\bibfnamefont {I.~G.}\ \bibnamefont {Savenko}},\ }\bibfield  {title} {\bibinfo {title} {{Strong-coupling theory of condensate-mediated superconductivity in two-dimensional materials}},\ }\href {https://doi.org/10.1103/PhysRevResearch.3.033166} {\bibfield  {journal} {\bibinfo  {journal} {Phys. Rev. Res.}\ }\textbf {\bibinfo {volume} {3}},\ \bibinfo {pages} {033166} (\bibinfo {year} {2021}{\natexlab{c}})}\BibitemShut {NoStop}%
\bibitem [{\citenamefont {Plyashechnik}\ \emph {et~al.}(2023)\citenamefont {Plyashechnik}, \citenamefont {Sokolik}, \citenamefont {Voronova},\ and\ \citenamefont {Lozovik}}]{Plyashechnik2023DoubleBogolon}%
  \BibitemOpen
  \bibfield  {author} {\bibinfo {author} {\bibfnamefont {A.~S.}\ \bibnamefont {Plyashechnik}}, \bibinfo {author} {\bibfnamefont {A.~A.}\ \bibnamefont {Sokolik}}, \bibinfo {author} {\bibfnamefont {N.~S.}\ \bibnamefont {Voronova}},\ and\ \bibinfo {author} {\bibfnamefont {Y.~E.}\ \bibnamefont {Lozovik}},\ }\bibfield  {title} {\bibinfo {title} {{Coupled system of electrons and exciton-polaritons: Screening, dynamical effects, and superconductivity}},\ }\href {https://doi.org/10.1103/PhysRevB.108.024513} {\bibfield  {journal} {\bibinfo  {journal} {Phys. Rev. B}\ }\textbf {\bibinfo {volume} {108}},\ \bibinfo {pages} {024513} (\bibinfo {year} {2023})}\BibitemShut {NoStop}%
\bibitem [{\citenamefont {Marsiglio}\ \emph {et~al.}(1988)\citenamefont {Marsiglio}, \citenamefont {Schossmann},\ and\ \citenamefont {Carbotte}}]{Marsiglio1988AnalyticCont}%
  \BibitemOpen
  \bibfield  {author} {\bibinfo {author} {\bibfnamefont {F.}~\bibnamefont {Marsiglio}}, \bibinfo {author} {\bibfnamefont {M.}~\bibnamefont {Schossmann}},\ and\ \bibinfo {author} {\bibfnamefont {J.~P.}\ \bibnamefont {Carbotte}},\ }\bibfield  {title} {\bibinfo {title} {{Iterative analytic continuation of the electron self-energy to the real axis}},\ }\href {https://doi.org/10.1103/PhysRevB.37.4965} {\bibfield  {journal} {\bibinfo  {journal} {Phys. Rev. B}\ }\textbf {\bibinfo {volume} {37}},\ \bibinfo {pages} {4965} (\bibinfo {year} {1988})}\BibitemShut {NoStop}%
\bibitem [{\citenamefont {M{\ae}land}\ and\ \citenamefont {Sudb{\o}}(2023{\natexlab{b}})}]{Maeland2023CC}%
  \BibitemOpen
  \bibfield  {author} {\bibinfo {author} {\bibfnamefont {K.}~\bibnamefont {M{\ae}land}}\ and\ \bibinfo {author} {\bibfnamefont {A.}~\bibnamefont {Sudb{\o}}},\ }\bibfield  {title} {\bibinfo {title} {{Exceeding the Chandrasekhar-Clogston limit in flat-band superconductors: A multiband strong-coupling approach}},\ }\href {https://doi.org/10.1103/PhysRevB.108.214511} {\bibfield  {journal} {\bibinfo  {journal} {Phys. Rev. B}\ }\textbf {\bibinfo {volume} {108}},\ \bibinfo {pages} {214511} (\bibinfo {year} {2023}{\natexlab{b}})}\BibitemShut {NoStop}%
\bibitem [{\citenamefont {Bruus}\ and\ \citenamefont {Flensberg}(2004)}]{BruusFlensberg}%
  \BibitemOpen
  \bibfield  {author} {\bibinfo {author} {\bibfnamefont {H.}~\bibnamefont {Bruus}}\ and\ \bibinfo {author} {\bibfnamefont {K.}~\bibnamefont {Flensberg}},\ }\href@noop {} {\emph {\bibinfo {title} {{Many-Body Quantum Theory in Condensed Matter Physics: An Introduction}}}}\ (\bibinfo  {publisher} {Oxford University Press},\ \bibinfo {address} {Oxford},\ \bibinfo {year} {2004})\BibitemShut {NoStop}%
\bibitem [{\citenamefont {Abrikosov}\ \emph {et~al.}(1963)\citenamefont {Abrikosov}, \citenamefont {Gorkov},\ and\ \citenamefont {Dzyaloshinski}}]{abrikosov}%
  \BibitemOpen
  \bibfield  {author} {\bibinfo {author} {\bibfnamefont {A.~A.}\ \bibnamefont {Abrikosov}}, \bibinfo {author} {\bibfnamefont {L.~P.}\ \bibnamefont {Gorkov}},\ and\ \bibinfo {author} {\bibfnamefont {I.~E.}\ \bibnamefont {Dzyaloshinski}},\ }\href@noop {} {\emph {\bibinfo {title} {Methods of Quantum Field Theory in Statistical Physics}}}\ (\bibinfo  {publisher} {Dover, New York},\ \bibinfo {year} {1963})\BibitemShut {NoStop}%
\bibitem [{\citenamefont {Migdal}(1958)}]{migdal1958interaction}%
  \BibitemOpen
  \bibfield  {author} {\bibinfo {author} {\bibfnamefont {A.~B.}\ \bibnamefont {Migdal}},\ }\bibfield  {title} {\bibinfo {title} {Interaction between electrons and lattice vibrations in a normal metal},\ }\href@noop {} {\bibfield  {journal} {\bibinfo  {journal} {Zh. Eksp. Teor. Fiz.}\ }\textbf {\bibinfo {volume} {34}},\ \bibinfo {pages} {1438} (\bibinfo {year} {1958})},\ \bibinfo {note} {[Sov. Phys. JETP \textbf{34}, 996 (1958)]}\BibitemShut {NoStop}%
\bibitem [{\citenamefont {Roy}\ \emph {et~al.}(2014)\citenamefont {Roy}, \citenamefont {Sau},\ and\ \citenamefont {Das~Sarma}}]{Migdal2D}%
  \BibitemOpen
  \bibfield  {author} {\bibinfo {author} {\bibfnamefont {B.}~\bibnamefont {Roy}}, \bibinfo {author} {\bibfnamefont {J.~D.}\ \bibnamefont {Sau}},\ and\ \bibinfo {author} {\bibfnamefont {S.}~\bibnamefont {Das~Sarma}},\ }\bibfield  {title} {\bibinfo {title} {{Migdal's theorem and electron-phonon vertex corrections in Dirac materials}},\ }\href {https://doi.org/10.1103/PhysRevB.89.165119} {\bibfield  {journal} {\bibinfo  {journal} {Phys. Rev. B}\ }\textbf {\bibinfo {volume} {89}},\ \bibinfo {pages} {165119} (\bibinfo {year} {2014})}\BibitemShut {NoStop}%
\bibitem [{\citenamefont {Vidberg}\ and\ \citenamefont {Serene}(1977)}]{Vidberg1977AnalyticCont}%
  \BibitemOpen
  \bibfield  {author} {\bibinfo {author} {\bibfnamefont {H.~J.}\ \bibnamefont {Vidberg}}\ and\ \bibinfo {author} {\bibfnamefont {J.~W.}\ \bibnamefont {Serene}},\ }\bibfield  {title} {\bibinfo {title} {{Solving the Eliashberg equations by means of $N$-point Pad{\ifmmode\acute{e}\else\'{e}\fi} approximants}},\ }\href {https://doi.org/10.1007/BF00655090} {\bibfield  {journal} {\bibinfo  {journal} {J. Low Temp. Phys.}\ }\textbf {\bibinfo {volume} {29}},\ \bibinfo {pages} {179} (\bibinfo {year} {1977})}\BibitemShut {NoStop}%
\bibitem [{\citenamefont {Linder}\ and\ \citenamefont {Balatsky}(2019)}]{Linder2019Oddw}%
  \BibitemOpen
  \bibfield  {author} {\bibinfo {author} {\bibfnamefont {J.}~\bibnamefont {Linder}}\ and\ \bibinfo {author} {\bibfnamefont {A.~V.}\ \bibnamefont {Balatsky}},\ }\bibfield  {title} {\bibinfo {title} {{Odd-frequency superconductivity}},\ }\href {https://doi.org/10.1103/RevModPhys.91.045005} {\bibfield  {journal} {\bibinfo  {journal} {Rev. Mod. Phys.}\ }\textbf {\bibinfo {volume} {91}},\ \bibinfo {pages} {045005} (\bibinfo {year} {2019})}\BibitemShut {NoStop}%
\bibitem [{\citenamefont {Sigrist}\ and\ \citenamefont {Ueda}(1991)}]{Sigrist}%
  \BibitemOpen
  \bibfield  {author} {\bibinfo {author} {\bibfnamefont {M.}~\bibnamefont {Sigrist}}\ and\ \bibinfo {author} {\bibfnamefont {K.}~\bibnamefont {Ueda}},\ }\bibfield  {title} {\bibinfo {title} {Phenomenological theory of unconventional superconductivity},\ }\href {https://doi.org/10.1103/RevModPhys.63.239} {\bibfield  {journal} {\bibinfo  {journal} {Rev. Mod. Phys.}\ }\textbf {\bibinfo {volume} {63}},\ \bibinfo {pages} {239} (\bibinfo {year} {1991})}\BibitemShut {NoStop}%
\bibitem [{\citenamefont {Allen}\ and\ \citenamefont {Dynes}(1975)}]{Allen1975Aug}%
  \BibitemOpen
  \bibfield  {author} {\bibinfo {author} {\bibfnamefont {P.~B.}\ \bibnamefont {Allen}}\ and\ \bibinfo {author} {\bibfnamefont {R.~C.}\ \bibnamefont {Dynes}},\ }\bibfield  {title} {\bibinfo {title} {{Transition temperature of strong-coupled superconductors reanalyzed}},\ }\href {https://doi.org/10.1103/PhysRevB.12.905} {\bibfield  {journal} {\bibinfo  {journal} {Phys. Rev. B}\ }\textbf {\bibinfo {volume} {12}},\ \bibinfo {pages} {905} (\bibinfo {year} {1975})}\BibitemShut {NoStop}%
\bibitem [{\citenamefont {Cui}\ \emph {et~al.}(2023)\citenamefont {Cui}, \citenamefont {Zeng}, \citenamefont {Cui}, \citenamefont {Yu},\ and\ \citenamefont {Yang}}]{Cui2023AMmagnonmodel}%
  \BibitemOpen
  \bibfield  {author} {\bibinfo {author} {\bibfnamefont {Q.}~\bibnamefont {Cui}}, \bibinfo {author} {\bibfnamefont {B.}~\bibnamefont {Zeng}}, \bibinfo {author} {\bibfnamefont {P.}~\bibnamefont {Cui}}, \bibinfo {author} {\bibfnamefont {T.}~\bibnamefont {Yu}},\ and\ \bibinfo {author} {\bibfnamefont {H.}~\bibnamefont {Yang}},\ }\bibfield  {title} {\bibinfo {title} {{Efficient spin Seebeck and spin Nernst effects of magnons in altermagnets}},\ }\href {https://doi.org/10.1103/PhysRevB.108.L180401} {\bibfield  {journal} {\bibinfo  {journal} {Phys. Rev. B}\ }\textbf {\bibinfo {volume} {108}},\ \bibinfo {pages} {L180401} (\bibinfo {year} {2023})}\BibitemShut {NoStop}%
\bibitem [{\citenamefont {Schrieffer}\ and\ \citenamefont {Wolff}(1966)}]{Schrieffer1966Wolff}%
  \BibitemOpen
  \bibfield  {author} {\bibinfo {author} {\bibfnamefont {J.~R.}\ \bibnamefont {Schrieffer}}\ and\ \bibinfo {author} {\bibfnamefont {P.~A.}\ \bibnamefont {Wolff}},\ }\bibfield  {title} {\bibinfo {title} {{Relation between the Anderson and Kondo Hamiltonians}},\ }\href {https://doi.org/10.1103/PhysRev.149.491} {\bibfield  {journal} {\bibinfo  {journal} {Phys. Rev.}\ }\textbf {\bibinfo {volume} {149}},\ \bibinfo {pages} {491} (\bibinfo {year} {1966})}\BibitemShut {NoStop}%
\bibitem [{\citenamefont {Aperis}\ \emph {et~al.}(2015)\citenamefont {Aperis}, \citenamefont {Maldonado},\ and\ \citenamefont {Oppeneer}}]{Aperis2015magneticField}%
  \BibitemOpen
  \bibfield  {author} {\bibinfo {author} {\bibfnamefont {A.}~\bibnamefont {Aperis}}, \bibinfo {author} {\bibfnamefont {P.}~\bibnamefont {Maldonado}},\ and\ \bibinfo {author} {\bibfnamefont {P.~M.}\ \bibnamefont {Oppeneer}},\ }\bibfield  {title} {\bibinfo {title} {{Ab initio theory of magnetic-field-induced odd-frequency two-band superconductivity in ${\mathrm{MgB}}_{2}$}},\ }\href {https://doi.org/10.1103/PhysRevB.92.054516} {\bibfield  {journal} {\bibinfo  {journal} {Phys. Rev. B}\ }\textbf {\bibinfo {volume} {92}},\ \bibinfo {pages} {054516} (\bibinfo {year} {2015})}\BibitemShut {NoStop}%
\bibitem [{\citenamefont {Schnell}\ \emph {et~al.}(2006)\citenamefont {Schnell}, \citenamefont {Mazin},\ and\ \citenamefont {Liu}}]{Schnell2006dwavephonon}%
  \BibitemOpen
  \bibfield  {author} {\bibinfo {author} {\bibfnamefont {I.}~\bibnamefont {Schnell}}, \bibinfo {author} {\bibfnamefont {I.~I.}\ \bibnamefont {Mazin}},\ and\ \bibinfo {author} {\bibfnamefont {A.~Y.}\ \bibnamefont {Liu}},\ }\bibfield  {title} {\bibinfo {title} {{Unconventional superconducting pairing symmetry induced by phonons}},\ }\href {https://doi.org/10.1103/PhysRevB.74.184503} {\bibfield  {journal} {\bibinfo  {journal} {Phys. Rev. B}\ }\textbf {\bibinfo {volume} {74}},\ \bibinfo {pages} {184503} (\bibinfo {year} {2006})}\BibitemShut {NoStop}%
\bibitem [{\citenamefont {Schrodi}\ \emph {et~al.}(2021)\citenamefont {Schrodi}, \citenamefont {Oppeneer},\ and\ \citenamefont {Aperis}}]{Schrodi2021dwavephonon}%
  \BibitemOpen
  \bibfield  {author} {\bibinfo {author} {\bibfnamefont {F.}~\bibnamefont {Schrodi}}, \bibinfo {author} {\bibfnamefont {P.~M.}\ \bibnamefont {Oppeneer}},\ and\ \bibinfo {author} {\bibfnamefont {A.}~\bibnamefont {Aperis}},\ }\bibfield  {title} {\bibinfo {title} {{Unconventional superconductivity mediated solely by isotropic electron-phonon interaction}},\ }\href {https://doi.org/10.1103/PhysRevB.104.L140506} {\bibfield  {journal} {\bibinfo  {journal} {Phys. Rev. B}\ }\textbf {\bibinfo {volume} {104}},\ \bibinfo {pages} {L140506} (\bibinfo {year} {2021})}\BibitemShut {NoStop}%
\bibitem [{\citenamefont {Skomski}(2003)}]{Skomski2003anisotropySOC}%
  \BibitemOpen
  \bibfield  {author} {\bibinfo {author} {\bibfnamefont {R.}~\bibnamefont {Skomski}},\ }\bibfield  {title} {\bibinfo {title} {{Nanomagnetics}},\ }\href {https://doi.org/10.1088/0953-8984/15/20/202} {\bibfield  {journal} {\bibinfo  {journal} {J. Phys.: Condens. Matter}\ }\textbf {\bibinfo {volume} {15}},\ \bibinfo {pages} {R841} (\bibinfo {year} {2003})}\BibitemShut {NoStop}%
\bibitem [{\citenamefont {Kamra}\ \emph {et~al.}(2020)\citenamefont {Kamra}, \citenamefont {Belzig},\ and\ \citenamefont {Brataas}}]{Kamra2020MagnonSqueezingRev}%
  \BibitemOpen
  \bibfield  {author} {\bibinfo {author} {\bibfnamefont {A.}~\bibnamefont {Kamra}}, \bibinfo {author} {\bibfnamefont {W.}~\bibnamefont {Belzig}},\ and\ \bibinfo {author} {\bibfnamefont {A.}~\bibnamefont {Brataas}},\ }\bibfield  {title} {\bibinfo {title} {{Magnon-squeezing as a niche of quantum magnonics}},\ }\href {https://doi.org/10.1063/5.0021099} {\bibfield  {journal} {\bibinfo  {journal} {Appl. Phys. Lett.}\ }\textbf {\bibinfo {volume} {117}},\ \bibinfo {pages} {090501} (\bibinfo {year} {2020})}\BibitemShut {NoStop}%
\bibitem [{\citenamefont {Wick}(1950)}]{Wick1950Oct}%
  \BibitemOpen
  \bibfield  {author} {\bibinfo {author} {\bibfnamefont {G.~C.}\ \bibnamefont {Wick}},\ }\bibfield  {title} {\bibinfo {title} {{The Evaluation of the Collision Matrix}},\ }\href {https://doi.org/10.1103/PhysRev.80.268} {\bibfield  {journal} {\bibinfo  {journal} {Phys. Rev.}\ }\textbf {\bibinfo {volume} {80}},\ \bibinfo {pages} {268} (\bibinfo {year} {1950})}\BibitemShut {NoStop}%
\end{thebibliography}%

\end{document}